\documentclass[english]{article}
\usepackage[a4paper]{geometry}

\usepackage[T1]{fontenc}
\usepackage[latin9]{inputenc}
\usepackage{babel}

\usepackage{color}
\usepackage{amsmath}
\usepackage{amssymb}
\usepackage{graphicx}
\usepackage{wasysym}
\usepackage{subfig}
\usepackage{dsfont}

\usepackage{authblk}
\usepackage{url}
\usepackage[ colorlinks=true
  ,urlcolor=blue
  ,anchorcolor=blue
  ,citecolor=blue
  ,filecolor=blue
  ,linkcolor=blue
  ,menucolor=blue
  ,pagecolor=blue
  ,linktocpage=true
  ,pdfproducer=medialab
  ,pdfa=true]{hyperref}

\makeatletter

\title{\bfseries The complete trans-series for conserved charges in integrable field theories}

\author[1]{Zolt\'an Bajnok}
\author[1]{J\'anos Balog}
\author[1,2]{Istv\'an Vona}

\affil[1]{\small HUN-REN Wigner Research Centre for Physics\\ Konkoly-Thege Mikl\'os u. 29-33, 1121 Budapest\\ Hungary}
\affil[2]{\small Roland E\"otv\"os University\\ P\'azm\'any P\'eter s\'et\'any 1/A, 1117 Budapest\\ Hungary}
\affil[ ]{\vskip 1pt \textit {bajnok.zoltan@wigner.hun-ren.hu\\ balog.janos@wigner.hun-ren.hu\\ vona.istvan@wigner.hun-ren.hu}}

\begin{document}

\maketitle

\begin{abstract}
We analyze the vacuum expectation values of conserved charges in two
dimensional integrable theories. We study the situations when the
ground-state can be described by a single integral equation with a
finite support: the thermodynamic limit of the Bethe ansatz equation.
We solve this integral equation by expanding around the infinite support
limit and write the expectation values in terms of an explicitly calculable
trans-series, which includes both perturbative and all non-perturbative
corrections. These different types of corrections are interrelated
via resurgence relations, which we all reveal. We provide explicit
formulas for a wide class of bosonic and fermionic models including
the $O(N)$ (super) symmetric nonlinear sigma and Gross-Neveu, the
$SU(N)$ invariant principal chiral and chiral Gross-Neveu models
along with the Lieb-Liniger and Gaudin-Yang models and the case of
the disk capacitor. With numerical analyses we demonstrate that the
laterally Borel resummed trans-series is convergent and reproduces
the physical result.
\end{abstract}

\tableofcontents

\section{Introduction}

Asymptotically free quantum field theories, including quantum chromodynamics,
suffer the asymptotic nature of their perturbative expansions \cite{Dyson:1952tj,Hurst:1952zh,Lipatov:1976ny}.
The asymptotically growing coefficients signal non-perturbative corrections,
which can originate from instantons or renormalons \cite{Coleman:1978ae,Brezin:1976wa,Beneke:1998ui,Bauer:2011ws}.
The instantons are related to semiclassical saddle points of the path
integral, while the renormalons do not have such an interpretation\footnote{It is conjectured that \lq bions' \cite{Dunne:2015ywa,Fujimori:2018kqp}
explain IR renormalons semi-classically, but they need a twisted
compactification of the model.
Recently there was an attempt to relate IR renormalons to the saddle points
of the quantum effective action \cite{Bhattacharya:2024hhh}.}. The physical observable is a trans-series, i.e. a double series in
the non-perturbative correctons multiplied by perturbative series.
In case of instantons it can be interpreted as the evaluation of the
path integral, when we sum over all multiinstanton saddles multiplied
with the expansion around each. Although, there is no similar picture
for renormalons, but it is expected that the trans-series, once resummed
using lateral Borel resummations, reproduce the physical value.
There are not many examples of explicitly computed transseries in asymptotically free quantum field theories.
These are mostly based on integrability or large N-expansion \cite{DiPietro:2021yxb,Nishimura:2021lno, Marino:2021six}.
Here
we would like to present a family of models, where such a solution
can be achieved. These results are the culmination of activities in
two dimensional integrable models in the last few years. 

Two dimensional integrable quantum field theories are useful toy models
of particle physics, where non-perturbative effects and strongly interacting
phenomena can be tested in simplified circumstances. The nonlinear
$O(N)$ sigma models, the various Gross-Neveu models, the principal
chiral models are similar to QCD in the sense that they are asymptotically
free in perturbation theory and exhibit a dynamically generated scale
at the quantum level \cite{Gross:1974jv,Polyakov:1975rr,Polyakov:1983tt,Hasenfratz:1990zz,Hasenfratz:1990ab,Forgacs:1991rs,Forgacs:1991ru,Forgacs:1991nk,Balog:1992cm,Hollowood:1994np,Evans:1996ah,Fateev:1994dp,Fateev:1994ai,Fendley:2000bw,Fendley:2001hc},
and see also the review \cite{Evans:1995dn}.  Their statistical physical
counterparts the Lieb-Liniger \cite{Lieb:1963rt} and Gaudin-Yang
models \cite{gaudin1967systeme,yang1967some} are paradigmic examples
where ideas and methods were developed the first time: their groundstate
energy was determined first in \cite{Lieb:1963rt} by analysing the
thermodynamic limit of the Bethe ansatz. The TBA method for relativistic
models was studied first in \cite{zamolodchikov1990thermodynamic}.
Some of these models appear also in real condensed matter systems,
such as cold atom experiments or physically realized in highly anisotropic
materials. Developing an all order trans-series weak coupling expansion
for their observables is of central interest for many communities. 

Integrable quantum field theories are described in terms of their
particle spectrum and scattering matrices. There are no particle production
in scatterings and multiparticle processes factorize into two particle
ones \cite{Zamolodchikov:1977nu}. In the simplest theories we have
only one particle type, which scatter on itself with a single function,
which is a phase. This phase can be used to formulate momentum quantization
in a finite volume, which is called the Bethe ansatz (or Bethe Yang)
equation. In the groundstate, momenta are totally filled below the
Fermi surface. The thermodynamic limit of the Bethe ansatz (TBA) leads
to an integral equation for the momentum density of the particles,
from which the groundstate energy density and the density of particles
can be calculated \cite{Lieb:1963rt,Samaj:2013yva}. In more complicated
systems with many particles and non-diagonal scatterings the ground-state
Bethe ansatz equations are much more involved. In many cases, however,
one can apply an external field coupled to a conserved charge, to
force only one (or few) types of particles to condense into the vacuum,
allowing the simplified analysis above \cite{Polyakov:1983tt,Hasenfratz:1990ab,Hasenfratz:1990zz,Forgacs:1991rs,Forgacs:1991nk,
Balog:1992cm, Hollowood:1994np,Evans:1994sv,Evans:1994sy,Evans:1995dn}.
In this simplified situation the main focus is the expansion of the
observables as functions of the Fermi surface. The expansion for small
Fermi surface is straightforward and convergent. The most interesting
case is at large Fermi surface as it corresponds to large external
fields, where asymptotic freedom can be exploited perturbatively and
the mass gap can be related to the dynamically generated scale. This
expansion, however, is only asymptotic and one has to supplement it
with non-perturbative correction and to build a trans-series eventually.
The aim of our present paper is to provide the complete trans-series
for various observables in this situation.

In the last five years there has been extensive activity and great
progress in the expansion of the linear TBA equations at weak coupling.
They were based on the pioneering result of Volin, who managed to
expand the integral equation in the $O(N)$ nonlinear sigma models
systematically \cite{Volin:2009wr,Volin:2010cq}. This method was
generalized for statistical models and for the circular plate capacitor
\cite{Marino:2019fuy,Reichert:2020ymc} and to integrable relativistic
quantum field theories \cite{Marino:2019eym}. Having enough perturbative
terms one can exploit resurgence theory \cite{Dingle1973:asy,Ecalle1981:res,Marino:2012zq,Dorigoni:2014hea,Dunne:2015eaa,Aniceto:2018bis,Serone:2024uwz}
to extract the leading (and a few subleading) non-perturbative corrections.
These were done on the statistical physics side for the Lieb-Liniger,
Gaudin-Yang and Hubbard models together with their generalizations
\cite{Marino:2019fuy,Marino:2019wra,Marino:2020dgc,Marino:2020ggm,PhysRevA.106.062216,PhysRevLett.130.020401}.
On the particle physics side the $O(N)$ non-linear sigma model, its
supersymmetric extension, the Gross-Neveu model and the principal
chiral models were analysed in \cite{Marino:2019eym,Abbott:2020mba,Abbott:2020qnl,Bajnok:2021dri,Bajnok:2021zjm,Bajnok:2022rtu}.
The origin of the obtained non-perturbative corrections were identified
as instantons and renormalons. These findings were further confirmed
by large $N$ calculations \cite{Marino:2021six,DiPietro:2021yxb}
and in certain sigma models by introducing a $\theta$-term \cite{Marino:2022ykm}.
A more systematic treatment to determine all non-perturbative correction
was initiated in \cite{Marino:2021dzn}, which identified the location
of renormalons, confronting with previous expectations. These analyses
were extended in \cite{Bajnok:2022xgx} for bosonic models, which
described all non-perturbative corrections in terms of a perturbatively
defined basis. The convergence of the trans-series was investigated
in \cite{Marino:2023epd} based on the leading perturbative term at
each non-perturbative order. Recent reviews on these subjects are
\cite{Reis:2022tni,Serone:2024uwz}. 

The aim of the present paper is to extend and complete these previous
investigations in various ways as follows: In section 2 we recall
how the thermodynamic limit of the Bethe ansatz equation leads to
a linear integral equation for the momentum density of the particles
and how it is related to the free energy density in an external field.
We then generalize the integral equation to incorporate higher spin
charges in two different ways: in the source terms corresponding to
generalized space-time evolutions as well as in the observables, which
are the expectation values of higher spin charges. These generalized
observables are not all independent and we summarize the various relations,
including differential equations between them. We close the section
by listing the various models we investigate with the explicit kernels
of the corresponding integral equations. In section 3 we reformulate
the integral equation based on the Wiener-Hopf method and solve it
in terms of a trans-series. Each non-perturbative correction is written
in terms of a perturbatively defined basis. The elements of the basis
are the perturbative parts of the generalized observables and are
related to each other by differential equations. The generalization
for higher order poles in the Wiener Hopf kernel is relegated to Appendix
\ref{sec:higher poles}. In section 4 we provide a graphical interpretation
of the non-perturbative corrections in terms of lattice paths and
investigate the resurgence relations between the various terms. We
also construct an infinite parameter trans-series and show that the
Stokes automorphism acts as shifts in these parameters. Section 5
contains the derivation of the explicit expressions for bosonic and
fermionic models, which are placed in Appendix \ref{sec:Coefficients}.
In section 6 we recall the many checks, which were already done to
test various parts of the trans-series solution. We also extend the
previous studies with the concrete analysis of the supersymmetric
nonlinear $O(7)$ sigma model. This model brings two new features:
the Stokes constants have real parts and the cuts on the Borel plane
are not logarithmic. In section 7 we investigate the convergence properties
of the trans-series numerically. We go beyond \cite{Marino:2023epd}
by analysing the convergence properties of the laterally Borel resummed
trans-series terms. In order to make contact with Lagrangian perturbation
theory, we provide formulas for the free energy density in the perturbative
running coupling in section 8. Details of the calculations are presented
in Appendix \ref{sec:Falpha}. Finally, we conclude in section 9 and
provide an outlook. 

\section{Groundstate energy densities in integrable models}

We analyze the groundstate energy density of integrable many-body
systems. We assume that the groundstate is formed by the condensation
of a single particle type. Multi-particle scatterings are factorized
into two particle scatterings with an explicitly known scattering
matrix, which satisfies the unitarity and crossing symmetry relations.
We assume that the scattering matrix is of a difference form. In the
non-relativistic setting this is the difference of the momenta, while
in the relativistic case the difference of the rapidities. 

In deriving the ground-state energy density we put $N$ particles
in a finite (but large) volume $L$. Demanding periodicity of the
wave function leads to the Bethe ansatz equation \cite{Lieb:1963rt}
whose logarithm takes the form 
\begin{equation}
p(\theta_{j})L-i\sum_{k:k\neq j}^{N}\log S(\theta_{j}-\theta_{k})=2\pi n_{j}
\end{equation}
where $\theta$ is a rapidity-like variable, in which the scattering
is of a difference form. The corresponding energy is the sum of one-particle
energies $E=\sum_{k}e(\theta_{k})$ and $e(\theta)$ follows from
the dispersion relation. We analyze systems in which only one particle
can occupy a given state, i.e. all $n_{j}$ -s are different and in
the groundstate they are completely filled between $-M$ and $M$.
In the thermodynamic limit, when $N\to\infty$ and $L\to\infty$,
one can introduce the rapidity density of particles $\chi(\theta)$
such that $\frac{L}{2\pi}\chi(\theta)d\theta$ is the number of particle
rapidities in the interval $[\theta,\theta+d\theta]$. In this limit
the Bethe ansatz equation leads to a linear integral equation for
the rapidity density
\begin{equation}
\chi(\theta)-\int_{-B}^{B}d\theta'K(\theta-\theta')\chi(\theta')=r(\theta)\quad;\quad\vert\theta\vert\leq B,\label{eq:TBA}
\end{equation}
where 
\begin{equation}
r(\theta)=\frac{dp(\theta)}{d\theta}\quad;\quad K(\theta)=-\frac{i}{2\pi}\frac{d\log S(\theta)}{d\theta}
\end{equation}
and $B$ is the analogue of the Fermi rapidity. The density $\rho=\lim_{L\to\infty}\frac{N}{L}$
and the groundstate energy density $\epsilon=\lim_{L\to\infty}\frac{E}{L}$
can be written as 
\begin{equation}
\rho=\int_{-B}^{B}\frac{d\theta}{2\pi}\chi(\theta)\quad;\qquad\epsilon=\int_{-B}^{B}\frac{d\theta}{2\pi}e(\theta)\chi(\theta)
\end{equation}
The aim of our paper is to develop a systematic and explicit large
$B$ expansion of these quantities, including all perturbative $B^{-1}$
, ($\ln B$) and non-perturbative $e^{-B}$ corrections. 

Thermodynamically, a more natural parameter is the density $\rho$,
rather than $B$ and by inverting the first relation, $B(\rho)$,
we can express the energy density in terms of $\rho$ as $\epsilon(\rho$).
In many cases the condensation of the particles is ensured by introducing
a large enough external field $h$ coupled to one of the conserved
global charges. Then the density is determined by minimising the free
energy density wrt. to the density 
\begin{equation}
{\cal F}(h)=\min_{\rho}(\epsilon(\rho)-h\rho)\label{eq:Fh}
\end{equation}
This relates the external field to the density as $h=\frac{d\epsilon(\rho)}{d\rho}$.
The free energy density ${\cal F}(h)$ and energy density $\epsilon(\rho)$
are related by Legendre transform. 

This powerful technique was initiated in \cite{japaridze1984exact,wiegmann1985exact}
and further elaborated in \cite{Hasenfratz:1990ab,Hasenfratz:1990zz,Forgacs:1991nk,Forgacs:1991rs,Balog:1992cm,Evans:1994sv,Evans:1994sy,Evans:1995dn}
mainly to relate the mass of the scattering particles to the dynamically
generated scale in perturbation theory to support the identification
between the scattering matrix and the Lagrangian description of the
various models.

In the following we generalize these observables and elaborate on
their relations. 

\subsection{Physical observables and their relations}

In solving Bethe ansatz systems in the thermodynamic limit we arrive
at a linear integral equation of the form (\ref{eq:TBA}) where the
kernel is a symmetric function, which is related to the logarithmic
derivative of the scattering matrix. The unknow function $\chi(\theta)$
is the rapidity density of the particles in the ground-state being
non-zero only below the Fermi rapidity, which is denoted by $B$.
This rapidity density depends also on $B$, but we suppress this (and
later any other) $B$-dependence in the notation. The various source
terms $r(\theta)$ correspond to different situations. In the relativistic
setting these sources take the form $\cosh(\alpha\theta)$ or $\sinh(\alpha\theta)$,
while in the non-relativistic case they are rational functions, which
can be expanded in powers of $\theta$. They are related to the expectation
values of higher spin charges \cite{Castro-Alvaredo:2016cdj,Bajnok:2019mpp}. 

In most of the paper we analyze the integral equation with the source
terms 
\begin{equation}
\chi_{\alpha}(\theta)-\int_{-B}^{B}d\theta'K(\theta-\theta')\chi_{\alpha}(\theta')=r_{\alpha}(\theta)=\cosh(\alpha\theta)\label{eq:TBAalpha}
\end{equation}
where $\alpha$ is a non-negative real number. The solution is a symmetric
function $\chi_{\alpha}(\theta)=\chi_{\alpha}(-\theta)$ . We will
show shortly that the case of the source $\bar{r}_{\alpha}(\theta)=\sinh(\alpha\theta)$
and the corresponding $\bar{\chi}_{\alpha}(\theta)$ can be recovered
easily from $\chi_{\alpha}(\theta)$. The rapidity density $\bar{\chi}(\theta)$
is anti-symmetric $\bar{\chi}_{\alpha}(\theta)=-\bar{\chi}_{\alpha}(-\theta)$.
The boundary values of the rapidity densities $\chi_{\alpha}(B)\equiv\chi_{\alpha}$
will play a special role in the following as both $\rho$ and $\epsilon$
can be expressed in terms of them. 

By generalizing the calculation in \cite{PhysRevLett.130.020401,PhysRevA.106.062216}
one can differentiate the integral equation twice and derive 
\begin{equation}
\left(\partial_{\theta}^{2}-\partial_{B}^{2}+2\frac{\dot{\chi}_{\alpha}}{\chi_{\alpha}}\partial_{B}-\alpha^{2}\right)\chi_{\alpha}(\theta)=0\label{eq:diffeqchin}
\end{equation}
together with the same equation for $\bar{\chi}_{\alpha}$. Here and
from now on we denote differentiation wrt. $B$ by a dot, i.e. $\dot{\chi}_{\alpha}=\frac{d\chi_{\alpha}(B)}{dB}$. 

The observables appearing in the physical applications are the ``moments''
of the rapidity densities 
\begin{equation}
{\cal O}_{\alpha,\beta}=\int_{-B}^{B}\frac{d\theta}{2\pi}\chi_{\alpha}(\theta)r_{\beta}(\theta)\label{eq:Oab}
\end{equation}
which are related to the Fourier transform of the rapidity density
$\chi_{\alpha}(\theta)$ as 
\begin{equation}
{\cal O}_{\alpha,\beta}=\frac{1}{2\pi}\int_{-\infty}^{\infty}d\theta\,\chi_{\alpha}(\theta)e^{i(i\beta)\theta}=\frac{1}{2\pi}\tilde{\chi}_{\alpha}(i\beta)
\end{equation}
These observables are functions of $B$ and are symmetric in $\alpha,\beta$,
see \cite{Bajnok:2022rtu}. By differentiating the observables we
can relate them to the boundary values of the rapidity densities
\begin{equation}
\dot{{\cal O}}_{\alpha,\beta}=\frac{1}{\pi}\chi_{\alpha}\chi_{\beta}\quad.\label{eq:Onmdot}
\end{equation}
 By taking the moments of the differential equation (\ref{eq:diffeqchin})
we can also obtain 
\begin{equation}
\ddot{{\cal O}}_{\alpha,\beta}-2\frac{\dot{\chi}_{\alpha}}{\chi_{\alpha}}\dot{{\cal O}}_{\alpha,\beta}+(\alpha^{2}-\beta^{2}){\cal O}_{\alpha,\beta}=0
\end{equation}
which, when combined with eq. (\ref{eq:Onmdot}) leads to 
\begin{equation}
\pi(\alpha^{2}-\beta^{2}){\cal O}_{\alpha,\beta}=\chi_{\beta}\dot{\chi}_{\alpha}-\chi_{\alpha}\dot{\chi}_{\beta}
\end{equation}
These imply that the combination 
\begin{equation}
\frac{\ddot{\chi}_{\alpha}}{\chi_{\alpha}}-\alpha^{2}=\frac{\ddot{\chi}_{\beta}}{\chi_{\beta}}-\beta^{2}=F\quad,\label{eq:chidot}
\end{equation}
does not depend either on $\alpha$ or $\beta$. We denote this combination
by $F$, which is a non-trivial function of $B$. Observe now that
by integrating the differential equations (\ref{eq:Onmdot},\ref{eq:chidot})
we can express all observables $\chi_{\alpha}$ and ${\cal O}_{\alpha,\beta}$
up to some integration constants in terms of $F$ which can be calculated
from any $\chi_{\alpha}$ say from $\chi_{0}$ or $\chi_{1}$. 

We note that all these equations are also true for the variables with
a bar. Even more, we can express the bar variables from the ones without
a bar. In doing so we compare the first order derivatives of the original
integral equation with the analogous one with bars. After taking moments
one can derive

\begin{equation}
\beta{\cal O}_{\alpha,\beta}+\alpha\bar{{\cal O}}_{\alpha,\beta}=\frac{1}{\pi}\chi_{\alpha}\bar{\chi}_{\beta}
\end{equation}
This allows to express the variables with bars as 
\begin{equation}
\bar{\chi}_{\alpha}=\frac{\alpha\pi{\cal O}_{0,\alpha}}{\chi_{0}}\quad;\qquad\bar{{\cal O}}_{\alpha,\beta}=\frac{\beta}{\alpha}\left(\frac{\chi_{\alpha}}{\chi_{0}}{\cal O}_{0,\beta}-{\cal O}_{\alpha,\beta}\right)\label{eq:barfromnobar}
\end{equation}

Let us mention finally, that in one of the typical applications the
integral equation describes the ground-state of an integrable relativistic
quantum field theory in an external field. The density and the energy
density in the ground-state are given by the observables 
\begin{equation}
\rho=m{\cal O}_{1,0}\quad;\quad\epsilon=m^{2}{\cal O}_{1,1}
\end{equation}
where $m$ is the mass of the scattering particles. The external field
can be obtained by minimising the free energy $m{\cal O}_{1,1}-h{\cal O}_{1,0}$,
which leads to 
\begin{equation}
h=m\frac{\chi_{1}}{\chi_{0}}\label{eq:hchi0overchi0}
\end{equation}
The free energy density turns out to be the observable with the bar:
\begin{equation}
{\cal F}=\epsilon-h\rho=m^{2}({\cal O}_{1,1}-\frac{\chi_{1}}{\chi_{0}}{\cal O}_{1,0})=-m^{2}\bar{{\cal O}}_{1,1}\label{eq:FhbarO}
\end{equation}

In the non-relativistic applications we also need moments of the $\chi_{0}(\theta)$
problem of the form 
\begin{equation}
{\cal O}_{0}^{(2k)}=\int_{-B}^{B}\frac{d\theta}{2\pi}\chi_{0}(\theta)\theta^{2k}=\left.\frac{d^{2k}{\cal O}_{0,\alpha}}{d\alpha^{2k}}\right|_{\alpha=0}\label{eq:nonrelobs}
\end{equation}
In the following we list both non-relativistic and relativistic models
which manifest the above setting and observables. 

\subsection{Non-relativistic models}

In the non-relativistic setting the scattering matrix depends on the
difference of the particles' momenta. We provide a non-exhaustive
list which leads to the settings above. 
\begin{itemize}
\item \textbf{Lieb-Liniger model.} This model is the simplest integrable
model consisting of bosonic spin-less particles which interact with
each other through a $\delta$-function interaction \cite{Lieb:1963rt}.
The scattering matrix depends on the difference of momenta and the
kernel takes the form
\begin{equation}
K(k)=\frac{2c}{k^{2}+c^{2}}
\end{equation}
 where $c$ is the strength of the interaction. The energy follows
from the non-relativistic dispersion $e(k)=\frac{k^{2}}{2}$. The
observables of the model are the density ${\cal O}_{0,0}$ and the
energy density ${\cal O}_{0}^{(2)}$ with the $r_{0}=1$ source term.
In the practical applications the observables $\gamma$ and $h(\gamma)$
are used instead, which are defined by \cite{Reis:2022tni}
\begin{equation}
\gamma=\frac{c}{\rho}=\frac{\pi}{{\cal O}_{0,0}},\qquad\qquad h(\gamma)=\frac{2m\varepsilon(\rho)}{\rho^{3}}=\frac{{\cal O}_{0}^{(2)}}{{\cal O}_{0,0}^{3}}.
\end{equation}
\item \textbf{Gaudin-Yang model. }This model describes the $\delta$-function
interaction of spin $\frac{1}{2}$ fermions \cite{gaudin1967systeme,yang1967some}.
Due to the inner degree of freedom the calculation of the ground-state
energy density requires to use the nested Bethe ansatz technique.
Nevertheless, the integral equations can be put into a one-component
form with the kernel \cite{Reis:2022tni}
\begin{equation}
K(k)=-\frac{2c}{k^{2}+c^{2}}
\end{equation}
Observe that this kernel has opposite sign compared to the Lieb-Liniger
model and behaves differently for $k=0$. The physical observables
are derived from the density and energy density with an $r_{0}=1$
source and can be written in terms of the general observables as
\begin{equation}
\gamma=\frac{c}{\rho}=\frac{\pi}{4\mathcal{O}_{0,0}},\qquad\qquad h(\gamma)=\frac{2m\varepsilon(\rho)}{\rho^{3}}=-\frac{\pi^{2}}{64\mathcal{O}_{0,0}^{2}}+\frac{1}{16}\,\frac{\mathcal{O}_{0}^{(2)}}{\mathcal{O}_{0,0}^{3}}.
\end{equation}
\item \textbf{Disk capacitor}. There is an interesting classical electrodynamical
problem, which leads to the very same integral equation. The history
and early results of this interesting problem is summarized in \cite{Samaj:2013yva}.
Consider a coaxial disk capacitor of radii $a$ and distance $d$,
which is charged either with the same or with opposite charges of
magnitude $Q$. The calculation of the capacity leads to Love equation
\cite{Love} with kernels 
\begin{equation}
K(k)=\pm\frac{2}{k^{2}+1}
\end{equation}
where the plus sign corresponds to the opposite ($+$), while the
minus to the same charge case ($-$). Note that these correspond to
the kernel of the Lieb-Liniger and Gaudin-Yang model respectively.
The capacity (in Gaussian units) in both cases is proportional to
the density with the $r_{0}=1$ source as
\begin{equation}
C^{(+)}=\frac{d}{\pi}{\cal O}_{0,0}\quad;\qquad C^{(-)}=\frac{4d}{\pi}{\cal O}_{0,0}.
\end{equation}
For interested readers, we provide the trans-series of the capacity
as a function of the ratio $s=d/(2\pi a)$ proportional to the separation
of the disks 
\begin{equation}
C^{(+)}=\frac{a^{2}}{4d}\left(C_{0}^{(+)}+s\sum_{k=1}^{\infty}C_{2k}^{(+)}e^{-2k(1/s+1)}\right);\;C^{(-)}=\frac{2a}{\pi}\left(C_{0}^{(-)}+s\sum_{k=1}^{\infty}C_{k}^{(-)}e^{-k(1/s+1)}\right),\label{eq:cap}
\end{equation}
where the coefficient series $C_{k}^{(\pm)}$ may be found up to a
few orders in Appendix \ref{sec:Coefficients}. 
\item Other statistical models. There are also other statistical models,
where similar integral equation appear. The Hubbard model at half
filling and the Kondo model in a magnetic field is described by the
Gaudin-Yang kernel \cite{Reis:2022tni,Samaj:2013yva}. The source
term in the Hubbard case is more complicated, while the integration
in the Kondo case goes from $-B$ to infinity. Since our generic method
does not apply directly to these cases we do not consider these models
in our paper. 
\end{itemize}

\subsection{Relativistic models}

In the relativistic case the scattering matrix depends on the difference
of the rapidities and the dispersion relation is $e(p)=\sqrt{p^{2}+m^{2}}$,
which reads in the rapidity variable as $p(\theta)=m\sinh\theta$
and $e(\theta)=m\cosh\theta$. We analyze non-diagonal scattering
theories, in which a large enough ``external'' field can ensure
that in the groundstate only one (or a few) type of particles condense
and the above setting can be applied. We list the models and kernels
as follows. More details can be found in \cite{Reis:2022tni} and
references therein. 
\begin{itemize}
\item \textbf{O(N) non-linear sigma model.} This model is an asymptotically
free quantum field theory of an $N$-component scalar field $\boldsymbol{\phi}=(\phi_{1},\dots,\phi_{N})$
with a unit length and Lagrangian 
\begin{equation}
{\cal L}=\frac{1}{2g_{0}^{2}}\partial_{\mu}\boldsymbol{\phi}\cdot\partial^{\mu}\boldsymbol{\phi}\quad;\qquad\boldsymbol{\phi}\cdot\boldsymbol{\phi}=1
\end{equation}
where $g_{0}$ is the bare coupling. This coupling is renormalized
and running, which is parametrized by the dynamically generated scale,
$\Lambda$, see Appendix \ref{sec:Falpha} for more details. Particles
form the vector representation of $O(N)$ which scatter on themselves
on a factorized way \cite{Zamolodchikov:1977nu}. By coupling an external
(magnetic) field to one of the $O(N)$ charges say to $Q_{12}$ (rotation
symmetry in the $12$ plane) only one type of particles condense into
the vacuum \cite{Hasenfratz:1990ab,Hasenfratz:1990zz}. The integral
equation for the rapidity density then has the following kernel 
\begin{equation}
K(\theta)=\int_{-\infty}^{\infty}\frac{d\omega}{2\pi}e^{-i\omega\theta}\tilde{K}(\omega)\quad;\qquad1-\tilde{K}(\omega)=\frac{1-e^{-2\pi\Delta\vert\omega\vert}}{1+e^{-\pi\vert\omega\vert}}
\end{equation}
where $\Delta=\frac{1}{N-2}$. Let us note that for $\Delta=1$, i.e.
for the $O(3)$ model the Fourier transform of the kernel simplifies
to 
\begin{equation}
\tilde{K}(\omega)=e^{-\pi\vert\omega\vert}
\end{equation}
which is the same as in the Lieb-Liniger model and in the oppositely
charged capacitor.
\item \textbf{Supersymmetric $O(N)$ non-linear sigma model}. The $O(N)$
non-linear sigma model has a supersymmetric extension \cite{PhysRevD.16.2991},
supplemented with an $N$ component Majorana fermion field $\boldsymbol{\psi}$,
which is orthogonal to the bosonic one $\boldsymbol{\phi}\cdot\boldsymbol{\psi}=0$
. The dynamics is governed by the Lagrangian 
\begin{equation}
{\cal L}=\frac{1}{2g_{0}^{2}}\left\{ \partial_{\mu}\boldsymbol{\phi}\cdot\partial^{\mu}\boldsymbol{\phi}+i\bar{\boldsymbol{\psi}}\cdot\partial\!\negmedspace\negmedspace\slash\boldsymbol{\psi}+\frac{1}{4}(\bar{\boldsymbol{\psi}}\cdot\boldsymbol{\psi})^{2}\right\} 
\end{equation}
The global symmetry commutes with the supersymmetry and after coupling
a magnetic field to any of the conserved charges still two particle
species condense into the vacuum. Nevertheless the two coupled integral
equations can be transformed into a one-component form with the kernel
\begin{equation}
1-\tilde{K}(\omega)=\frac{\left(1+e^{-(1-2\Delta)\pi\vert\omega\vert}\right)\left(1-e^{-2\pi\Delta\vert\omega\vert}\right)}{\left(1+e^{-\pi\vert\omega\vert}\right)^{2}}
\end{equation}
with $\Delta=\frac{1}{N-2}$. 
\item \textbf{$SU(N)$ principal model}. In this model the field $U(x,t)$
takes values on the group manifold $SU(N)$ and the Lagrangian 
\begin{equation}
{\cal L}=\frac{1}{2g_{0}^{2}}{\rm Tr}(U^{-1}\partial_{\mu}UU^{-1}\partial^{\mu}U)
\end{equation}
has $su(N)\oplus su(N)$ symmetry. The spectrum consists of the fundamental
particle, which transforms wrt. the fundamental representation, and
its boundstates \cite{WIEGMANN1984173}.
By coupling an external field to a specific $su(N)$
conserved charge one can ensure the condensation of only one particle
species in the vacuum. The corresponding kernel can be extracted from
\begin{equation}
1-\tilde{K}(\omega)=\frac{(1-e^{-2\pi\Delta\vert\omega\vert})(1-e^{-2\pi(1-\Delta)\vert\omega\vert})}{1-e^{-2\pi\vert\omega\vert}}
\end{equation}
with $\Delta=\frac{1}{N}$. There could be other charge choices, which
lead to a similar structure, but we do not analyse them here. 
\item \textbf{$O(N)$ Gross-Neveu model}. The model describes the dynamics
of an $N$-component Majorana fermion $\boldsymbol{\psi}$ with Lagrangian
\cite{Gross:1974jv}
\begin{equation}
{\cal L}=\frac{1}{2g_{0}^{2}}\left\{ i\bar{\boldsymbol{\psi}}\cdot\partial\!\negmedspace\negmedspace\slash\boldsymbol{\psi}+\frac{1}{4}(\bar{\boldsymbol{\psi}}\cdot\boldsymbol{\psi})^{2}\right\} 
\end{equation}
The external field is coupled to one of the conserved $O(N)$ charges.
The corresponding kernel is simply 
\begin{equation}
1-\tilde{K}(\omega)=\frac{1+e^{-2\pi(\frac{1}{2}-\Delta)\vert\omega\vert}}{1+e^{-\pi\vert\omega\vert}}
\end{equation}
where $\Delta=\frac{1}{N-2}$. 
\item \textbf{$SU(N)$ chiral Gross-Neveu model. }This model formulates
the theory of an $N$-component complex fermion field via the Lagrangian
\cite{BERG1978125,KURAK1979289}
\begin{equation}
{\cal L}=\frac{1}{2g_{0}^{2}}\left\{ i\bar{\boldsymbol{\psi}}\cdot\partial\!\negmedspace\negmedspace\slash\boldsymbol{\psi}+\frac{1}{4}(\bar{\boldsymbol{\psi}}\cdot\boldsymbol{\psi})^{2}-\frac{1}{4}(\bar{\boldsymbol{\psi}}\cdot\gamma_{5}\boldsymbol{\psi})^{2}\right\} 
\end{equation}
The kernel can be written as 
\begin{equation}
1-\tilde{K}(\omega)=\frac{1-e^{-2\pi(1-\Delta)\vert\omega\vert}}{1-e^{-2\pi\vert\omega\vert}}
\end{equation}
where $\Delta=\frac{1}{N}$ . 
\end{itemize}
We note that in $O(N)$ symmetric models $\Delta=\frac{1}{N-2}$,
while in the $SU(N)$ symmetric ones $\Delta=\frac{1}{N}$. 

\section{Solving the integral equation \label{sec:Solving}}

In this section we provide a systematic solution of the integral equation
with the symmetric sources (\ref{eq:TBAalpha}).

\subsection{Wiener-Hopf method}

If the integral equation were defined on the whole line, we could
easily solve it by Fourier transformation, namely by inverting $1-\tilde{K}(\omega)$.
In contrast, the problem is defined only on the $[-B,B]$ interval,
thus we should use the Wiener-Hopf technique \cite{japaridze1984exact,Hasenfratz:1990ab,Hasenfratz:1990zz,Marino:2021dzn,Bajnok:2022rtu}.
The main idea is to extend the equations for the whole line, by introducing
an unknown function and then use the specific analytic properties
of the Fourier transform of a function defined on the interval, or
half line, to separate and extract the needed variables. 

In practice, we extend the integral equation for the whole line 
\begin{equation}
\chi_{\alpha}(\theta)-\int_{-\infty}^{\infty}d\theta'K(\theta-\theta')\chi_{\alpha}(\theta')=r_{\alpha}(\theta)+R(\theta)+L(\theta)
\end{equation}
with $L(\theta)=R(-\theta)$. Since $\chi_{\alpha}(\theta)$ is non-zero
only on the interval $[-B,B]$, we had to introduce an unknown function
$R(\theta)$, vanishing for $\theta<B$, to make the equation correct
for all $\theta$. Actually, we have a freedom in introducing this
function and it is technically simpler to modify slightly the source
as well $r_{\alpha}(\theta)=\Theta(-\theta+B)\frac{e^{\alpha\theta}}{2}+\Theta(\theta+B)\frac{e^{-\alpha\theta}}{2}$
, where $\Theta$ is the Heaviside step function. It is chosen outside
the interval such a way that it has a well-defined and simple Fourier
transform. We thus solve the so defined integral equation by Fourier
transform, i.e by inverting $1-\tilde{K}$. In separating $\tilde{\chi}_{\alpha}$
from $\tilde{R}$ their analytic properties are crucial, so we decompose
\begin{equation}
(1-\tilde{K}(\omega))^{-1}=G_{+}(\omega)G_{-}(\omega)\label{eq:WHdecomp}
\end{equation}
into two factors: one analytic in the lower, $G_{-}(\omega)$, and
one, $G_{+}(\omega)$, on the upper half plane. This can be done additively
by acting with the integral projectors on $\ln(1-\tilde{K}):$ 
\begin{equation}
\ln G_{\pm}(\omega)=\mp\int_{-\infty}^{\infty}\frac{d\omega'}{2\pi i}\frac{\ln(1-\tilde{K}(\omega'))}{\omega'-(\omega\pm i0)}
\end{equation}
 Since the kernel is symmetric we have $G_{-}(\omega)=G_{+}(-\omega)$.
The trick to solve the integral equation is to partially invert the
$1-\tilde{K}$ operator: 
\begin{equation}
\frac{\tilde{\chi}_{\alpha}(\omega)}{G_{+}(\omega)}e^{i\omega B}=e^{i\omega B}G_{-}(\omega)\left(\tilde{r}_{\alpha}(\omega)+\tilde{R}(\omega)+\tilde{L}(\omega)\right)
\end{equation}
and then project the equation on the upper and lower analytical pieces.
The lower analytic part results in an integral equation for the unknown
$\tilde{R}(\omega)$ only. Then using the solution for $\tilde{R}(\omega)$
we can express $\tilde{\chi}_{\alpha}(\omega)$ from the upper analytic
part of the equation, which finally provides the observables ${\cal O}_{\alpha,\beta}=\frac{1}{2\pi}\tilde{\chi}_{\alpha}(i\beta)$. 

Let us see how we can determine $\tilde{R}(\omega)$. The equation
takes a slightly simpler form for the unknown 
\begin{equation}
X_{\alpha}(\omega)=\frac{2e^{-(\alpha+i\omega)B}G_{+}(\omega)\tilde{R}(\omega)}{G_{+}(i\alpha)}+\frac{G_{+}(\omega)}{G_{+}(i\alpha)}\frac{1}{(\alpha+i\omega)}
\end{equation}
as it simplifies to 
\begin{equation}
X_{\alpha}(i\kappa)+\int_{-\infty}^{\infty}\frac{e^{2i\omega B}\sigma(\omega)X_{\alpha}(\omega)}{\kappa-i\omega}\frac{d\omega}{2\pi}=\frac{1}{\alpha-\kappa}\quad,\label{eq:intX}
\end{equation}
where we assumed that $\alpha>0$. The $\alpha=0$ case requires a
special care and deserved a separate analyses, whis was detailed in
\cite{Bajnok:2024qro}. Here we will recover the corresponding observables
by solving the differential equations (up to some integration constants,
which we fix from Volin's method). The combination $\sigma(\omega)=\frac{G_{-}(\omega)}{G_{+}(\omega)}$
is called the Wiener-Hopf kernel, which has singularities on the upper
half plane. In the cases we consider here the singularities are located
on the positive imaginary line. These singularities include a cut
and poles, whose locations $i\kappa_{l}$ we label with $\,l=1,2,\dots$. 

In order to write a closed system of equations one deforms the contour
to surround the imaginary line\footnote{We note that to achieve an analytical continuation for negative $B$
we can deform the contour to surround the negative imaginary line.
See \cite{Marino:2021dzn} for the implementation in the Gross-Neveu
case and \cite{Samaj:2013yva} for the Kondo problem. }: coming down on the left and going up on the right. Around the poles
we have to calculate a typically different half residue on the two
sides. In order to avoid this inconvenience one can deform the cut
a bit (say) left of the imaginary line and pick up only one type of
residues \cite{Marino:2021dzn,Bajnok:2022rtu}. As a result we are
left with the integral of the jump 
\begin{equation}
\delta\sigma(\kappa)=\frac{1}{2i}(\sigma(i\kappa-0)-\sigma(i\kappa+0))
\end{equation}
and the residues ${\rm -res}_{\kappa=\kappa_{l}}\sigma(i\kappa+0)e^{-2B\kappa}=d_{\kappa_{l}}$
multiplying $q_{\alpha,\kappa_{l}}=X_{\alpha}(i\kappa_{l})$ and the
explicit pole of $X_{\alpha}(i\kappa)$ at $\kappa=\alpha$ coming
with residue $\sigma(i\alpha+0)e^{-2B\alpha}=d_{\alpha}$. Altogether,
the equation takes the form 
\begin{align}
X_{\alpha}(i\kappa)+\frac{d_{\alpha}}{\kappa+\alpha}+\sum_{l=1}^{\infty}\frac{q_{\alpha,\kappa_{l}}d_{\kappa_{l}}}{\kappa+\kappa_{l}} & +\int_{C_{+}}e^{-2B\kappa'}\frac{\delta\sigma(\kappa')X_{\alpha}(i\kappa')}{\kappa+\kappa'}\frac{d\kappa'}{\pi}=\frac{1}{\alpha-\kappa}\quad,
\end{align}
In evaluating the residues we assumed that all poles $\kappa_{l}$
are distinct and $\alpha\neq\kappa_{l}$. The more general case of
coinciding and higher order poles is investigated in Appendix A. 

We are interested in the large $B$ expansion of our observable. By
rescaling the integration variable $\kappa'$ with $B$ one can see
that there are perturbative $B^{-1}$ and non-perturbative $e^{-B}$
corrections. In the typical applications one also encounters $\ln B$
terms as well. In order to avoid those, one can introduce a running
coupling $v$ \cite{Forgacs:1991rs} 
\begin{equation}
2B=v^{-1}-a\ln v+L\label{eq:runningcoupling}
\end{equation}
with some model dependent $a$ and arbitrary constant $L$. They have
to be chosen such a way that after rescaling the integration variable
as $\kappa'=vy$ the appearing kernel $e^{-y}e^{-vy(-a\ln v+L)}\delta\sigma(vy)\equiv e^{-y}{\cal A}(y)$
has a power-series expansion in $v$ without any $\ln v$ term: ${\cal A}(y)=\sum_{j=0}^{\infty}v^{j}\alpha_{j}(y)$.
In all the cases we analyze here, this can be achieved. In the rescaled
variable $Q_{\alpha}(x)=X_{\alpha}(ivx)$ the integral equation takes
the form 
\begin{align}
Q_{\alpha}(x)+\frac{d_{\alpha}}{\alpha+vx}+\sum_{l=1}^{\infty}\frac{q_{\alpha,\kappa_{l}}d_{\kappa_{l}}}{\kappa_{l}+vx} & +\int_{C_{+}}\frac{e^{-y}{\cal A}(y)Q_{\alpha}(y)}{x+y}\frac{dy}{\pi}=\frac{1}{\alpha-vx}\quad,\label{eq:Qeq}
\end{align}
where the non-perturbative corrections are encoded in $d_{\kappa}$.
The residues $q_{\alpha,\kappa_{l}}$ are also unknowns, which can
be calculated by substituting the integral equation (\ref{eq:Qeq})
at the positions $Q_{\alpha}(\frac{\kappa_{l}}{v})=q_{\alpha,\kappa_{l}}$.
This closes the system of equations. Once the variables $Q_{\alpha}(x)$
including $q_{\alpha,\kappa_{l}}$ are determined, the observable
${\cal O}_{\alpha,\beta}$ can be written ( \cite{Bajnok:2022rtu})
as
\begin{equation}
{\cal O}_{\alpha,\beta}=\frac{e^{(\alpha+\beta)B}}{4\pi}G_{+}(i\alpha)G_{+}(i\beta)W_{\alpha,\beta}\label{eq:Onm}
\end{equation}
with 
\begin{align}
W_{\alpha,\beta} & =\frac{1}{\alpha+\beta}+\sum_{l=1}^{\infty}\frac{q_{\alpha,\kappa_{l}}d_{\kappa_{l}}}{\beta-\kappa_{l}}+\frac{d_{\alpha}}{\beta-\alpha}+d_{\beta}Q_{\alpha}(\beta/v)+\frac{v}{\pi}\int_{C_{+}}\frac{e^{-x}{\cal A}(x)Q_{\alpha}(x)}{\beta-vx}dx\quad,
\end{align}
where we assumed that $\alpha,\beta\neq\kappa_{l}$ and $\alpha\neq\beta$.
The case $\alpha=\beta$ can be recovered by carefully analysing the
$\beta\to\alpha$ limit, see later. The boundary value of the rapidity
density  \cite{Bajnok:2022rtu} for 
\begin{equation}
\chi_{\alpha}=\frac{e^{\alpha B}}{2}G_{+}(i\alpha)w_{\alpha}\label{eq:chin}
\end{equation}
can be obtained from the limit $w_{\alpha}\equiv W_{\alpha,\infty}=\lim_{\beta\to\infty}\beta W_{\alpha,\beta}$
giving
\begin{equation}
w_{\alpha}=1+d_{\alpha}+\sum_{l=1}^{\infty}q_{\alpha,\kappa_{l}}d_{\kappa_{l}}+\frac{v}{\pi}\int_{C_{+}}e^{-x}{\cal A}(x)Q_{\alpha}(x)dx
\end{equation}
for $\alpha\text{\ensuremath{\neq0}. }$ 

In the case of $\alpha=0$ the analogous quantities $W,w$ are introduced
as follows 
\begin{align}
{\cal O}_{0,\beta} & =\frac{1}{2\pi}G_{+}(i\beta)e^{\beta B}W_{0,\beta}\quad;\quad\beta>0\\
{\cal O}_{0,0} & =\frac{1}{\pi}W_{0,0}\quad;\qquad\chi_{0}=w_{0}
\end{align}
With these normalizations they satisfy the following system of differential
equations 
\begin{align}
(\alpha+\beta)W_{\alpha,\beta}+\dot{W}_{\alpha,\beta} & =w_{\alpha}w_{\beta}\label{eq:de1}\\
(\alpha^{2}-\beta^{2})W_{\alpha,\beta} & =(\alpha-\beta)w_{\alpha}w_{\beta}+w_{\beta}\dot{w}_{\alpha}-w_{\alpha}\dot{w}_{\beta}\label{eq:de2}\\
2\alpha\dot{w}_{\alpha}+\ddot{w}_{\alpha} & =Fw_{\alpha}\label{eq:de3}
\end{align}
for $\alpha,\beta\geq0$. They can be used to calculate the solutions
for the $\alpha=0$ problem, which otherwise would require a separate
treatment. Indeed, we can calculate $F$ from $w_{1}$ and solve the
equation (\ref{eq:de3}) for $w_{0}$ and (\ref{eq:de1}) for $W_{0,0}$
with some unknown constants, which should be fixed by other method,
say by comparing to the Volin's method. Note that the derivatives
are wrt. $B$ and not $v$. 

\subsection{Trans-series expansion}

In the following we construct a systematic expansion in the perturbative
parameter $v$ and in the non-perturbative parameter $\nu=e^{-2B}$.
The solution will be given in terms of a trans-series, where each
non-perturbative term has a perturbative expansion. We start with
the perturbative solution of (\ref{eq:Qeq}), i.e. we neglect all
terms containing $\nu$-s. The perturbative part of $Q_{\alpha}(x)$
will be denoted by $P_{\alpha}(x)$, which is understood as a power
series in $v$. It satisfies 
\begin{equation}
P_{\alpha}(x)+\int_{C_{+}}\frac{e^{-y}{\cal A}(y)P_{\alpha}(y)}{x+y}\frac{dy}{\pi}=\frac{1}{\alpha-vx}\quad.
\end{equation}
The parameter $v$ appears on the rhs. as well as in the kernel ${\cal A}(y)$.
The appropriately chosen running coupling $v$ ensures that the equation
has a regular power series expansion in $v$ without any $\ln v$
terms. The first few orders can be explicitly solved iteratively \cite{Marino:2021dzn,Bajnok:2022rtu,Bajnok:2024qro},
but we will not need their explicit form. What is relevant for us
is that, although the equation were defined originally for $\alpha>0$,
the perturbative solution can be extended for negative $\alpha$ as
well. This extension is not symmetric for $\alpha\leftrightarrow-\alpha$. 

Interestingly, the non-perturbative terms in (\ref{eq:Qeq}) show
up as source terms of the same type as the source of the perturbative
part, thus we can express $Q_{\alpha}(x)$ in terms of the perturbative
solution as 
\begin{equation}
Q_{\alpha}(x)=P_{\alpha}(x)+d_{\alpha}P_{-\alpha}(x)+\sum_{l=1}^{\infty}q_{\alpha,\kappa_{l}}d_{\kappa_{l}}P_{-\kappa_{l}}(x)\quad,\label{eq:Qn}
\end{equation}
In determining $q_{\alpha,\kappa_{s}}=Q_{\alpha}(\kappa_{s}/v)$ we
introduce\footnote{In the limit $\alpha\to-\beta$ the function $A_{\alpha,\beta}$ develops
a pole, which we removed in the definition for $\alpha=-\beta$. } 
\begin{equation}
A_{\alpha,\beta}=\begin{cases}
P_{\alpha}(-\beta/v)=\frac{1}{\alpha+\beta}+v\int_{C_{+}}\frac{e^{-y}{\cal A}(y)P_{\alpha}(y)}{\beta-vy}\frac{dy}{\pi} & \text{for}\quad\beta\neq-\alpha\\
-v\int_{C_{+}}\frac{e^{-y}{\cal A}(y)P_{\alpha}(y)}{\alpha+vy}\frac{dy}{\pi} & {\rm for}\quad\beta=-\alpha
\end{cases}\label{eq:Aab}
\end{equation}
The quantities $A_{\alpha,\beta}$ are defined as perturbative power
series in $v$. Although it is not obvious from the definition, but
they are symmetric in $\alpha$ and $\beta$ as $A_{\alpha,\beta}=A_{\beta,\alpha}$
and can be extended for any real (non-zero) values. They form a basis,
with which we obtain a closed system of linear equations for the unknowns
$q_{\alpha,\kappa_{s}}$ 
\begin{equation}
q_{\alpha,\kappa_{s}}-\sum_{l=1}^{\infty}q_{\alpha,\kappa_{l}}d_{\kappa_{l}}A_{-\kappa_{l},-\kappa_{s}}=A_{\alpha,-\kappa_{s}}+d_{\alpha}A_{-\alpha,-\kappa_{s}}\equiv s_{\alpha,-\kappa_{s}}\label{eq:qn}
\end{equation}
This is an infinite linear matrix equation for the infinite vector\footnote{Note that $\alpha$ is a label not an index, which labels the various
integral equations with rhs. $r_{\alpha}(\theta)=\cosh(\alpha\theta)$. } $\boldsymbol{q}_{\alpha}=\{q_{\alpha,\kappa_{s}}\}_{s=1,2,\dots}$
with source term $\boldsymbol{s}_{\alpha}=\{s_{\alpha,-\kappa_{s}}\}=\{A_{\alpha,-\kappa_{s}}+d_{\alpha}A_{-\alpha,-\kappa_{s}}\}$
of the form 
\begin{equation}
\mathbf{q}_{\alpha}(\mathbb{\mathbf{I}}-\mathbf{DA})=\mathbf{s}_{\alpha}\quad;\qquad\mathbf{A}_{s,l}=A_{-\kappa_{s},-\kappa_{l}}\quad;\qquad\mathbf{D}_{s,l}=i\delta_{sl}d_{l}
\end{equation}
The solution can be written 
\begin{equation}
\mathbf{q}_{\alpha}=\mathbf{s}_{\alpha}(\mathbb{\mathbf{I}}-\mathbf{DA})^{-1}=\mathbf{s}_{\alpha}(\mathbf{I}+\mathbf{DA}+(\mathbf{DA})^{2}+\dots)
\end{equation}
which takes the form 
\begin{equation}
q_{\alpha,\kappa_{l}}=s_{\alpha,-\kappa_{l}}+\sum_{\{l_{1},l_{2},\dots\}}s_{\alpha,-\kappa_{l_{1}}}d_{\kappa_{l_{1}}}A_{-\kappa_{l_{1}},-\kappa_{l_{2}}}d_{\kappa_{l_{2}}}A_{-\kappa_{l_{2}},-\kappa_{l_{3}}}\dots d_{\kappa_{l_{N}}}A_{-\kappa_{l_{N}},-\kappa_{l}}\quad,\label{eq:qnmsol}
\end{equation}
Clearly, at each non-perturbative order in $\nu=e^{-2B}$ only a finite
number of terms contributes. We can organize the expansion as a trans-series
in this non-perturbative parameter. At each order we have products
of the $A_{-\kappa_{l},-\kappa_{s}}$ functions, which are formal
power series in $v$. Once we calculated the trans-series form of
$q_{\alpha,\kappa_{l}}$ we can write the trans-series for the observable
as 
\begin{align}
W_{\alpha,\beta} & =s_{\alpha,\beta}+d_{\beta}s_{\alpha,-\beta}+\sum_{l=1}^{\infty}q_{\alpha,\kappa_{l}}d_{\kappa_{l}}s_{\beta,-\kappa_{l}}=s_{\alpha,\beta}+d_{\beta}s_{\alpha,-\beta,}+\mathbf{q}_{\alpha}\mathbf{D}\mathbf{s}_{\beta}=\label{eq:Wnm}\\
 & =A_{\alpha,\beta}+d_{\alpha}A_{-\alpha,\beta}+d_{\beta}(A_{\alpha,-\beta}+d_{\alpha}A_{-\alpha,-\beta})+\mathbf{s}_{\alpha}\mathbb{{\cal A}}\,\mathbf{s}_{\beta}\nonumber 
\end{align}
where we introduced the compact notation 
\begin{equation}
\mathbf{{\cal A}}=(\mathbb{\mathbf{I}}-\mathbf{DA})^{-1}\mathbf{D}=(\mathbf{D}+\mathbf{DAD}+\mathbf{DADAD}+\dots)\label{eq:AA}
\end{equation}
which is manifestly symmetric as $\mathbf{A}$ is . Observe that the
basic building block $A_{\alpha,\beta}$ is nothing but the perturbative
part of our generic observable $W_{\alpha,\beta}$. The boundary value
of the rapidity density can be expressed as 
\begin{align}
w_{\alpha} & =a_{\alpha}+d_{\alpha}a_{-\alpha}+\sum_{l=1}^{\infty}q_{\alpha,\kappa_{l}}d_{\kappa_{l}}a_{-\kappa_{l}}\quad,\label{eq:wn}\\
 & =a_{\alpha}+d_{\alpha}a_{-\alpha}+\mathbf{q}_{\alpha}\mathbf{D}\mathbf{a}=a_{\alpha}+d_{\alpha}a_{-\alpha}+\mathbf{s}_{\alpha}\mathbf{{\cal A}}\,\mathbf{a}\nonumber 
\end{align}
where $a_{\alpha}\equiv A_{\alpha,\infty}:=\lim_{\beta\to\infty}\beta A_{\alpha,\beta}$
and $\mathbf{a}=\{a_{-\kappa_{l}}\}$. These formulas can be considered
as the formal extensions of the $W_{\alpha,\beta}$ expressions for
$\beta=\infty$ by defining $d_{\infty}=0$ which is very natural
as $d_{\beta}\propto e^{-2B\beta}$. 

In order to have a complete solution we need to calculate the perturbatively
defined $A_{\alpha,\beta}$-s. Since they are the perturbative parts
of the $W,w$ quantities they must satisfy the following differential
equations
\begin{align}
(\alpha+\beta)A_{\alpha,\beta}+\dot{A}_{\alpha,\beta} & =a_{\alpha}a_{\beta}\label{eq:de1-1}\\
(\alpha^{2}-\beta^{2})A_{\alpha,\beta} & =(\alpha-\beta)a_{\alpha}a_{\beta}+a_{\beta}\dot{a}_{\alpha}-a_{\alpha}\dot{a}_{\beta}\label{eq:de2-1}\\
2\alpha\dot{a}_{\alpha}+\ddot{a}_{\alpha} & =fa_{\alpha}\label{eq:de3-1}
\end{align}
for all $\alpha,\beta$ (including also zero) and we denoted the parturbative
part of $F$ by $f$. These equations can be used to calculate all
the perturbative observables from a single one only. For example,
Volin's algorithm calculates $A_{1,1}$ recursively at any perturbative
order. This perturbative series then can be used to determine $a_{1}$
using eq. (\ref{eq:de1-1}). Then eq. (\ref{eq:de3-1}) determines
$f$ from which $a_{\alpha}$ can be integrated. Then eq. (\ref{eq:de1-1})
or (\ref{eq:de2-1}) can be used to determine the generic $A_{\alpha,\beta}.$
We elaborate this procedure and present explicit formulas later in
all the cases. 

\section{Structural result}

In this section we provide a universal structural result for all the
observables together with their graphical representations and investigate
their resurgence properties \cite{Dingle1973:asy,Ecalle1981:res,Marino:2012zq,Dorigoni:2014hea,Dunne:2015eaa,Aniceto:2018bis,Serone:2024uwz}. 

\subsection{Trans-series for all observables \label{subsec:Trans-series}}

\begin{figure}
\begin{centering}
\includegraphics[width=10cm]{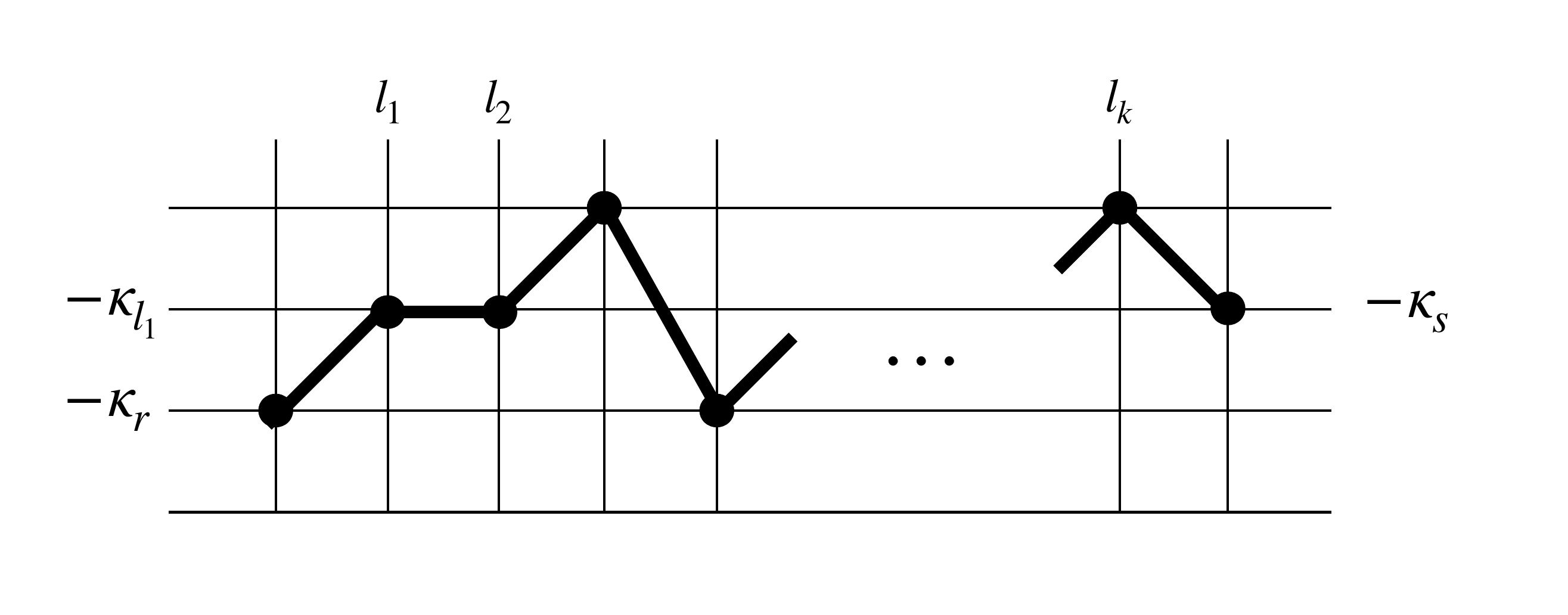}
\par\end{centering}
\caption{Graphical representation of a path appearing in the matrix element
${\cal A}_{-\kappa_{r},-\kappa_{s}}$. It connects the point $-\kappa_{r}$
to $-\kappa_{s}$ going through the points $-\kappa_{l_{j}}$. Each
vertex $-\kappa_{l}$ comes with a factor $d_{\kappa_{l}}$ and a
link between $-\kappa_{l}$ and $-\kappa_{j}$ with a factor $A_{-\kappa_{l},-\kappa_{j}}$. }
\label{fig:path}
\end{figure}

As we emphasized the expressions for $W_{\alpha,\beta}$ are valid
for $\alpha,\beta>0$ such that $\alpha,\beta,\kappa_{l}$ are all
distinct. Formally we can put either $\alpha$ or $\beta$ to $\infty$
(take the limit $W_{\alpha,\infty}\equiv\lim_{\beta\to\infty}\beta W_{\alpha,\beta}=w_{\alpha}$),
in which case we obtain the formulas for $w_{\alpha}$. In order to
obtain formulas for $\alpha$ or $\beta$ equal to zero we can use
either the differential equations or solve directly the integral equations
with a source $r_{0}=1$. The latter approach was taken in \cite{Bajnok:2024qro}.
In directly solving the differential equation we have to determine
$F$ first, then obtain $w_{0}\equiv W_{0,\infty}$, from which $W_{\alpha,0}$
and $W_{0,0}$ can be calculated. Integration constants can be fixed
from Volin's approach. All these trans-series can be written in terms
of the fundamental matrix (\ref{eq:AA}). This symmetric matrix has
matrix elements
\begin{equation}
{\cal A}_{-\kappa_{r},-\kappa_{s}}=d_{\kappa_{r}}(\delta_{r,s}+\sum_{\{l_{1},l_{2},\dots\}}A_{-\kappa_{r},-\kappa_{l_{1}}}d_{\kappa_{l_{1}}}A_{-\kappa_{l_{1}},-\kappa_{l_{2}}}d_{\kappa_{l_{2}}}\dots d_{\kappa_{l_{N}}}A_{-\kappa_{l_{N}},-\kappa_{s}}d_{\kappa_{s}})\label{eq:calAA}
\end{equation}
and can be represented as the sum of contributing lattice paths leaving
from the point $-\kappa_{r}$ and arriving at $-\kappa_{s}$ . Each
vertex at $-\kappa_{l}$ comes with a non-perturbative factor $d_{\kappa_{l}}$
involving a Stokes constant together with the appropriate power of
the non-perturbative expansion parameter $\nu^{\kappa_{l}}$. For
a link between the vertex $-\kappa_{l}$ and $-\kappa_{j}$ we multiply
with the factor $A_{-\kappa_{l},-\kappa_{j}}$. See figure \ref{fig:path}
for the graphical representation. 

Surprisingly this matrix governs the non-perturbative corrections
of all observables. It has two indices, which can couple to physical
indices and we can define the dressed version of any perturbative
object as 
\begin{equation}
\hat{A}_{\alpha,\beta}=A_{\alpha,\beta}+\sum_{r,s}A_{\alpha,-\kappa_{r}}{\cal A}_{-\kappa_{r},-\kappa_{s}}A_{-\kappa_{s},\beta}\label{eq:barAab}
\end{equation}
where $\alpha$ and $\beta$ can take any values including zero and
infinity. This quantity can be represented as a sum for paths starting
from the 'perturbative' index $\alpha$ going through all the non-perturbative
indices (via ${\cal A}$) and finally arriving at the perturbative
index $\beta$. The perturbative part of the expression corresponds
to the direct path from $\alpha$ to $\beta$. 

With this dressed quantity the observables take the form 
\begin{align}
W_{\alpha,\beta} & =\hat{A}_{\alpha,\beta}+d_{\alpha}\hat{A}_{-\alpha,\beta}+d_{\beta}\hat{A}_{\alpha,-\beta}+d_{\alpha}d_{\beta}\hat{A}_{-\alpha,-\beta}\label{eq:Wab}
\end{align}
This is literally true for $\alpha\neq\beta$ and $\alpha,\beta\neq0,\infty$.
For the exceptional values we found that we can formally take $d_{0}=0=d_{\infty}$. Thus,
we can spell out for the zero index 
\begin{equation}
W_{0,\alpha}=\hat{A}_{0,\alpha}+d_{\alpha}\hat{A}_{0,-\alpha}\quad\alpha>0\quad;\quad W_{0,0}=\hat{A}_{0,0}
\end{equation}
 For the case $\beta=\infty$ we recall that $A_{\alpha,\infty}=a_{\alpha}$
thus we can define $\hat{a}_{\alpha}\equiv\hat{A}_{\alpha,\infty}$
and write
\begin{equation}
w_{\alpha}=\hat{a}_{\alpha}+d_{\alpha}\hat{a}_{-\alpha}\quad\alpha>0\quad;w_{0}=\hat{a}_{0}
\end{equation}
Finally, the $\alpha$ independent $F$ has a trans-series of the
form 
\begin{equation}
F=f+2\partial_{B}(\sum_{l}d_{\kappa_{l}}\hat{a}_{-\kappa_{l}}a_{-\kappa_{l}})=2\dot{\hat{A}}_{\infty,\infty}
\end{equation}
where $A_{\infty,\infty}=\lim_{\alpha\to\infty}(\alpha(a_{\alpha}-1))$.
Observe that $d_{\kappa_{l}}$ has also a $B$-dependence, which should
be differentiated, too. These results can be obtained from solving
the differential equations (\ref{eq:de1}-\ref{eq:de3}) for the non-perturbative
part and by exploiting that the perturbative building blocks satisfy
the differential equations (\ref{eq:de1}-\ref{eq:de3-1}). Although
we obtained the results differently, we have checked explicitly that
the solutions satisfy the differential equations (\ref{eq:de1}-\ref{eq:de3}),
if the building blocks satisfy (\ref{eq:de1}-\ref{eq:de3-1}). The
final result for $F$ follows also from the fact that $\lim_{\alpha\to\infty}w_{\alpha}=\lim_{\alpha\to\infty}a_{\alpha}=1$
and implies that 
\begin{equation}
f=2\dot{A}_{\infty,\infty}=2\lim_{\alpha,\beta\to\infty}\alpha\beta\dot{A}_{\alpha,\beta}=2\lim_{\alpha\to\infty}\alpha\dot{a}_{\alpha}
\end{equation}

The coinciding limit in $W_{\alpha,\beta}$, when $\beta\to\alpha\neq\kappa_{r}$
is regular and can be obtained by analysing the poles in the leading
contributions of $A_{\alpha,-\beta}$ and $A_{-\alpha,\beta}$ in
(\ref{eq:Wab}). They cancel each other and differentiate the prefactors
as
\begin{equation}
\lim_{\beta\to\alpha}\frac{d_{\alpha}-d_{\beta}}{\beta-\alpha}=-\partial_{\alpha}d_{\alpha}=\left(2B-\partial_{\alpha}\ln\sigma(i\alpha+0)\right)d_{\alpha},
\end{equation}
Thus in the limit, we arrive at
\begin{equation}
W_{\alpha,\alpha}=\lim_{\beta\to\alpha}W_{\alpha,\beta}=\hat{A}_{\alpha,\alpha}-\partial_{\alpha}d_{\alpha}+2d_{\alpha}\hat{A}_{\alpha,-\alpha}+d_{\alpha}^{2}\hat{A}_{-\alpha,-\alpha}.\label{eq:Waa}
\end{equation}
Note that in the case when $\sigma(i\alpha+0)$ vanishes itself, we
have simply $\partial_{\alpha}d_{\alpha}=i\sigma'(i\alpha+0)e^{-2B}$,
where the prime means differentiation w.r.t. to the argument. In the
third term the perturbative part of the dressed object $\hat{A}_{\alpha,-\alpha}$
is understood as in (\ref{eq:Aab}). 

Similar cancellation happens if $\alpha$ coincides with some of the
$\kappa_{r}$-s, while $\beta$ is not. The building block $\hat{A}_{\alpha,\beta}$
itself is singular in the limit $\alpha\to\kappa_{r}$ due to the
$A_{\alpha,-\kappa_{r}}$ factor in (\ref{eq:barAab}). Its combination,
however with $d_{\alpha}\hat{A}_{\alpha,-\beta}$ is regular and the
limit can be taken. To see this, one expands $d_{\alpha}$ around
$\kappa_{r}$ as 
\begin{equation}
d_{\alpha}=\frac{-d_{\kappa_{r}}}{\alpha-\kappa_{r}}+\bar{d}_{\kappa_{r}}+\dots
\end{equation}
and notes that the singular part of $A_{\alpha,-\kappa_{r}}$ is of
the form $\frac{1}{\alpha-\kappa_{r}}$. By cancelling this singularity
we arrive at the result 
\begin{align}
W_{\kappa_{r},\beta}=\lim_{\alpha\to\kappa_{r}}W_{\alpha,\beta}= & \hat{A}_{\kappa_{r},\beta}+d_{\beta}\hat{A}_{\kappa_{r},-\beta}+\left[\bar{d}_{\kappa_{r}}\hat{A}_{-\kappa_{r},\beta}+d_{\kappa_{r}}\left(\partial_{\kappa}\hat{A}_{\kappa,\beta}\right)\Big\vert_{\kappa=-\kappa_{r}}\right]\nonumber \\
 & +d_{\beta}\left[\bar{d}_{\kappa_{r}}\hat{A}_{-\kappa_{r},-\beta}+d_{\kappa_{r}}\left(\partial_{\kappa}\hat{A}_{\kappa,-\beta}\right)\Big\vert_{\kappa=-\kappa_{r}}\right]
\end{align}
where for the dressed expressions $\hat{A}_{-\kappa_{r},\beta}$ in
the sum (\ref{eq:barAab}) the appearing $A_{\kappa_{r},-\kappa_{r}}$
is understood in the sense of (\ref{eq:Aab}). 

In these two special cases higher order poles appear in the Wiener-Hopf
method. They can also show up in more complicated models, where the
Wiener -Hopf kernel itself has higher order poles. We generalize our
analysis for these cases in Appendix (\ref{sec:higher poles}). 

\subsection{Resurgence properties}

In this section we apply the theory of resurgence for our problem
\cite{Dingle1973:asy,Ecalle1981:res,Marino:2012zq,Dorigoni:2014hea,Dunne:2015eaa,Aniceto:2018bis,Serone:2024uwz}.
The expressions for the observables presented so far are trans-series,
i.e. a double series in the perturbative coupling $v$ and in the
non-perturbative scale $\nu=e^{-2B}=v^{a}e^{-\frac{1}{v}-L}$. Each
perturbative series is an asymptotic series, written in terms of the
basis $A_{\alpha,\beta}$, which has only a formal meaning so far.
In order to connect this series to the physical solution of the integral
equation we have to use lateral Borel resummation. 

For a power series $\Psi(v)=\sum_{n=0}\psi_{n}v^{n}$ lateral Borel
resummation is understood as 
\begin{equation}
S^{\pm}(\Psi(v))=v^{-1}\int_{0}^{\infty e^{\pm i0}}e^{-s/v}\hat{\Psi}(s)ds\quad;\quad\hat{\Psi}(s)=\sum_{n=0}^{\infty}\frac{\psi_{n}}{n!}s^{n}.
\end{equation}
Since the perturbative coefficients grow asymptotically as 
\begin{equation}
\psi_{n}=\sum_{k=0}\Gamma(n-\lambda-k)\phi_{k}c^{k-n}+\dots\label{eq:asympsi}
\end{equation}
 (for some parameter $\lambda$) the Borel transformed function $\hat{\Psi}(s)$
has a $(c-s)^{\lambda}$ type singularity on the real line 
\begin{equation}
\hat{\Psi}(s)=c^{-\lambda}\sum_{k\geq0}\phi_{k}\Gamma(-\lambda-k)\left(c-s\right)^{\lambda+k}+\dots\label{eq:Psicut}
\end{equation}
To avoid this cut we have to integrate a bit above/below the real
line. Depending on the choice, this procedure gives two different
results, which differ in non-perturbative corrections, some of which
are even real. By assuming real perturbative coefficients, the leading
imaginary part of the lateral resummations is of the form 
\begin{equation}
\mp i\pi c^{-\lambda}v^{\lambda}e^{-c/v}\sum_{k=0}\phi_{k}v^{k}\label{eq:imagtrans}
\end{equation}
and the difference of these two are related to the alien derivative
of $\Psi$ at $c$, denoted by $\Delta_{c}\Psi$, in the following
way
\begin{equation}
S^{+}(e^{-c/v}\Delta_{c}\Psi(v))=S^{+}(\Psi)-S^{-}(\Psi)+\dots=-2\pi ic^{-\lambda}e^{-c/v}S^{+}(\sum_{k=0}\phi_{k}v^{\lambda+k}).
\end{equation}
The Stokes automorphism contains the exponentiated alien derivatives
and relates the two lateral resummations:
\begin{equation}
S^{-}(\Psi)=S^{+}(\mathfrak{S}^{-1}\Psi)
=S^{+}(e^{-\sum_{c}e^{-\frac{c}{v}}\Delta_{c}}\Psi),
\end{equation}
where the sum goes over all possible discontinuities on the real
positive line. Typically, the physical result is given by the median
resummation 
\begin{equation}
S_{{\rm med}}(\Psi)=S^{+}(\mathfrak{S}^{-1/2}\Psi)=S^{+}(e^{-\frac{1}{2}\sum_{c}\Delta_{c}e^{-\frac{c}{v}}}\Psi).
\label{media}  
\end{equation}

Although it does not follow directly from our derivation but we found
that the physical value of ${\cal O}_{\alpha,\beta}$ in (\ref{eq:Oab})
can be obtained from the trans-series (\ref{eq:Wab}) by
applying the $S^{+}$ lateral resummation: 
\begin{equation}
{\cal O}_{\alpha,\beta}=\frac{e^{(\alpha+\beta)B}}{4\pi}G_{+}(i\alpha)G_{+}(i\beta)S^{+}(W_{\alpha,\beta}).
\label{mcon}
\end{equation}
This is our main assumption, which we cannot prove but in what follows we
study its consequences. We will find that these consequences are nice and
self-consistent and thus make (\ref{mcon}) very plausible.

If (\ref{mcon}) provides the physical value, which we can obtain by solving the
integral equations numerically, this
in particular implies that $S^{+}(W_{\alpha,\beta})$ is real.
From the Wiener-Hopf solution of the
integral equation this result is very natural as we integrated on
the left of the imaginary line and picked up the residues $d_{\kappa_{l}}=-{\rm res}_{\kappa=\kappa_{l}}\sigma(i\kappa+0)e^{-2B\kappa_{l}}$
and $d_{\alpha}=\sigma(i\alpha+0)e^{-2B\alpha}$. Had we integrated
on the right we would have picked up the $-0$ residues, which
results in complex conjugation of the $d_{\alpha},d_{\kappa_{l}}$ expressions.
In most of the cases these $d$-s are purely imaginary and complex
conjugation actually means changing their signs. We can separate
the real and imaginary parts as 
\begin{equation}
d_{\alpha}=(iS_{\alpha}+\hat{S}_{\alpha})\nu^{\alpha}\quad;\quad d_{\kappa_{l}}=(iS_{\kappa_{l}}+\hat{S}_{\kappa_{l}})\nu^{\kappa_{l}}\quad;\quad\nu=e^{-2B}=v^{a}e^{-\frac{1}{v}-L}
\end{equation}
where we also separated the Stokes constants from the non-perturbative
scale $\nu$. 

The cancellation of the imaginary part of $S^{+}(W_{\alpha,\beta})$
at the non-perturbative order $\nu^{\alpha}$ implies that 
\begin{equation}
\Delta_{\alpha}A_{\alpha,\beta}=2iS_{\alpha}A_{-\alpha,\beta},\label{eq:Deltaalpha}
\end{equation}
valid for any $\beta$, including $0$ and $\infty$. The dotted
alien derivative is understood in the running coupling $v$ as: $\dot{\Delta}_{n}=\nu^{n}\Delta_{n}$,
where $[\dot{\Delta}_{n},\partial_{B}]=0$. Thus it has an extra $v^{a}$
factor compared to the standard definition. Observe that only the
imaginary part of $d_{\alpha}$ appears in this relation. Note also
that $\Delta_{\alpha}^{2}A_{\alpha,\beta}=0$. Similar argumentations
lead to 
\begin{equation}
\Delta_{\kappa_{l}}A_{\alpha,\beta}=2iS_{\kappa_{l}}A_{\alpha,-\kappa_{l}}
A_{-\kappa_{l},\beta}\label{eq:DeltaAnm}
\end{equation}
valid again for generic indices including $0$ and $\infty$. Note that this
result and the rest of this subsection already follows from the reality of
$S^{+}(W_{\alpha,\beta})$, which is only the first half of our main assumption
(\ref{mcon}).
We have checked that the above alien derivative relations are compatible with
the differential
equations (\ref{eq:de1}-\ref{eq:de3-1}). These alien derivatives
can be extracted also from the asymptotic behaviour of the pertubative
coefficients of $A_{\alpha,\beta}$. As they behave continuously in
$\alpha,\beta$ the relation (\ref{eq:DeltaAnm}) is valid also for
negative $\alpha$ and $\beta$ including the values $-\kappa_{l}$
. Clearly $\Delta_{\kappa_{l}}\Delta_{\kappa_{j}}$ is non-zero but
it equals to $\Delta_{\kappa_{j}}\Delta_{\kappa_{l}}$. This is a
speciality of our setting and can be easily checked by noting how
$\Delta_{\kappa_{l}}$ acts on a path, i.e. on a chain of $A$-operators:
$A_{\alpha,-\kappa_{l_{1}}}A_{-\kappa_{l_{1}},-\kappa_{l_{2}}}\dots A_{-\kappa_{l_{N}},\beta}.$
The Leibnitz rule implies that the result is a sum of terms in which
each $A_{-\kappa_{l_{j}},-\kappa_{l_{j+1}}}$ (including also $\alpha$
and $\beta$) is differentiated using the rule (\ref{eq:DeltaAnm}).
Graphically it means that the result is a sum of path in which we
break up each link and insert a new node $-\kappa_{l}$ in between,
i.e. $A_{-\kappa_{l_{j}},-\kappa_{l_{j+1}}}\to A_{-\kappa_{l_{j}},-\kappa_{l}}A_{-\kappa_{l},-\kappa_{l_{j+1}}}$
. This property enables us to define a multi-parameter trans-series
of the formal variables $\{\sigma_{\alpha},\sigma_{\kappa_{l}}\}\equiv\{\sigma\}$
by replacing $d_{\kappa_{l}}$ with $\sigma_{\kappa_{l}}\nu^{\kappa_{l}}$
and $d_{\alpha}$ with $\sigma_{\alpha}\nu^{\alpha}$ : 
\begin{equation}
W_{\alpha,\beta}(\{\sigma\})=W_{\alpha,\beta}(d_{\kappa}\to\sigma_{\kappa}\nu^{\kappa})
\end{equation}
where $\kappa=\alpha$ or $\kappa=\kappa_{l}$. We should keep in
mind that $\sigma_{\alpha}$ is fermionic in the sense that $\sigma_{\alpha}^{2}=0$.
The pointed alien derivative $\dot{\Delta}_{\kappa}=\nu^{\kappa}\Delta_{\kappa}$
acts as $-2iS_{\kappa}$ times differentiation wrt. $\sigma_{\kappa}$:
\begin{equation}
\dot{\Delta}_{\kappa}W_{\alpha,\beta}(\{\sigma\})=-2iS_{\kappa}\partial_{\sigma_{\kappa}}W_{\alpha,\beta}(\{\sigma\})\quad.
\end{equation}
Since the Stokes automorphism is the exponentiation of the alien derivatives,
it acts on the trans-series parameters by a shift 
\begin{align}
\mathfrak{S}W_{\alpha,\beta}(\{\sigma\}) & =e^{\sum_{\kappa}\dot{\Delta}_{\kappa}}W_{\alpha,\beta}(\{\sigma\})=e^{-\sum_{\kappa}2iS_{\kappa}\partial_{\sigma_{\kappa}}}W_{\alpha,\beta}(\{\sigma\})=W_{\alpha,\beta}(\{\sigma_{\kappa}\to\sigma_{\kappa}-2iS_{\kappa}\}).
\end{align}
Similarly,
%The ambiguity free median resummation is given by
\begin{equation}
\frak{\mathfrak{S}}^{-\frac{1}{2}}W{}_{\alpha,\beta}(\{\sigma\})=W_{\alpha,\beta}(\{\sigma_{\kappa}\to\sigma_{\kappa}+iS_{\kappa}\}).
\end{equation}
In other words, the full trans-series
solution of the Wiener-Hopf problem can be obtained by choosing
the Stokes constants for the multiparameter trans-series as $\sigma_{\kappa}=\hat{S}_{\kappa}$ and applying $\frak{\mathfrak{S}}^{-\frac{1}{2}}$:
\begin{equation}
W_{\alpha,\beta}=\frak{\mathfrak{S}}^{-\frac{1}{2}}W_{\alpha,\beta}(\{\hat S\}).
\label{compres1}
\end{equation}  
This implies that for problems when $\hat{S}_{\kappa}=0$ (i.e. the
residues are purely imaginary)\footnote{These include the $O(N)$  models for
$N>3$, the $SU(N)$ principal models, the Lieb-Liniger and Gaudin-Yang models,
and both cases of the disk capacitor problems.}, the Stokes constants can be
put to zero and the trans-series is simply given by
\begin{equation} 
%S^{+}(W_{\alpha,\beta})=S_{\mathrm{med}}(A_{\alpha,\beta})
W_{\alpha,\beta}=\frak{\mathfrak{S}}^{-\frac{1}{2}}A_{\alpha,\beta}.
\label{compres2}
\end{equation}
In these models the perturbative $A_{\alpha,\beta}$ \lq\lq knows'' everything
about the full trans-series $W_{\alpha,\beta}$ (complete resurgence).

Performing the lateral resummation, using the median resummation  defined by
(\ref{media}), we obtain
\begin{equation} 
S^{+}(W_{\alpha,\beta})=S_{\mathrm{med}}(W_{\alpha,\beta}(\{\hat S\})
\label{Splusphys}
\end{equation}
and for models with $\hat S_\kappa=0$ simply
\begin{equation} 
S^{+}(W_{\alpha,\beta})=S_{\mathrm{med}}(A_{\alpha,\beta}).
\label{Splusphys2}
\end{equation}

The second part of our main assumption states that (\ref{Splusphys}) is not
only real but gives also the physical value of $W_{\alpha,\beta}$. We have
verified this for $O(N)$ models numerically very precisely (see section 7 and
\cite{Bajnok:2024qro}) and also for the SUSY $O(7)$ case (see subsection 6.2).

The lateral resummation $S^{-}$ corresponds to $\sigma_{k}\to\sigma_{k}-iS_{k}$,
which indeed corresponds to the alternative integrations in the contour
deformations as we anticipated before. 

\section{Explicit formulae \label{sec:Explicit}}

In this section we provide explicit formulae for the various models.
These involve the perturbative functions $A_{\alpha,\beta}$ and the
non-perturbative information $\kappa_{l}$. We do it separately for
the bosonic and fermionic models. For the $O(N)$ symmetric models
$\Delta=\frac{1}{N-2}$ and the Wiener-Hopf decomposition of the kernel
has $a=1-2\Delta$ universally (see later), while for $SU(N)$ symmetric
models $\Delta=\frac{1}{N}$. 

\subsection{Bosonic models }

The bosonic models include the $O(N)$ symmetric sigma model, its
supersymmetric extension and the $SU(N)$ principal chiral models.
In these models the Wiener-Hopf decomposition (\ref{eq:WHdecomp})
has a square root singularity at the origin
\begin{equation}
G_{+}(i\kappa)=\frac{1}{\sqrt{\kappa}}H(\kappa)e^{\frac{a}{2}\kappa\ln\kappa+\frac{b}{2}\kappa}
\end{equation}
and the corresponding Wiener-Hopf kernel is 
\begin{equation}
\sigma(i\kappa\pm0)=e^{-a\kappa\ln\kappa-b\kappa}\frac{H(-\kappa)}{H(\kappa)}\left(\mp i\cos(\frac{a\pi\kappa}{2})-\sin(\frac{a\pi\kappa}{2})\right)\quad,\label{eq:sigma}
\end{equation}
It has a cut and poles on the positive imaginary line. By introducing
the running coupling (\ref{eq:runningcoupling}), the discontinuity
function can be made free of $\ln v$ terms 
\begin{equation}
{\cal A}(x)=e^{-vx(L-a\ln v)}\delta\sigma(vx)=\cos(a\pi vx/2)e^{-avx(\ln x+q)+\sum_{k=1}^{\infty}z_{2k+1}(vx)^{2k+1}}\quad,\label{eq:calA}
\end{equation}
where $\ln(H(-\kappa)/H(\kappa))=\sum_{k=0}^{\infty}z_{2k+1}\kappa^{2k+1}$
and the linear term in the exponent is $aq=L+b-z_{1}$. This parametrization
is valid for all the models we consider. Using the product representation
of the Gamma functions one can show that $z_{k>1}$ is proportional
to $\zeta_{k}$ in a model-dependent way. The constant $L$ parametrizes
the various running couplings, which will be chosen in a convenient
way. The choice of gauge for the coefficients presented in Appendix
\ref{sec:Coefficients} is the one where $q=\gamma_{E}+2\ln2$ universally
for each bosonic model.

The perturbative basis can be determined explicitly for all the models
following Volin's method \cite{Volin:2009wr,Volin:2010cq}. This method
matches two different parametrizations of the resolvent 
\begin{equation}
R_{\alpha}(\theta)=\int_{-B}^{B}\frac{\chi_{\alpha}(\theta')}{\theta-\theta'}d\theta'
\end{equation}
The first is in coordinate space, valid in the middle of the $[-B,B]$
interval 
\begin{equation}
R_{\alpha}(\theta)=\sum_{n,m=0}^{\infty}\sum_{k=0}^{n+m}\frac{\sqrt{B}c_{n,m,k}\bigl(\frac{\theta}{B})^{h(k)}}{B^{m-n}(\theta^{2}-B^{2})^{n+\frac{1}{2}}}\left[\ln\frac{\theta-B}{\theta+B}\right]^{k}
\end{equation}
with $h(k)=k\,\mathrm{mod}\,2$, while the other is for the Laplace
transform,
\begin{equation}
\hat{R}_{\alpha}(s)=\int_{-i\infty+0}^{i\infty+0}\frac{dz}{2\pi i}e^{sz}R_{\alpha}(B+z/2)\quad;\quad z=2(\theta-B)
\end{equation}
 valid at the edge, near $B$. The resolvent is related to the Fourier
transform of the rapidity density and can be parametrized with the
leading order Wiener-Hopf solution 
\begin{equation}
\hat{R}_{\alpha}(s)=2e^{-2Bs}\tilde{\chi}(2is)=\frac{1}{2}G_{+}(2is)G_{+}(i\alpha)e^{\alpha B}\left[\frac{1}{s+\frac{\alpha}{2}}+\frac{1}{Bs}\sum_{n,m=0}^{\infty}\frac{Q_{n,m}}{B^{n+m}s^{n}}\right]
\end{equation}
 Matching the two parametrizations for large $B$ in the overlapping
region determines both sets of coefficients. Technically, one expands
the bulk solution for large $B$ in the limit when $z$ is kept fixed
and maps a given power $z^{\beta}$ to $s^{-\beta-1}/\Gamma(-\beta)$
in order to compare with the expansion of $\hat{R}_{\alpha}(s)$.
The $\ln^{k}(\frac{\theta-B}{\theta+B})$ term can be written as $\frac{d^{k}}{dx^{k}}(\frac{\theta-B}{\theta+B})^{x}\vert_{x=0}$,
which implies that $\ln B$ appears only in the combination $\ln^{k}(4Bs)$.
The analogous dependence can be written for $\hat{R}_{\alpha}(s)$
as $\exp(as\ln4Bs-as\ln B/B_{0})$, which enables to compare directly
the powers of $\ln4Bs$ treated as an independent expansion parameter.
As the result of the $\ln B/B_{0}$ term the coefficients $c_{n,m,k}$
and $Q_{n,m}$ acquire $\ln B$ dependence. Since in the free energy
this $\ln B$ dependence disappears due to the renormalization group
behaviour, Volin chose the convenient value $\ln B=\ln B_{0}$ to
speed up the calculations. In our case, however, we are interested
in ${\cal O}_{\alpha,\beta}$ as the function of the running coupling
$v$. The $\ln B$ terms can be expressed in terms of $v$ and $\ln v$
as $\ln2B=-\ln v+\ln(1+v(\ln v+L))$ when expanded in $v$. We have
already shown that in this running coupling the $\ln v$ dependence
disappears. We can thus freely choose the value $\ln v=-\ln2B_{0}$
together with the arbitrary parameter $L$ as $L=\ln2B_{0}$, such
that Volin's expansion in $1/2B$ becomes our expansion in $v$. 

Originally, Volin performed the calculation for the $O(N)$ model
and obtained $A_{1,1}$. Later the method was extended for many other
bosonic models \cite{Marino:2019eym,Marino:2019fuy,Marino:2019wra,Marino:2020dgc,Marino:2020ggm,Reis:2022tni,Reichert:2020ymc}.
The $O(N)$ model , however, is generic enough and we can extract
the $A_{1,1}$ perturbative series as the function of $a$ and $z_{2k+1}$
. We can then use the differential equations to determine $a_{1}$
and $f$, which then leads to $a_{\alpha}$ and finally $A_{\alpha,\beta}$.
In Appendix (\ref{sec:Coefficients}) we present the first few orders
in terms of the generic parameters $a$ and $z_{2k+1}$. 

In order to apply the perturbative formulas for the various models
we have to specify these parameters. The information on the non-perturbative
corrections is encoded in the location of the poles $\kappa_{l}$
which we also list. 
\begin{itemize}
\item $O(N)$ sigma model
\begin{equation}
H(\kappa)=\frac{1}{\sqrt{\Delta}}\frac{\Gamma(1+\Delta\kappa)}{\Gamma(\frac{1}{2}+\frac{\kappa}{2})}
\end{equation}
$a=1-2\Delta,$ $b=2\Delta(1-\ln\Delta)-(1+\ln2)$ and the kernel
is described by $z_{1}=-(1-2\Delta)\gamma_{E}-2\ln2$ and 
\begin{equation}
z_{2k+1}=2\frac{\zeta_{2k+1}}{2k+1}(\Delta^{2k+1}-1+2^{-2k-1})
\end{equation}
for $k>0$. The running coupling (\ref{eq:runningcoupling}) is defined
with the specific choice $L=-b-4\Delta\ln2$. The zeros of $\sigma(i\kappa)$
are located $N$-independently at the positions $\kappa=2l-1$, while
its poles are at $\kappa=l(N-2)$, where $l\in\mathbb{N}$. This implies
that $\kappa_{l}=l\kappa_{1}$ with $\kappa_{1}=N-2$ for $N$ even
and $\kappa_{1}=2N-4$ for $N$ odd \cite{Marino:2021dzn,Bajnok:2022rtu}.
\item Principal chiral model
\begin{equation}
H(\kappa)=\frac{1}{\sqrt{2\pi\Delta(1-\Delta)}}\frac{\Gamma(1+\Delta\kappa)\Gamma(1+(1-\Delta)\kappa)}{\Gamma(1+\kappa)}
\end{equation}
and $a=0$, while $b=-2\Delta\ln\Delta-2(1-\Delta)\ln(1-\Delta)$.
In order to use the generic form we need the replacements $z_{1}=0$
and
\begin{equation}
z_{2k+1}=2\frac{\zeta_{2k+1}}{2k+1}\left(-1+\Delta^{2k+1}+(1-\Delta)^{2k+1}\right)
\end{equation}
for $k>0$, while the running coupling is defined with $L=-b$. The
poles of $\sigma(i\kappa)$ again form a lattice $\kappa_{l}=l\kappa_{1}$
with $\kappa_{1}=\frac{N}{N-1}$ and $l\in\mathbb{N}$. This model
is very similar to the $O(4)$ model, which is the $SU(2)$ case here.
\item supersymmetric $O(N)$ sigma model
\begin{equation}
H(\kappa)=\frac{1}{\sqrt{\Delta}}\frac{\Gamma(1+\Delta\kappa)\Gamma(\frac{1}{2}+\frac{(1-2\Delta)\kappa}{2})}{\Gamma(\frac{1}{2}+\frac{\kappa}{2})^{2}}
\end{equation}
and $a=1$, while $b=-(1+2\Delta)\ln2-2\Delta\ln\Delta-1-(1-2\Delta)\ln(1-2\Delta)$
and $z_{1}=-\gamma_{E}-2(1+2\Delta)\ln2$ with
\begin{equation}
z_{2k+1}=2\frac{\zeta_{2k+1}}{2k+1}\left(\Delta^{2k+1}-2+2^{-2k}+(1-2\Delta)^{2k+1}(1-2^{-2k-1})\right)\quad.
\end{equation}
The running coupling is defined with $L=-b-4\Delta\ln2$. The first
case is $N=5$ for which $\kappa_{1}=6$. For $N>5$ we have to distinguish
between the even and odd cases. We have the same set of poles as for
the $O(N)$ models and additionally $\mu_{l}=\frac{N-2}{N-4}(2l+1)$
for $N$ even with $l=0,1,2,\ldots$, while for odd $N$ the values
$l=(N-4)s+(N-5)/2$ with $s=0,1,\dots$ have to be left out. 
\item Lieb-Liniger model and the disk capacitor with opposite charges 
\begin{equation}
H(\kappa)=\frac{1}{\sqrt{\pi}}\Gamma\left(1+\frac{\kappa}{2}\right)
\end{equation}
$a=-1,$  $b=1+\ln2$ and the kernel is described by $z_{1}=\gamma_{E}$
and
\begin{equation}
z_{2k+1}=\frac{\zeta_{2k+1}}{2k+1}2^{-2k}
\end{equation}
for $k>0$. The running coupling (\ref{eq:runningcoupling}) is defined
with the specific choice $L=-3\ln2-1$. The non-perturbative corrections
are located at $\kappa_{l}=2l$ with $l\in\mathbb{N}$ \cite{Bajnok:2022rtu}.
\end{itemize}

\subsection{Fermionic models}

These models include the $O(N)$ Gross-Neveu model, the $SU(N)$ chiral
Gross-Neveu model, the Gaudin-Yang model and the disk capacitor with
the same charges. In these fermionic models we do not have any square
root singularity at the origin in the Wiener-Hopf decompostion (\ref{eq:WHdecomp})
\begin{equation}
G_{+}(i\kappa)=H(\kappa)e^{\frac{a}{2}\kappa\ln\kappa+\frac{b}{2}\kappa}
\end{equation}
The Wiener-Hopf kernel takes a slightly different form than in the
bosonic models
\begin{equation}
\sigma(i\kappa\pm0)=e^{-a\kappa\ln\kappa-b\kappa}\frac{H(-\kappa)}{H(\kappa)}\left(\mp i\sin(\frac{a\pi\kappa}{2})+\cos(\frac{a\pi\kappa}{2})\right)\quad,\label{eq:sigma-1}
\end{equation}
but the running coupling (\ref{eq:runningcoupling}) can be introduced
the same way, such that he discontinuity function is free of $\ln v$
terms 
\begin{equation}
{\cal A}(x)=e^{vx(a\log v-L)}\delta\sigma(vx)=\sin(a\pi vx/2)e^{-avx(\log x+q)+\sum_{k=1}^{\infty}z_{2k+1}(vx)^{2k+1}}\quad,\label{eq:calA-1}
\end{equation}
where the $z_{2k+1}$ terms are related to the expansion of $\ln(H(-\kappa)/H(\kappa))$
and are proportional to $\zeta_{2k+1}$ in a model dependent way.
The value of $L=-b+z_{1}+aq$ was chosen in Appendix \ref{sec:Coefficients}
such that $q=\gamma_{E}$ for the fermionic models, universally. Volins
method can be adapted also for the fermonic case \cite{Marino:2019eym,Reis:2022tni},
with the bulk ansatz 
\begin{equation}
R_{\alpha}(\theta)=\sum_{m=0}^{\infty}\sum_{n=1}^{\infty}\sum_{k=0}^{m+n}\frac{c_{n,m,k}\left(\frac{\theta}{B}\right)^{h(k-1)}}{B^{m-n}\left(\theta^{2}-B^{2}\right)^{n}}\left[\ln\frac{\theta-B}{\theta+B}\right]^{k}.
\end{equation}
where $h(k)=k\,\mathrm{mod}\,2$. The ansatz, valid at the edge for
the Laplace transform follows from the leading order Wiener-Hopf solution
\begin{equation}
\hat{R}_{\alpha}(s)=\frac{1}{2}G_{+}(2is)G_{+}(i\alpha)e^{\alpha B}\left[\frac{1}{s+\frac{\alpha}{2}}+\frac{1}{Bs}\sum_{m=0}^{\infty}\sum_{n=0}^{m}\frac{Q_{n,m-n}}{B^{m}s^{n}}\right]
\end{equation}
 The calculation goes as before: the matching of the two representations
via Laplace transform determines the series expansion of $A_{1,1}$
in the running coupling $v$ in terms of $a$ and $z_{2k+1}$. Using
the differential equations we can extend this solution for the $a_{n},f,A_{n,m}$
quantities, which are displayed in Appendix (\ref{sec:Coefficients}).
To specify the results for the various models we list the corresponding
values here. 
\begin{itemize}
\item $O(N)$ Gross-Neveu model
\begin{equation}
H(\kappa)=\frac{\Gamma(\frac{1}{2}+\frac{(1-2\Delta)\kappa}{2})}{\Gamma(\frac{1}{2}+\frac{\kappa}{2})}
\end{equation}
and $a=2\Delta$ while $b=-2\Delta(1+\ln2)-(1-2\Delta)\ln(1-2\Delta)$.
The kernel ${\cal A}(x)$ is described by $z_{1}=-2\Delta(\gamma_{E}+2\ln2)$
and
\begin{equation}
z_{2k+1}=2\frac{\zeta_{2k+1}}{2k+1}\left((1-2\Delta)^{2k+1}-1\right)\left(1-2^{-2k-1}\right)
\end{equation}
for $k>0$, while the running coupling is $L=-b-4\Delta\ln2$. The
non-perturbative corrections are encoded into the pole positions $\kappa_{l}=\frac{N-2}{N-4}(2l+1)$
for even $N$ with $l=0,1,2,\ldots$, while for odd $N$ we need to
leave out the $l=(N-4)s+\frac{N-5}{2}$ values with $s=0,1,\dots$.
For $N=5$ all $l$-s are exceptional, such that there are no poles
at all. 
\item $SU(N)$ chiral Gross-Neveu model 
\begin{equation}
H(\kappa)=\frac{1}{\sqrt{(1-\Delta)}}\frac{\Gamma(1+(1-\Delta)\kappa)}{\Gamma(1+\kappa)}
\end{equation}
and $a=2\Delta$, while $b=-2\Delta-2(1-\Delta)\ln(1-\Delta)$. The
generalized parameters are $z_{1}=-2\Delta\gamma_{E}$ and
\begin{equation}
z_{2k+1}=2\frac{\zeta_{2k+1}}{2k+1}\left((1-\Delta)^{2k+1}-1\right)
\end{equation}
for $k>0$. The choice of the running coupling leads to $L=-b$. The
poles are located at $\kappa_{l}=\frac{N}{N-1}l$ with $l\in\mathbb{N}$,
except $l=s(N-1)$ with $s=1,2\dots$. For $N=2$ all $l$s are exceptional,
so there are no poles at all. 
\item Gaudin-Yang model and the disk capacitor with the same charges
\begin{equation}
H(\kappa)=\frac{1}{\sqrt{2\pi}}\Gamma\left(\frac{1}{2}+\frac{\kappa}{2}\right)
\end{equation}
and $a=-1$, while $b=1+\ln2$. The $z$-parameters are $z_{1}=\gamma_{E}+2\ln2$
and
\begin{equation}
z_{2k+1}=2\frac{\zeta_{2k+1}}{2k+1}\left(1-2^{-2k-1}\right)
\end{equation}
for $k>0$. The choice of the running coupling leads to $L=\ln2-1$,
and the poles are located as $\kappa_{l}=2l+1$ with $l=0,1,2,\ldots$
in this case.\footnote{Note that although the distance of poles from each other is in general
$2$, as the closest pole to the origin is at $\kappa_{0}=1$, all
powers of $e^{-2B}$ will appear in the trans-series of the capacity
$C^{(-)}$ in (\ref{eq:cap}).} 
\end{itemize}

\section{Checking the trans-series solutions}

The trans-series solutions (\ref{eq:Wab},\ref{eq:Waa}) are understood
as laterally Borel resummed with the $S^{+}$ prescription. This procedure
introduces imaginary parts of the trans-series terms, which should
be cancelled, implying non-trivial resurgence relations between the
various non-perturbative corrections. As we explained before the asymptotic
behaviour of the perturbative coefficients (\ref{eq:asympsi}) is
related to the non-perturbative corrections as (\ref{eq:imagtrans}).
This asymptotic growth can be extracted either numerically, or in
certain cases analytically, and can be directly compared to the\emph{
imaginary} parts of the higher order trans-series terms. In order to
test the \emph{real} parts one has to compare the (numeric) solution
of the integral equation to the laterally Borel resummed trans-series
order by order. In the following we review these two types of checks
in the various models. 

\subsection{Asymptotic relations and numerical analysis}

As the results for the bosonic and fermionic models can be derived
from the $O(N)$ sigma and Gross-Neveu models the investigations focused
on these cases. The basic observable is the ground-state energy density
${\cal O}_{1,1}$ the density ${\cal O}_{1,0}$ and the free energy
$\bar{{\cal O}}_{1,1}$. These observables were thoroughly investigated
in the simplest $O(4)$ model in a series of papers \cite{Abbott:2020mba,Abbott:2020qnl,Bajnok:2021dri}.
By calculating a large number (2000) of perturbative coefficients
numerically the asymptotic behaviour (\ref{eq:asympsi}) was identified
over a hundred digits precision and was compared to (\ref{eq:imagtrans}).
Similar analysis was performed, although not so extensively, for the
$O(6)$ and $O(7)$ models in \cite{Marino:2021dzn}. By treating
Volin's method analytically in \cite{Bajnok:2021dri} the relation
(\ref{eq:Deltaalpha}) was exactly derived for the $O(4)$ model.
Similar analysis for the Gross-Neveu and Gaudin-Yang model was done
in \cite{Marino:2021dzn}. In all of these models the perturbative
series determined through the resurgence relations all the non-perturbative
corrections. The $O(3)$ model is exceptional among the $O(N)$ sigma
models having instantons. Indeed, in this model the asymptotical behaviour
of the perturbative series determine only part of the trans-series
by resurgence relations \cite{Marino:2021dzn,Bajnok:2021zjm,Bajnok:2024qro},
depending on the observable and the coupling there could be independent
one, two or infinitely many instanton sectors, unseen by the imaginary
part relations. To test them we need to do a direct numerical comparison. 

As we already pointed out, the asymptotic relations test only the
imaginary parts of the neighboring non-perturbative corrections in
the trans-series. Thus it does not test the complete trans-series
with non-vanishing $\hat{S}_{k}$ Stokes constants and in cases with
instanton sectors such as in the $O(3)$ model. In order to check
our solution in these cases one has to solve numerically the integral
equation (\ref{eq:TBA}) very precisely and subtract the laterally
Borel resummed trans-series terms order by order. Such analysis for
the $O(3)$ and $O(4)$ models were done in \cite{Abbott:2020qnl,Abbott:2020mba,Bajnok:2024qro,Bajnok:2021zjm}.
The integral equation was solved on the basis of Chebisev polynomials
with high precision, while the lateral Borel resummation was performed
using the diagonal Pade approximant. It was found that by subtracting
each resummed non-perturbative term the deviation from the numerical
solution decreased to the order of the next non-perturbative correction. 

All these checks are demonstrating the correctness of the trans-series
solutions. However, none of the investigated cases involved a non-vanishing
Stokes constant $\hat{S}_{k}$. In the following we investigate such
a case in detail. 

\subsection{The supersymmetric $O(7)$ sigma model}

This model was chosen to represent a case where the Stokes constant
$\hat{S}_{\kappa}$ is non-zero, and the cuts on the Borel-plane are
not logarithmic. We investigated the energy density in detail
\begin{equation}
\epsilon=m^{2}e^{2B}\frac{G_{+}^{2}(i)}{8\pi}2W_{1,1}\quad;\quad\frac{G_{+}^{2}(i)}{8\pi}=\frac{\sqrt[5]{25+11\sqrt{5}}\pi}{4\ 30^{3/5}e}.
\end{equation}
As a first step we used Volin's algorithm to generate $N_{\text{max}}=200$
perturbative coefficients up to $\apprge2200$ digits of precision
in the running coupling:

\begin{equation}
2A_{1,1}=\sum_{n=0}^{N_{\max}}\psi_{n}v^{n}\qquad,\quad2B=\frac{1}{v}-\ln v+1+\frac{3}{5}\ln6-\ln5
\end{equation}
We aimed at testing the leading non-perturbative correction of $W_{1,1}$
which has its root in the closest singularity of $\sigma(i\kappa+0)$
at $\kappa=\mu_{0}=5/3$ and takes the form 
\begin{equation}
W_{1,1}=A_{1,1}+d_{5/3}A_{1,-5/3}^{2}+\ldots,\quad d_{5/3}=(iS_{\frac{5}{3}}+\hat{S}_{\frac{5}{3}})\nu^{\frac{5}{3}}=e^{i\frac{2\pi}{3}}\frac{16\sqrt[3]{\frac{2}{5}}e^{5/3}\pi}{75\ 3^{5/6}\Gamma\left(\frac{2}{3}\right)^{2}}\nu^{\frac{5}{3}}.\label{eq:W11corr}
\end{equation}
 The non-perturbative scale in the running coupling reads as $\nu^{5/3}=\frac{1}{6}\left(\frac{5}{e}\right)^{5/3}\times v^{5/3}e^{-\frac{5}{3v}}$,
where the fractional power of $v$ indicates a corresponding type
of singularity on the Borel plane at $\kappa=5/3$. To check whether
the imaginary part of our trans-series would indeed cancel the ambiguity
in the $S^{+}$ Borel integral, we investigated the analytic structure
of the generalized Borel transform (\ref{eq:Psicut}). After approximating
the Borel-transform $\hat{\Psi}(s)\approx\sum_{n=0}^{N_{\max}}\frac{\psi_{n}}{n!}s^{n}$
via the diagonal $(N_{\max}/2,N_{\max}/2)$ Pade-approximant of the
finite sum we changed variables $s(w)=\kappa-w^{3}$such that the
cut at $s=\kappa$ gets opened up. Using the formulas (\ref{eq:asympsi},\ref{eq:Psicut},\ref{eq:imagtrans}) we could identify that $c=b=5/3$ and arrive at 
\begin{equation}
\hat{\Psi}(s(w))=(5/3)^{-5/3}\sum_{k=0}\phi_{k}\Gamma(-5/3-k)w^{5+3k}.
\end{equation}
That is, after the mapping $s(w)$, and expanding the result around
$w=0$ we can read off the $\phi_{k}$-s as every $3^{\text{rd}}$
coefficient in the Taylor expansion, starting from the $5^{\text{th}}$.
In order to catch the correct analytical structure at $\kappa$ we
need to calculate another Pade-approximant as an intermediate step
in terms of $w$ at some $w=w_{*}$ that is closer (yet not identical)
to the original expansion point $s=0$, that is, $w=\sqrt[3]{5/3}\approx1.186$.
We chose the rational value $w_{*}=\frac{119}{100}$ and then its
Taylor expansion gave us the ratios of $\phi_{k}$ to $\phi_{0}$
as 
\begin{align}
\sum\phi_{k}/\phi_{0}v^{k}=1 & +(1/5\pm9\cdot10^{-17})v-(2/5\pm5\cdot10^{-12})v^{2}+1.10733333(3\!\pm\!5)v^{3}\nonumber\\
	&-4.43892(85\!\pm\!21)v^{4} +(23.1652\pm0.0004)v^{5}-(143.009\pm0.031)v^{6}\nonumber\\
	&+(1024.1\pm1.0)v^{7}-(8334.92\pm0.06)v^{8}\nonumber \\
 & +(76046\pm207)v^{9}-(7.69\pm0.05)\times10^{5}\cdot v^{10}+O(v^{11})\label{eq:series10}
\end{align}
where the errors indicate the magnitude of deviation from the expected
result $A_{1,-5/3}^{2}(1-5/3)^{2}$. Even for the last coefficient
the relative error is less than one percent. The leading coefficient
was measured to be
\begin{equation}
\phi_{0}=-0.76346272674279197(30\!\pm\!13).
\end{equation}
The magnitude of the relative deviation from the theoretical value
is of the order of $10^{-18}$, where the latter comes by requiring
ambiguity cancellation in the Borel-resummed trans-series:
\begin{equation}
\frac{2\nu^{5/3}S_{5/3}}{(1-5/3)^{2}}\overset{!}{=}\pi(5/3)^{-5/3}v^{5/3}e^{-\frac{5}{3v}}\phi_{0}^{\text{(theor)}}\quad\Rightarrow\quad\phi_{0}^{\text{(theor)}}=-\frac{10\cdot2^{1/3}}{\Gamma^{2}(-1/3)}.
\end{equation}

To check whether our result also gives the real part of the residues
$\hat{S}_{5/3}$ correctly, we compared the real part of the lateral-Borel
resummation of $A_{1,1}$ to precision numerics of the original TBA
integral equation (\ref{eq:TBAalpha}). The method we used for the
latter is based on an expansion of the solution $\chi_{\alpha}(\theta)$
on the basis of even Chebyshev polynomials as explained in \cite{Abbott:2020qnl} for the $O(4)$ model. However, as for the
supersymmetric $O(N)$ models only the Fourier transform of the kernel
$\tilde{K}(\omega)$ can be written explicitly, we used a numerical
approximation\footnote{We evaluated a numerical inverse Fourier transform of $\tilde{K}(\omega)$
to determine $K(\theta)$ at $5000$ adaptively chosen points in $\theta$
space (instead of slicing up the interval $[-B,B]$ uniformly, we
divided the range of $K(\theta)$ into equal intervals to chose more
points where the function changes rapidly).} of $K(\theta)$, that heavily limited the precision of this technique
compared to the $O(3),O(4)$ cases \cite{Bajnok:2024qro}. We call
the (dimensionless) result of these numerics as $\epsilon_{\text{TBA}}\equiv\epsilon m^{-2}$
and think of it as an approximation of the exact physical value.

\begin{figure}[!h]
\centering{}\subfloat[Energy density $\epsilon/m^{2}$]{\begin{centering}
\includegraphics[height=0.25\textheight]{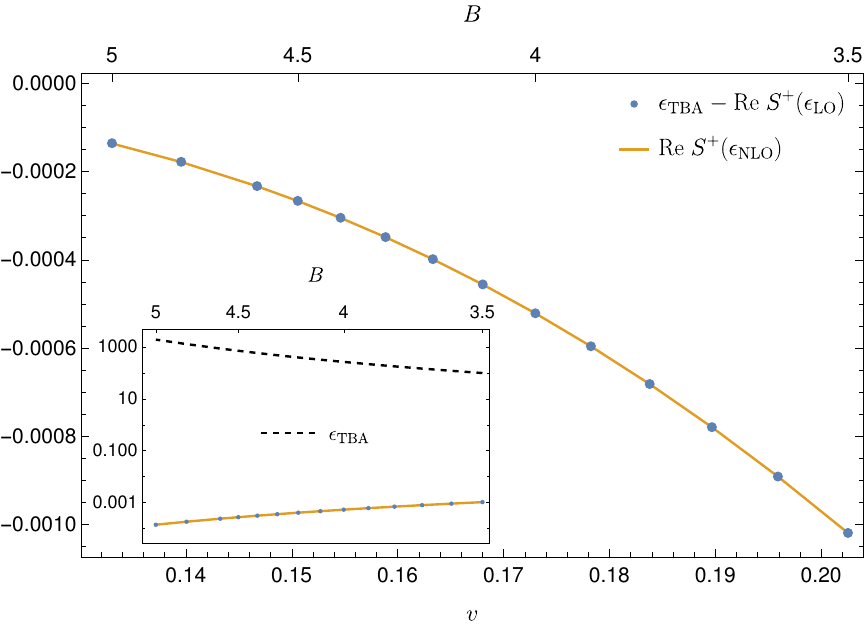}
\par\end{centering}
\label{subfig:a}}\subfloat[Normalized energy density $\hat{\epsilon}$]{\begin{centering}
\includegraphics[height=0.25\textheight]{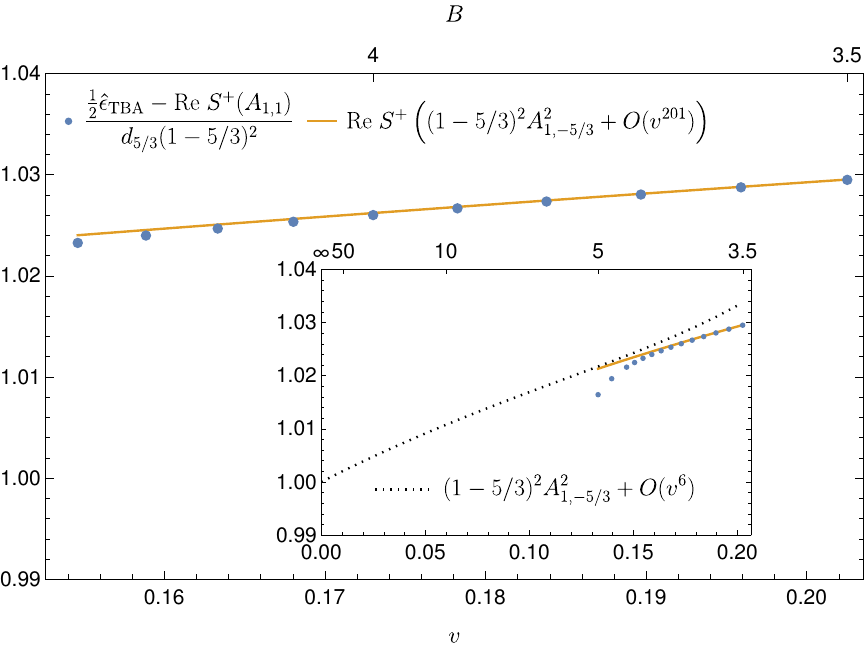}
\par\end{centering}
\label{subfig:b}}\caption{Comparison of the difference between precision numerics of the TBA
and the lateral Borel resummation of the perturbative part for the
energy density $\epsilon$. Left: The figure shows that the difference
agrees with the resummation of the subleading term in the trans-series.
The inset shows the magnitude of the difference w.r.t. $\epsilon/m^{2}$
itself, on a logarithmic scale. Right: The same difference with the
exponential factors removed. The black dotted line in the inset shows
the series (\ref{eq:series10}) truncated at $v^{5}$, around the
optimal truncation for the $v\approx0.15-0.2$ range, while the blue
dots and the orange line correspond to the same data sets as those
shown on the main figure.}
\label{fig:cmpTBA}
\end{figure}

The difference of the resummation and the numerics is then the resummation
of the correction term in (\ref{eq:W11corr}):
\begin{equation}
\epsilon_{\text{TBA}}-S^{+}(\epsilon_{\text{LO}})=S^{+}(\epsilon_{\text{NLO}})+\ldots,\quad\begin{cases}
\epsilon_{\text{LO}}\equiv e^{2B}\frac{G_{+}^{2}(i)}{4\pi}A_{1,1}\\
\epsilon_{\text{NLO}}\equiv e^{2B}\frac{G_{+}^{2}(i)}{4\pi}d_{5/3}A_{1,-5/3}^{2}
\end{cases}
\end{equation}
and as we already checked the equality of the imaginary parts on both
sides, we only have to compare the real parts. We performed the numerical
resummations on the Pade-approximants of the Borel-transforms for
both $\epsilon_{\text{LO}}$ and $\epsilon_{\text{NLO}}$. The results
are shown in Figure \ref{fig:cmpTBA}. The difference is of several
orders of magnitude smaller compared to the physical value $\epsilon$
in the given range of the running coupling, and it exactly appears
to agree with the resummation of $\epsilon_{\text{NLO}}$. Working
with the normalized quantity $\hat{\epsilon}$ shows good agreement
for a certain range (see Subfigure \ref{subfig:b}), yet it reveals
a discrepancy that starts around $v\apprle0.15$. However, this side
of the range corresponds to larger $B$ values, and the TBA numerics
we compare to tends to be less reliable for increasing $B$. As explained
above, it is also less precise compared to the numerics in the $O(4)$
case, due to the numeric approximation of the kernel $K(\theta)$.
Thus this deviation can be attributed to the unreliability of the
TBA's numerical solution, rather than the incorrectness of the analytic
solution. 

In summarising, we can say that that the resurgence relations are
satisfied for the imaginary part of the non-perturbative correction,
while its real part was confirmed with high precision numerical solution
of the integral equation. 

\section{Convergence of the trans-series}

\begin{figure}[!h]
\centering{}\subfloat[Perturbative coefficients of the $O(4)$ model]{\centering{}\includegraphics[width=0.49\textwidth]{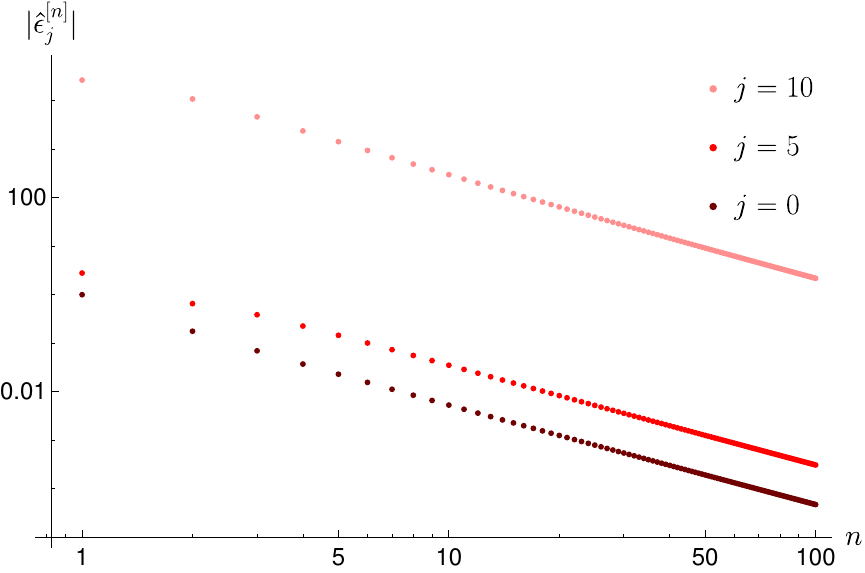}\label{subfig:power}}\subfloat[Comparison of Borel-resummed trans-series terms for $O(N)$ models]{\centering{}\includegraphics[width=0.49\textwidth]{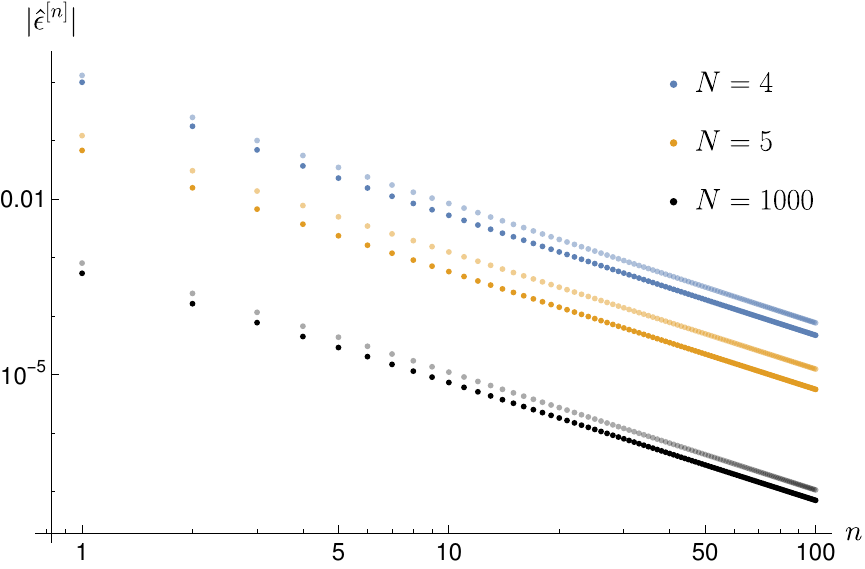}\label{subfig:Borelpower}}\caption{Left: Trans-asymptotics of the leading and higher order perturbative
coefficients in the $O(4)$ model. Note the universal power-law decay.
Right: Magnitude of the coefficients as a function of $n$, color
coded for different $N$s. The faint dots are the Pade-Borel-resummations
$\hat{\epsilon}^{[n]}$ at the highly non-perturbative value $B=0.01$
based on $12$ perturbative coefficients, while the opaque ones are
the leading perturbative coefficients $\hat{\epsilon}_{0}^{[n]}$
 (i.e. the $B\to\infty$ or $v\to0$ limits) for comparison.}
\end{figure}

In this section we study the trans-asymptotics of the trans-series.,
i.e. we would like to understand the convergence properties of the
individually Borel resummed trans-series terms. We focus on the energy
density $\hat{\epsilon}=2W_{1,1}$ in the $O(N)$ sigma models for
$N\geq4$ and use (\ref{eq:Waa}) for $\alpha=1$. Since for these
models $\sigma(i\kappa+0)$ vanishes at $\kappa=1$ the terms with
$d_{1}$ are absent, while $M\equiv-2\frac{\mathrm{d}}{\mathrm{d}\kappa}\sigma(i\kappa+0)\vert_{\kappa=1}$
shows up as 
\begin{equation}
\hat{\epsilon}=2W_{1,1}=M\nu+2\hat{A}_{1,1}\qquad,\quad M=-2e\left(\frac{\Delta}{e}\right)^{2\Delta}\frac{\Gamma(1-\Delta)}{\Gamma(1+\Delta)}e^{i\pi\Delta}\label{eq:M}
\end{equation}
The term $M$ is a constant in ${\cal O}_{1,1}$ and its real part
is related to the bulk energy density \cite{Marino:2021dzn,Bajnok:2022rtu}.
Since the poles of $\sigma(i\kappa+0)$ are equally spaced with $\kappa_{1}$,
the full trans-series can be organized as 
\begin{equation}
\hat{A}_{1,1}=\sum_{n=0}^{\infty}A_{1,1}^{[n]}(\nu^{\kappa_{1}})^{n},
\end{equation}
where $A_{1,1}^{[n]}$-s are asymptotic perturbative series in $v$
combined from the building blocks $A_{-n\kappa_{1},-m\kappa_{1}}$
and Stokes-constants $S_{n\kappa_{1}}$. Generating these $A_{1,1}^{[n]}$-s
up to high orders and keeping a sufficient number of their perturbative
terms can be achieved, by a simple recursion, as discussed briefly
in \cite{Bajnok:2024qro}. Typically, we choose a cutoff $N_{\text{max}}$
and keep the perturbative coefficients up to that order. Then at every
step we multiply two power-series, best done numerically, by convolution.
The recursive procedure goes as follows: 
\begin{align}
A_{1,1}^{[0]} & =A_{1,1}, & A_{1,1}^{[n]} & =\sum_{l=1}^{n}iS_{2l\kappa_{1}}A_{1,-\kappa_{l}}(v)q_{l,n-l}(v)+O(v^{N_{\max}+1}),
\end{align}
where the quantities $q_{l,k}$ start from $A_{-\kappa_{s},1}$ and
proceed as 
\begin{align}
q_{s,0}(v) & =A_{-\kappa_{s},1}, & q_{s,n}(v) & =\sum_{l=1}^{n}iS_{\kappa_{l}}A_{-\kappa_{s},-\kappa_{l}}(v)q_{l,n-l}(v)+O(v^{N_{\max}+1}).
\end{align}
 Once many $A_{1,1}^{[n]}$ coefficients are calculated we can check
their convergence properties in $n$. We also would like to see that
the lateral Borel resummation of the trans-series converges to the
physical value

\begin{equation}
\hat{\epsilon}_{\text{phys}}=S^{+}(\hat{\epsilon})=M\nu+2\sum_{n=0}^{\infty}S^{+}(A_{1,1}^{[n]})\nu^{\kappa_{1}n}.
\end{equation}
As a first step we investigate how the various perturbative coefficients
of $2A_{1,1}^{[n]}\sim\sum_{j=0}\hat{\epsilon}_{j}^{[n]}v^{j}$ behave
as the function of $n$. We thus fix $j$ and analyze the $n$-dependence
of $\hat{\epsilon}_{j}^{[n]}$. For $j=0,5,10$ the results are shown
on subfigure \ref{subfig:power} in the $O(4)$ model. Surprisingly,
each of these expansion coefficients decrease at the same rate, approximately
as $\propto n^{-2}$. This implies that by summing up the non-perturbative
correction first, we obtain at each perturbative order an $\text{Li}_{2}\left(\nu^{2}\right)$
behaviour, signaling a convergence radius of $1$. We observed the
same behaviour for the other $O(N>4)$ models in accord with \cite{Marino:2023epd}.
We then wanted to improve this analysis by resumming the perturbative
terms. The results for the Borel-Pade resummations $\hat{\epsilon}^{[n]}\equiv2S^{+}(A_{1,1}^{[n]})$
with $12$ terms at each non-perturbative order is represented on
Subfigure \ref{subfig:Borelpower}.

\begin{figure}[h]
\begin{centering}
\includegraphics[width=0.65\textwidth]{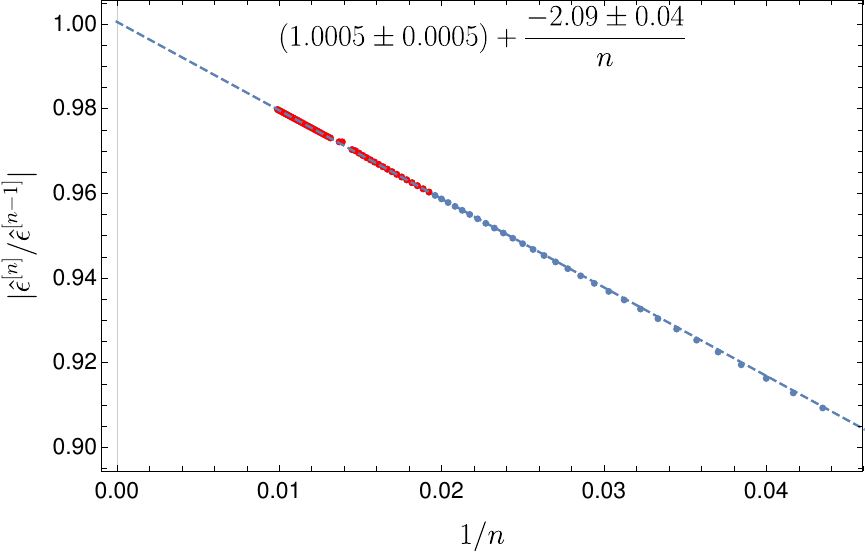}
\caption{Plot to estimate the convergence radius. The ratios of consecutive coefficients plotted as a function
of $1/n$  to estimate the convergence radius numerically. The intersection
of the fit line with the $y$-axis is at $1/R$.  For the fit, we
only used the red dots.  The integration contour was chosen as the
half-infinite line with acute angle $\varphi=3/5$ to the $x$-axis
for each integration.}
\label{fig:dombsykes}
\end{centering}
\end{figure}

In order to measure the convergence radius more precisely and to check
whether the sum would indeed converge to the physical value, we focused
on the $O(4)$ model. The value $B=0.1$ was chosen to evaluate the
Borel integrals as in this case the perturbative and non-perturbative
orders are at the same magnitude
\begin{equation}
2B=v^{-1}-2\ln2\quad\Rightarrow\quad v\simeq0.6304,\quad\nu^{2}\simeq0.6703,
\end{equation}
which are far from being practically perturbative, and close to the
extreme $B=0$ - that is $v_{\max}=\frac{1}{2\ln2}\simeq0.7213$ and
$\nu_{\max}^{2}=1$ - point. Choosing a smaller $B$ value, however,
would have decreased the precision of the lateral resummations considerably.
The latter were performed with $50$ perturbative coefficients via
the Pade-Borel technique, and up to $n=100$  for each $\hat{\epsilon}^{[n]}$.
The decay of the absolute value of the coefficients turned out to
be approximately $\propto n^{-2}$, and the convergence radius, $R$,
is thus estimated to be $1$, within error ($R=0.9995\pm0.0005$ from
Figure \ref{fig:dombsykes}). Note that having convergence radius
$1$ in $\nu=e^{-2B}$ implies convergence for all physical $B$. 

\begin{figure}[h]
\begin{centering}
\includegraphics[width=0.95\textwidth]{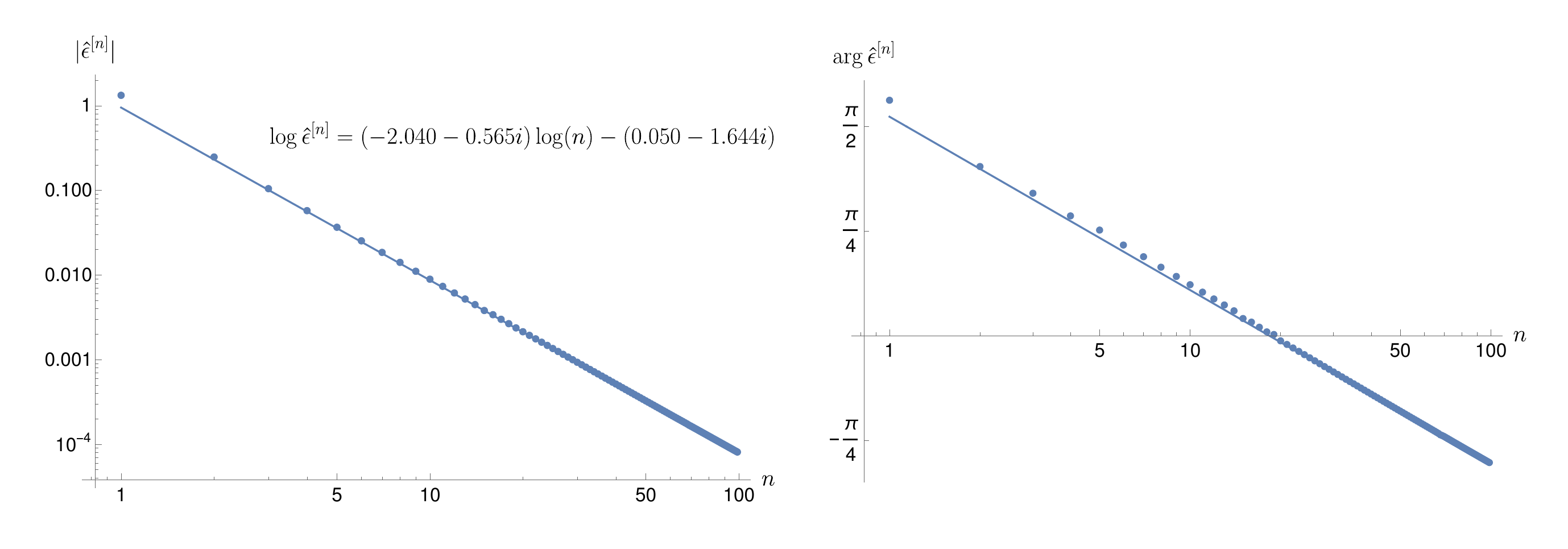}
\caption{The absolute value and complex argument of the coefficients as a function of $n$, together with their fit.}
\label{fig:decay}
\end{centering}
\end{figure}

To compare with the physical value we computed the latter via high-precision
numerics based on the Chebyshev-polinomials \cite{Bajnok:2024qro}
\begin{equation}
\hat{\epsilon}_{\text{phys}}(B=0.1)=0.430450507\ldots
\end{equation}
whose relative precision was estimated to be of the order of $10^{-78}$.
To resum the whole trans-series we took the following approach: we
calculated the first few $\hat{\epsilon}^{[n]}$-s up to $n=6$ more
precisely, based on $1000$ perturbative coefficients. We made a crude
estimate of their error by comparing them to a lateral resummation
taken for almost the same number of coefficients as described in \cite{Bajnok:2024qro}.
The sum truncated at $n=6$ differed from the physical value as
\begin{equation}
\hat{\epsilon}_{\text{phys}}-M\nu-\sum_{n=0}^{6}\hat{\epsilon}^{[n]}\nu^{2n}=0.002012(41\pm15)+i0.0014(36\pm12),\label{eq:diffstart}
\end{equation}
where $M=-2i$. Next we made an estimate on the contribution of the
sum's tail. On Figure \ref{fig:decay} we fitted the asymptotics
of the coefficients as $\epsilon^{[n]}\sim e^{p}n^{-q}$ with complex
parameters $p=-0.04(96\pm24)+i1.64(45\pm19)$ and $q=2.040(2\pm6)+i0.565(4\pm4)$.

Then the rest of the sum can be written as 
\begin{equation}
\delta\hat{\epsilon}\equiv e^{p}\sum_{n=7}^{\infty}n^{-q}\nu^{2n}=e^{p}\left(\text{Li}_{q}\left(\nu^{2}\right)-\sum_{n=1}^{6}n^{-q}\nu^{2n}\right)=0.0020(5\pm6)+i0.0010(3\pm9)\label{eq:difftail}
\end{equation}
where we evaluated the polylogarithm function $\text{Li}_{s}\left(x\right)=\sum_{n=1}^{\infty}n^{-s}x^{n}$
at complex order $s$. In (\ref{eq:diffstart}) and (\ref{eq:difftail})
the real parts agree within error, while the imaginary parts are of
the same magnitude. Subtracting the two equations from each other
we can conclude, that the relative deviation from the physical value
is$(\hat{\epsilon}_{\text{phys}}-\text{Re}S_{+}(\hat{\epsilon}))/\hat{\epsilon}_{\text{phys}}\approx8.8\times10^{-5}.$ 

This analysis demonstrates very clearly that the trans-series is convergent
with radius $R=1$ once the asymptotic behaviour of each non-perturbative
correction is resummed laterally. 

\section{Trans-Series of the free-energy density }

In this section we use our previous formulas to express the free energy
density $\mathcal{F}$ (\ref{eq:Fh},\ref{eq:FhbarO}) in relativistic
models as the function of the external field $h$, (\ref{eq:hchi0overchi0}).
The dependence on this external field can be encoded in the running
coupling $\alpha$. We simplify the analysis for an equidistant family
of poles $\kappa_{l}=\kappa_{1}l$ and assume that $\sigma(i)=0$.\footnote{Even in cases where the poles have exceptions or have a lattice where
$\kappa_{l}=\text{const.}\cdot(2l+1)$ with $l=0,1,2,\ldots$ the
following formulas can be still used in the following way: assuming
poles at $\kappa_{l}=\kappa_{1}l$ with $l\in\mathbb{N}$ where $\kappa_{1}$
is the closest pole and simply switching off the Stokes constants
$S_{\kappa_{l}},\hat{S}_{\kappa_{l}}$ where in reality no pole appears.} This simplification ensures that non-perturbative corrections go
in powers of $e^{-2B\kappa_{1}}$, and excludes the extra sectors
proportional to $d_{\alpha},d_{\beta}$ in formulas (\ref{eq:Wab})
leaving us only with a single extra term $M$, see (\ref{eq:M}). 

We can unify the results for all models by introducing the parameter
$\delta$, which is $1$ for the bosonic and $0$ for the fermionic
ones. The running coupling can be defined as 
\begin{equation}
\frac{1}{\alpha}+\frac{\delta-a}{2}\ln\alpha=\ln\frac{h}{\Lambda}
\end{equation}
 where $\Lambda$ is a dimensionful parameter, which is proportional
to the dynamically generated scale and the mass $m$ of the particles,
but otherwise can be freely chosen. This choice ensures that the trans-series
expansion of ${\cal F}(\alpha)$ is free of $\ln\alpha$ terms. The
non-perturbative structure of the free energy density looks as
\begin{equation}
\mathcal{F}=-\frac{h^{2}H^{2}(0)}{2\pi}\left(\frac{\pi}{2\alpha}\right)^{\delta}\left\{ F_{0}(\alpha)-MC_{\delta}^{2}\left(\frac{m}{\Lambda}\right)^{2}\alpha^{a}e^{-2/\alpha}+\sum_{l=1}^{\infty}F_{l}(\alpha)\left[C_{\delta}^{2}\left(\frac{m}{\Lambda}\right)^{2}\alpha^{a}e^{-2/\alpha}\right]^{\kappa_{1}l}\right\} ,\label{eq:Fbosandfer}
\end{equation}
where $C_{\delta}=\frac{e^{\frac{b}{2}}}{2}\frac{H(1)}{H(0)}\left(\sqrt{\frac{2}{\pi}}\right)^{\delta}$
and $F_{l}(\alpha)$-s are perturbative asymptotic series in $\alpha$,
all starting with $1+O(\alpha)$. The latters are presented up to
$l=2$ in Appendix (\ref{sec:Coefficients}). They are parametrized
in terms of the model parameters, the Stokes constants $S_{\kappa_{l}},\hat{S}_{\kappa_{l}}$,
the distance of the poles $\kappa_{1}$ and the constant
\begin{equation}
y_{1}=-z_{1}-a\left(\gamma_{E}+(1+2\delta)\ln2\right)-2\ln(C_{\delta}\frac{m}{\Lambda})
\end{equation}
This is the only place - except from the $m/\Lambda$ prefactors -
where (\ref{eq:Fbosandfer}) depends on to the choice of $\Lambda$.
We have checked that (\ref{eq:Fbosandfer}) are in complete agreement
with formulas (4.20) and (3.46) in \cite{Marino:2021dzn}, respectively.
These results are derived in Appendix (\ref{sec:Falpha}). 

\section{Conclusion}

In this paper we extended and completed the previous approaches to
solve the linear TBA equation (\ref{eq:TBAalpha}) and calculate the generalized
observables (\ref{eq:Onm}) in terms of trans-series. All trans-series are written in terms of lattice paths with perturbative building blocks
$A_{\alpha,\beta}$, which can be calculated using Volin's algorithm
and the differential equations (\ref{eq:de1-1},\ref{eq:de2-1},\ref{eq:de3-1}).
We provided explicit and universal formulas separately for bosonic
and fermionic models in Appendix \ref{sec:Coefficients}. Their parametrizations
in case of the nonlinear $O(N)$ sigma models, the various Gross-Neveu
models, the principal chiral models and their non-relativistic counterparts
the Lieb-Liniger and Gaudin-Yang models along with the disk capacitor
can be found in section \ref{sec:Explicit}. These trans-series provide
the right physical answer once they are resummed laterally by integrating
the Borel transform a bit above the real line. Imaginary parts cancel
due to the various resurgence relations, which we determined through
the alien derivatives of the basic building blocks $A_{\alpha,\beta}$,
which take a particularly simple form. We tested these results numerically
by exploiting the asymptotic relations between different non-perturbative
sectors and also by comparing directly the laterally Borel resummed
expressions to the numerical solution of the integral equation. Subtracting
more and more trans-series terms the correction always decreased to
the order of the next non-perturbative correction. This confirms all
the previous terms and provides a strong support of our result. On
the way we also shown that the various resummed trans-series terms
can be summed leading to a convergent series for all $B$.

In summary, although we cannot analytically prove but we found convincing
evidence that the laterally Borel resummed trans-series is 
convergent and converges to the physical solution. We can, however,
analytically study the consequences of this fundamental assumption and found
very compact formulas: (\ref{eq:Deltaalpha}) and (\ref{eq:DeltaAnm})
for the alien derivatives, (\ref{compres1}) and (\ref{compres2}) for the
representation of the full trans-series,  (\ref{Splusphys}) and
(\ref{Splusphys2}) for the Borel resummation. The simplicity and elegance of
these results confirm the correctness of our main assumption.

In order to make contact with asymptotically free perturbation theory we expressed
the free energy density in terms of the field theoretic running coupling.
Our result provides explicitly calculable trans-series solutions for
a large class of observables in integrable quantum field theories
and are unique in this respect. The main observables we analyzed were
the expectation values of conserved charges. The determined non-perturbative
corrections are expected to be related mostly to renormalons \cite{Marino:2021dzn,Marino:2021six,Marino:2022ykm},
while in certain cases to instantons \cite{Marino:2021dzn,Bajnok:2021zjm,Bajnok:2024qro}.
It would be nice to extend our analysis to other observables such
as two point functions, expectation values of condensates where a
clearer connection to renormalons through the operator product expansion
is available. Steps into this direction were already made in \cite{Marino:2024uco,Liu:2024omb}. 

In this work we focused on the relativistic observables (\ref{eq:Onm}).
In the statistical physical applications, however the non-relativistic
moments (\ref{eq:nonrelobs}) are more relevant. We have a work in progress
to specify our results for that situation. 

Here we analyzed a large class of models, where the ground-state energy can be described
by a single integral equation. There are  many other models with this property, including  Fendley's 
coset sigma models \cite{Fendley:2001hc}, or deformations of sigma models \cite{Ashwinkumar:2025wad} 
just to name a few. We see no obstacles to applying our general procedure to these cases as well. 

The $O(6)$ nonlinear sigma model plays a special role in the AdS/CFT
duality as it governs the excitations of a folded string spinning
in AdS in the dual description of the cusp anomalous dimension \cite{Bajnok:2008it}.
Part of the anomalous dimension of the cusped Wilson loop in the large
spin and twist limit can be captured by the groundstate energy density
of this sigma model. The other part is related to the cusp anomalous
dimension, whose leading resurgence properties were analysed in \cite{Aniceto:2015rua,Dorigoni:2015dha}
and were recently extended to higher orders in \cite{Bajnok:2024bqr}
based on novel methods \cite{Bajnok:2024epf,Bajnok:2024ymr}. It would
be interesting to understand how our $O(6)$ results supplement the
result in \cite{Bajnok:2024bqr}. 

\subsection*{Acknowledgments}

We thank Gerald Dunne, Marco Serone, Ramon Miravitllas Mas, Tomas
Reis, Marcos Marino the useful discussions. This work was supported
by the NKFIH research Grant K134946. The work of IV was also partially
supported by the NKFIH excellence grant TKP2021-NKTA-64.

\appendix

\section{Coinciding and higher order poles in the Wiener Hopf solution\label{sec:higher poles}}

In this Appendix we generalize our solution of the Wiener Hopf integral
equation for higher order poles of the integrand in (\ref{eq:intX}).
We thus relax the assumption we made in Section \ref{sec:Solving}
that $\alpha,\beta\neq\kappa_{l}$, and also that the singularities
of $\sigma(\omega)$ at $\omega=i\kappa_{l}$ are simple poles. The
first case appears when calculating the moments $\mathcal{O}_{\alpha,\beta}$
for $\alpha,\beta$ coinciding with $\kappa_{l}$. For higher order
poles in $\sigma(\omega)$ the motivation comes from models such as
Fendley's $SU(N)/SO(N)$ coset sigma models \cite{Fendley:2000bw,Marino:2022ykm}.
There $\sigma(\omega)$ has two families of poles $\kappa_{l}=\frac{(2l-1)N}{N-2}$
and $\kappa'_{l}=\frac{lN}{2}$ and a single family of zeros $\kappa_{l}=(2l-1)$
for $l=1,2,3,\ldots$. It can thus happen that the poles coincide
and not cancelled by any zero. 

For the more general case of higher order poles we have to return
to (\ref{eq:intX}) and repeat the subsequent steps more carefully.
Let us denote the set of positions for pole-like singularities in
$\sigma(i\kappa+0)$ as $\mathcal{K}\equiv\{\kappa_{l}\}_{l=1}^{\infty}$
then the union $\mathcal{K}\cup\{\alpha\}$ covers all places where
we need to take residues for both $\alpha=\kappa_{l}$ and $\alpha\ne\kappa_{l}$
cases. These residues show up as source terms in the integral equation
\begin{equation}
Q_{\alpha}(x)+\int_{C_{+}}\frac{e^{-y}\mathcal{A}(y)Q_{\alpha}(y)}{x+y}\frac{dy}{\pi}=\frac{1}{\alpha-vx}+\sum_{\kappa\in\mathcal{K}\cup\{\alpha\}}\mathop{\text{res}}_{\ \kappa'=\kappa}\frac{g(\kappa')}{\kappa'+vx}.\label{eq:Qres}
\end{equation}
and we made a shorthand for 
\begin{equation}
g(\kappa)\equiv e^{-2\kappa B}\sigma(i\kappa+0)Q_{\alpha}(\kappa/v).
\end{equation}
 Let us assume, that at a given $\kappa$ the highest order singularity
of $g(\kappa)$ is an $N_{\kappa}$-order pole (both the kernel $\sigma(i\kappa+0)$
and the unknown function $Q_{\alpha}(\kappa/v)$ may contribute).
We can then expand $g(\kappa')$ as $g(\kappa')=\sum_{n=0}^{N_{\kappa}-1}\frac{g_{-n-1}(\kappa)}{(\kappa'-\kappa)^{n+1}}+\dots$
, where $g_{-n-1}(\kappa)$-s are the strengths of its poles and thus
the residue on the r.h.s. of (\ref{eq:Qres}) looks like
\begin{equation}
\mathop{\text{res}}_{\ \kappa'=\kappa}\frac{g(\kappa')}{\kappa'+vx}=-\sum_{n=0}^{N_{\kappa}-1}\frac{g_{-n-1}(\kappa)}{((-\kappa)-vx)^{n+1}}.
\end{equation}
In this case the non-perturbative source terms are not the same type
as the source term in the perturbative part $P_{\alpha}(x)$. However,
they appear exactly in the same form as the source terms in its $\alpha$
expansion $P_{\alpha+\delta}(x)=\sum_{n=0}^{N_{\kappa}-1}P_{\alpha}^{[n]}(x)\delta^{n}+\mathcal{\mathcal{O}}(\delta^{N_{\kappa}}),$
as the coefficient functions $P_{\alpha}^{[n]}(x)$ satisfy the integral
equation
\begin{equation}
P_{\alpha}^{[n]}(x)+\int_{C_{+}}\frac{e^{-y}\mathcal{A}(y)P_{\alpha}^{[n]}(y)}{x+y}\frac{dy}{\pi}=\frac{(-1)^{n}}{(\alpha-vx)^{n+1}}.
\end{equation}
Thus (\ref{eq:Qres}) is solved by the ansatz
\begin{equation}
Q_{\alpha}(x)=P_{\alpha}(x)-\sum_{\kappa\in\mathcal{K}\cup\{\alpha\}}\sum_{n=0}^{N_{\kappa}-1}g_{-n-1}(\kappa)(-1)^{n}P_{-\kappa}^{[n]}(x)
\end{equation}
where the second term can be rewritten as a residue
\begin{equation}
Q_{\alpha}(x)=P_{\alpha}(x)-\sum_{\kappa\in\mathcal{K}\cup\{\alpha\}}\mathop{\text{res}}_{\ \kappa'=\kappa}g(\kappa')P_{-\kappa'}(x).\label{eq:Qfunctional}
\end{equation}
Using the residue we do not have to introduce the $g_{-n}$ coefficients
explicitly and the solution can be written into a compact form. Note,
however that $g_{-n}$ contains the expansion of $Q_{\alpha}$ around
$\kappa$, which are unknowns to be determined. We thus have to formulate
equations for these unknowns. This can be done in a compact way by
noting that $P_{\alpha}(x)=A_{\alpha,-vx}$ : 

\begin{equation}
Q_{\alpha}(x/v)=A_{\alpha,-x}-\sum_{\kappa\in\mathcal{K}\cup\{\alpha\}}\mathop{\text{res}}_{\ \kappa'=\kappa}Q_{\alpha}(\kappa'/v)e^{-2\kappa'B}\sigma(i\kappa'+0)A_{-\kappa',-x}\label{eq:Xeq}
\end{equation}
We can think of this equation as a linear operator acting on the unknowns
$Q_{\alpha}(x/v)$: 
\begin{equation}
Q_{\alpha}(x/v)\left(\mathds{1}-\mathds{A}\right)=A_{\alpha,-x}
\end{equation}
where $\mathds{A}$ acts on an arbitrary test function $\varphi(x)$
to the left as 
\begin{equation}
\varphi(x)\mathds{A}=\sum_{\kappa\in\mathcal{K}\cup\{\alpha\}}\mathop{\text{res}}_{\ \kappa'=\kappa}\varphi(\kappa')d(\kappa')A_{-\kappa',-x},
\end{equation}
with $d(\kappa')\equiv-e^{-2\kappa'B}\sigma(i\kappa'+0)$. Note that
here the poles at $\alpha\to-\beta$ in $A_{\alpha,\beta}$ has to
be taken into accout. We may solve (\ref{eq:Xeq}) iteratively, via
the Neumann-series of the operator $\mathds{A}$:
\begin{align}
Q_{\alpha}(x/v) & =\sum_{n=0}^{\infty}A_{\alpha,-x}\mathds{A}^{n}=A_{\alpha,-x}+\sum_{\kappa^{(1)}\in\mathcal{K}\cup\{\alpha\}}\mathop{\text{res}}_{\ \kappa'=\kappa^{(1)}}A_{\alpha,-\kappa'}d(\kappa')A_{-\kappa',-x}\nonumber \\
 & +\sum_{\substack{\kappa^{(2)}\in\mathcal{K}\cup\{\alpha\}\\
\kappa^{(1)}\in\mathcal{K}
}
}\mathop{\text{res}}_{\ \kappa''=\kappa^{(2)}}\mathop{\text{res}}_{\ \kappa'=\kappa^{(1)}}A_{\alpha,-\kappa''}d(\kappa'')A_{-\kappa'',-\kappa'}d(\kappa')A_{-\kappa',-x}+\ldots.\label{eq:Xsol}
\end{align}
The general observables then can be expressed as 
\begin{equation}
W_{\alpha,\beta}=\frac{1}{\alpha+\beta}+\frac{v}{\pi}\int_{C_{+}}\mathrm{d}x\ \frac{e^{-x}\mathcal{A}(x)Q_{\alpha}(x)}{\beta-vx}+\sum_{\kappa\in\mathcal{K}\cup\{\alpha,\beta\}}\mathop{\text{res}}_{\ \kappa'=\kappa}\frac{g(\kappa')}{\kappa'-\beta},
\end{equation}
and after substituting the ansatz (\ref{eq:Qfunctional}), and carefully
analyzing all possible situations (that is, whether $\alpha,\beta$
coincides with each other or any of the positions in $\mathcal{K}$)
we arrive at 
\begin{equation}
W_{\alpha,\beta}=A_{\alpha,\beta}-\sum_{\kappa\in\mathcal{K}\cup\{\alpha,\beta\}}\mathop{\text{res}}_{\ \kappa'=\kappa}g(\kappa')A_{-\kappa',\beta}.
\end{equation}
Via the iterative solution (\ref{eq:Xsol}) it expands to the manifestly
$\alpha\leftrightarrow\beta$ symmetric formula
\begin{align}
W_{\alpha,\beta}=A_{\alpha,\beta}+\sum_{n=1}^{\infty}\sum_{\substack{\kappa^{(1)},\ldots,\kappa^{(n)}\\
\in\mathcal{K}\cup\{\alpha,\beta\}
}
}\mathop{\text{res}}_{\ x_{1}=\kappa^{(1)}}\ldots\mathop{\text{res}}_{\ x_{n}=\kappa^{(n)}}A_{\alpha,-x_{1}}d(x_{1})A_{-x_{1},-x_{2}}d(x_{2})\ldots A_{-x_{n},\beta},
\end{align}
where in the sums we can drop the explicit $\{\alpha,\beta\}$ from
the union, except for $\kappa^{(1)},\kappa^{(n)}$. That is, we have
$\kappa^{(2)},\ldots,\kappa^{(n-1)}\in\mathcal{K}$, while $\kappa^{(1)}\in\mathcal{K}\cup\{\alpha\}$
and $\kappa^{(n)}\in\mathcal{K}\cup\{\beta\}$ with the only exception
for $n=1$, where $\kappa^{(1)}\in\mathcal{K}\cup\{\alpha,\beta\}$.

For the limit $w_{\alpha}\equiv W_{\alpha,\infty}=\lim_{\beta\to\infty}\beta W_{\alpha,\beta}$
we can separate $\beta$ as an arbitrary real parameter $\beta\notin\mathcal{K}\cup\{\alpha\}$
and take the separate residues. This will lead to terms proportional
to $d(\beta)$ that vanishes in the $\beta\to\infty$ limit. The remaining
terms are

\begin{equation}
w_{\alpha}=a_{\alpha}+\sum_{n=1}^{\infty}\sum_{\kappa^{(1)}\in\mathcal{K}\cup\{\alpha\}}\sum_{\kappa^{(2)},\ldots,\kappa^{(n)}\in\mathcal{K}}\mathop{\text{res}}_{\ x_{1}=\kappa^{(1)}}\ldots\mathop{\text{res}}_{\ x_{n}=\kappa^{(n)}}A_{\alpha,-x_{1}}d(x_{1})A_{-x_{1},-x_{2}}d(x_{2})\ldots a_{-x_{n}}.
\end{equation}
For the special case of $\alpha,\beta\notin\mathcal{K}$, where still
higher order poles might be present at the locations $\mathcal{K}$
in the kernel $\sigma(i\kappa+0)$, the result has a very similar
structure as the one presented in Subsection (\ref{subsec:Trans-series}),
and we only have to replace the dressed quantity $\hat{A}_{\alpha,\beta}$
with 
\begin{equation}
\hat{A}_{\alpha,\beta}\to A_{\alpha,\beta}+\sum_{n=1}^{\infty}\sum_{\kappa^{(1)},\ldots,\kappa^{(n)}\in\mathcal{K}}\mathop{\text{res}}_{\ x_{1}=\kappa^{(1)}}\ldots\mathop{\text{res}}_{\ x_{n}=\kappa^{(n)}}A_{\alpha,-x_{1}}d(x_{1})A_{-x_{1},-x_{2}}d(x_{2})\ldots A_{-x_{n},\beta}.
\end{equation}
We did not consider the $\alpha=0$ or $\beta=0$ and $\alpha,\beta=0$
limits of the moments directly in this calculation, in principle they
should be determined by the differential equations.

Note that due to the poles in the above expressions being higher order
in general, taking the residues will differentiate both the perturbative
building blocks and the exponential factors $e^{-2B\kappa}$ possibly
multiple times. The latter will contribute via powers of $2B$, that
is, $\ln v$ terms will appear in the $v$ language.

This solution was deduced in an abstract way and we present it only
for the completeness of our method. However, it goes beyond the aim
of this work to perform numerical or analytical checks in concrete
problems where such a situation would appear.

\section{Calculation of the free-energy density \label{sec:Falpha}}

In this Appendix we calculate the trans-series of
the free energy density ${\cal F}$ in the running coupling $\alpha$,
which is defined through the external field $h$. First we derive
the trans-series of ${\cal F}$ in the running coupling $v$, then
we change to the coupling to $\alpha$. Finally, we recall how the
relation between the mass gap and the dynamically generated scale
can be obtained. 

\subsection{Free energy density in the coupling $v$ }

The free energy density is defined as the Legrende transform of the
groundstate energy density (\ref{eq:Fh},\ref{eq:FhbarO}) in the
presence of an external field $h$, (\ref{eq:hchi0overchi0}). Although
it is straightforward to evaluate the trans-series solution of each
building block, we sketch an alternative route here based on the
trans-series of the observables with bars. This approach can have
other applications and completes the solution of the integral equations
with various sources. The integral equation with a $\sin\alpha\theta$
source can be solved analogously to the
$\cos\alpha\theta$ case. Functions here are anti-symmetric rather than symmetric
and the perturbative part satisfies the integral equation 
\begin{equation}
\bar{P}_{\alpha}(x)-\int_{C_{+}}\frac{e^{-y}\mathcal{A}(y)\bar{P}_{\alpha}(y)}{x+y}\frac{dy}{\pi}=\frac{1}{\alpha-vx},
\end{equation}
which, formally, can be obtained from the symmetric case via flipping
the sign of the kernel $\mathcal{A}(x)\to-\mathcal{A}(x)$. We define
the perturbative basis for $\beta\neq-\alpha$ as $\bar{A}_{\alpha,\beta}=\bar{P}_{\alpha}(-\beta/v)$,
while the $\alpha=-\beta$ pole is removed again as in 
(\ref{eq:Aab}). Analogously to ((\ref{eq:Onm}),(\ref{eq:chin}))
the normalized quantities
\begin{align}
\bar{\mathcal{O}}_{\alpha,\beta} & =\frac{e^{(\alpha+\beta)B}}{4\pi}G_{+}(i\alpha)G_{+}(i\beta)\bar{W}_{\alpha,\beta}\quad\alpha,\beta>0\\
\bar{\chi}_{\alpha} & =\frac{e^{\alpha B}}{2}G_{+}(i\alpha)\bar{w}_{\alpha},\quad\alpha>0
\end{align}
can be obtained in the form
\begin{align}
\bar{W}_{\alpha,\beta} & =\hat{\bar{A}}_{\alpha,\beta}-d_{\alpha}\hat{\bar{A}}_{-\alpha,\beta}-d_{\beta}\hat{\bar{A}}_{\alpha,-\beta}+d_{\alpha}d_{\beta}\hat{\bar{A}}_{-\alpha,-\beta}\label{eq:WbarTS}\\
\bar{w}_{\alpha} & =\hat{\bar{a}}_{\alpha}-d_{\alpha}\hat{\bar{a}}_{-\alpha}\label{eq:wbarTS}
\end{align}
with the dressed quantities 
\begin{equation}
\hat{\bar{A}}_{\alpha,\beta}=\bar{A}_{\alpha,\beta}+\sum_{r,s}\bar{A}_{\alpha,-\kappa_{s}}\mathcal{\bar{A}}_{-\kappa_{r},-\kappa_{s}}\bar{A}_{-\kappa_{s},\beta},\quad\hat{\bar{a}}_{\alpha,\beta}\equiv\hat{\bar{A}}_{\alpha,\infty}.
\end{equation}
Here $\mathcal{\bar{A}}_{-\kappa_{r},-\kappa_{s}}$ has the same structure
as ${\cal A}_{-\kappa_{r},-\kappa_{s}}$ in (\ref{eq:calAA}), but
one has to replace $A_{-\kappa_{l_{1}},-\kappa_{l_{2}}}\to\bar{A}_{-\kappa_{l_{1}},-\kappa_{l_{2}}}$and
$d_{\kappa_{l}}\to-d_{\kappa_{l}}$.

In summary, the trans-series of the observables with a bar can be
obtained simply by replacing every perturbative building block with
its bar version, and flipping the sign of all $d_{\kappa_{l}}$
and $d_{\alpha}$. The perturbative building blocks can be directly
calculated from (\ref{eq:barfromnobar}) as 
\begin{align}
\bar{A}_{\alpha,\beta} & =\frac{\beta}{\alpha}\left(\frac{a_{\alpha}}{a_{0}}A_{0,\beta}-A_{\alpha,\beta}\right)\quad;\quad\bar{a}_{\alpha}
=\alpha\frac{A_{0,\alpha}}{a_{0}}.\label{eq:barAab-1}
\end{align}
These formulas can be used to write ${\cal F}=-m^{2}\bar{{\cal O}}_{1,1}$
as the function of the running coupling $v$. 

\subsection{Free energy density in the coupling $\alpha$ }

We now would like to express $\mathcal{F}$ in terms of a coupling
$\alpha$, which resums $\ln h$ and all higher logarithmic $\ln\ln h\dots$
terms and is defined by
\begin{equation}
\frac{1}{\alpha}+\xi\ln\alpha=\ln\frac{h}{\Lambda}.\label{eq:alphacoupling}
\end{equation}
Here $\xi$ needs to be fixed so that the change of variable $v\to\alpha$
does not introduce $\ln\alpha$ terms in the trans-series for ${\cal F}$.
The parameter $\Lambda$ is an arbitrary dimensionful parameter that
is proportional to the dynamically generated scale and the mass of
the particles. To calculate the free energy density in powers of $\alpha$
and $e^{-1/\alpha}$ , we use the same procedure as in \cite{Bajnok:2024qro}.
At first we express $\alpha(v)$ as a trans-series in $v$, then substitute
it into a a trans-series ansatz for $\mathcal{F}(\alpha)$, which
we fix by requiring that it agrees with $\mathcal{F}(v)$, i.e. $\mathcal{F}(\alpha)\vert_{\alpha=\alpha(v)}\overset{!}{=}\mathcal{F}(v).$
In writing $\alpha(v)$ we introduce $\hat{h}$ as a trans-series
whose perturbative expansion starts with $\hat{h}=1+O(v)$: 
\begin{equation}
\frac{h}{m}=\frac{\chi_{1}}{\chi_{0}}=\frac{e^{B}G_{+}(i)}{2}\frac{w_{1}}{w_{0}}\equiv e^{B}C_{\delta}(\sqrt{2v})^{\delta}\hat{h}\label{eq:hchi}
\end{equation}
where 
\begin{equation}
C_{\delta}\equiv\frac{e^{\frac{b}{2}}}{2}\frac{H(1)}{H(0)}\left(\sqrt{\frac{2}{\pi}}\right)^{\delta},\quad\hat{h}\equiv H(0)\left(\frac{\sqrt{\pi}}{\sqrt{v}2}\right)^{\delta}\frac{w_{1}}{w_{0}}.\label{eq:Cdelta}
\end{equation}
To express the trans-series of $\alpha$ in terms of $v$ we need
to solve (\ref{eq:alphacoupling}), that leads to the following equation:
\begin{equation}
\frac{2}{\alpha}+2\xi\ln\alpha=\frac{1}{v}+L+(\delta-a)\ln v+\delta\ln2+2\ln C_{\delta}+2\ln\hat{h}.
\end{equation}
where we used the definition of the running coupling (\ref{eq:runningcoupling})
to substitute $B$. Since at leading order $\alpha=2v+O(v^{2})$,
we need
\begin{equation}
\xi=\frac{\delta-a}{2}
\end{equation}
to drop out $\ln v$ and eventually $\ln\alpha$ terms. Then, as every
term in $\hat{h}$ is a power series in $v$ (and the non-perturbative
parameter $\nu$) log terms will not appear in $\alpha$. It is convenient
to introduce the ratio 
\begin{equation}
Y\equiv\frac{\alpha}{2v}=1+y_{1}v+O(v)\quad;\quad y_{1}=-L-a\ln2-2\ln\left(C_{\delta}\frac{m}{\Lambda}\right)\label{eq:y1}
\end{equation}
as then we have to solve
\begin{equation}
1+Y\left[v\left(2\xi\ln Y+y_{1}-2\ln\hat{h}\right)-1\right]=0\label{eq:Yeq}
\end{equation}
which can be done in an iterative manner, taking an ansatz for $Y$,
expanding it in $v$ and $\nu$ and fixing its coefficients order
by order. Alternatively, one may also differentiate both sides of
(\ref{eq:alphacoupling}) w.r.t. $B$ to obtain
\begin{equation}
\frac{d}{dB}\left(\frac{1}{\alpha}+\xi\ln\alpha\right)=\frac{\dot{h}}{h}=\frac{\bar{\chi}_{1}}{\chi_{1}}
\end{equation}
where we used (\ref{eq:hchi}). Changing variables to $v$ leads to
\begin{equation}
\left(1-(\delta-a)vY\right)\left(v\frac{dY}{dv}+Y\right)w_{1}=\bar{w}_{1}(1+av)Y^{2}\label{eq:Ydiff-1}
\end{equation}
with $v\frac{d}{dv}=v\partial_{v}+\nu\left(v^{-1}+a\right)\partial_{\nu}$,
where we also used that $\dot{v}=-2v^{2}(1+av)^{-1}$. The coefficient
$y_{1}$ cannot be fixed from (\ref{eq:Ydiff-1}) alone, we have to
resort to (\ref{eq:Yeq}) to arrive at (\ref{eq:y1}). 

In order to express ${\cal F}$ in terms of $\alpha$ as a trans-series
we need an appropriate non-perturbative expansion parameter. It can
be defined as
\begin{equation}
\lambda\equiv C_{\delta}^{2}\left(\frac{m}{h}\right)^{2},\label{eq:invlambda}
\end{equation}
such that 
\begin{equation}
\tilde{\lambda}=\lambda\alpha^{\delta}=C_{\delta}^{2}\left(\frac{m}{\Lambda}\right)^{2}e^{-2/\alpha}\alpha^{a}=Y^{\delta}\hat{h}^{-2}\nu=\nu\cdot(1+O(v)).\label{eq:lambdadef}
\end{equation}
 Note that $\tilde{\lambda}$ is proportional to $\nu$, but expanding
it in $v$ and $\nu$ also introduces an infinite trans-series due
to the prefactors. We might also collect the prefactors in ${\cal F}$
\begin{equation}
\mathcal{F}=-m^{2}\frac{e^{2B}}{4\pi}G_{+}^{2}(i)\bar{W}_{1,1}\equiv-h^{2}k_{\delta}^{2}\alpha^{-\delta}\hat{\mathcal{F}},
\end{equation}
where the constant $k_{\delta}$ and the normalized trans-series $\hat{\mathcal{F}}$
turns out to be
\begin{equation}
k_{\delta}^{2}=\frac{1}{2\pi}\left(\frac{\pi}{2}\right)^{\delta}H^{2}(0),\quad\hat{\mathcal{F}}=2Y^{\delta}\hat{h}^{-2}\bar{W}_{1,1}=1+O(v).
\end{equation}

To solve the equation for $Y(v,\nu)$ and then substitute it into
$\hat{\mathcal{F}}(\alpha(v,\nu),\tilde{\lambda}(v,\nu))$ we need
ans\"atze for these trans-series. To simplify the situation we restrict
ourselves to a single family of poles of $\sigma$: $\kappa_{l}=\kappa_{1}l$.
Then the expansion goes only in powers of $\nu_{1}\equiv\nu^{\kappa_{1}}$,
except for the explicit sectors proportional to powers of $d_{1}\propto\nu$
in ((\ref{eq:Waa}),(\ref{eq:wbarTS})). Note however, that in most
of the cases listed in section \ref{sec:Explicit} the terms proportional
to $d_{1}$ vanish due to $\sigma(i+0)=0$. The only term remaining
proportional to $\nu$ is the single constant term with $\sigma'(i+0)$
coming from the coinciding limit of $W_{1,1}$ or $\bar{W}_{1,1}$,
that can be transformed separately. Considering then only this simpler
scenario we can make an ansatz:
\begin{align}
Y(v,\nu) & =\sum_{l=0}^{\infty}Y_{l}(v)\nu_{1}^{l} & Y_{l}(v) & \sim\sum_{k=0}^{\infty}y_{k,l}v^{k}
\end{align}
and for the free energy itself, with $\tilde{\lambda}_{1}\equiv\tilde{\lambda}^{\kappa_{1}}$
\begin{align}
\hat{\mathcal{F}}(\alpha,\tilde{\lambda}) & =-M\tilde{\lambda}+\sum_{l=0}^{\infty}F_{l}(\alpha)\tilde{\lambda}_{1}^{l} & F_{l}(\alpha) & \sim\sum_{k=0}^{\infty}f_{k,l}\alpha^{k}\label{eq:Fansatz}
\end{align}
where $M=-2i\sigma'(i+0)$ is a constant, related to the bulk energy
density. There are only positive powers of $\alpha$ in the ansatz
(\ref{eq:Fansatz}), since when matching the two trans-series
\begin{equation}
\hat{\mathcal{F}}(\alpha=2vY,\tilde{\lambda}=Y^{\delta}\hat{h}^{-2}\nu)\overset{!}{=}2Y^{\delta}\hat{h}^{-2}\bar{W}_{1,1}
\end{equation}
the powers of $\tilde{\lambda}_{1}=\nu_{1}\cdot(1+O(v)$) will not
introduce any additional powers of $v$, and the expression on the
r.h.s. contains only positive powers of $v$. The final result then
takes the form 
\begin{equation}
\mathcal{F}=-\frac{h^{2}H^{2}(0)}{2\pi}\left(\frac{\pi}{2}\right)^{\delta}\left\{ -M\lambda+\sum_{l=0}^{\infty}F_{l}(\alpha)\alpha^{\delta(\kappa_{1}l-1)}\lambda^{\kappa_{1}l}\right\}, 
\end{equation}
where $\lambda$ is defined in (\ref{eq:invlambda})
and is independent of the choice of scale $\Lambda$. Changing the latter then only affects
the perturbative expansions $F_{l}(\alpha)\alpha^{\delta(\kappa_{1}l-1)}$.
That is, to transform the formula to a coupling $\alpha'$ defined
analogously to (\ref{eq:alphacoupling}) with $\Lambda'$, we need
to first relate the two couplings by solving 
\begin{align}
X(\alpha')\left[1-\alpha'\left(\xi\ln X(\alpha')+\ln\frac{\Lambda}{\Lambda'}\right)\right] & =1\label{eq:Xeq-1}
\end{align}
for $X\equiv\frac{\alpha}{\alpha'}$, then substituting $\alpha\to\alpha'X(\alpha')$
in $\mathcal{F}$. Note however that (\ref{eq:Xeq-1}) is purely pertubative.
An alternative way is to simply change the $y_{1}$ parameter defined
in (\ref{eq:y1}) to 
\begin{equation}
y_{1}'=y_{1}-2\ln\frac{\Lambda}{\Lambda'}
\end{equation}
in formulas of $F_{l}(\alpha)$. Finally, expressed as a trans-series
of $\alpha$, the free energy density is

\begin{equation}
\mathcal{F}=-\frac{h^{2}H^{2}(0)}{2\pi}\left(\frac{\pi}{2}\right)^{\delta}\left\{ -MC_{\delta}^{2}\alpha^{a-\delta}e^{-2/\alpha}+\sum_{l=0}^{\infty}C_{\delta}^{2\kappa_{1}l}F_{l}(\alpha)\alpha^{a\kappa_{1}l-\delta}e^{-2\kappa_{1}l/\alpha}\right\} ,
\end{equation}
where $C_{\delta}$ is defined in (\ref{eq:Cdelta}). 

For completeness, we present the coefficients of $F_{l}(\alpha)$-s
for the bosonic and fermionic cases separately in
Appendix \ref{sec:Coefficients}
up to $l=2$ and up to the $\alpha^{2}$  perturbative order .

\subsection{Relation to the mass-gap}

In this subsection we recall how the massgap can be obtained by comparing
the Wiener-Hopf solution to the results of standard perturbative field theory.

In asymptotically free field theories physical quantities can be calculated
perturbatively in terms of the renormalized coupling $\alpha_X(\mu)$. Here
$\mu$ is the renormalization scale and the $\mu$-dependence of the coupling
is governed by the renormalization group (RG) equation
\begin{equation}
\mu\frac{{\rm d}\alpha_X(\mu)}{{\rm d}\mu}=\beta(\alpha_X(\mu)).  
\end{equation}
The two leading terms in the expansion of the beta function   
\begin{equation}
\beta(z)=-\beta_0z^2-\beta_1z^3+O(z^4)  
\end{equation}
are renormalization scheme independent. Let us introduce the subtracted inverse
of the beta function by
\begin{equation}
{\cal D}(z)=\frac{1}{\beta(z)}+\frac{1}{\beta_0 z^2}-\frac{\xi}{z},\qquad
\xi=\frac{\beta_1}{\beta_0^2}.  
\end{equation}
This function has regular small $z$ expansion. Defining
\begin{equation}
f(z)=\frac{1}{\beta_0z}+\xi\ln(\beta_0z)
+\int_0^z{\cal D} (z')
{\rm d}z'
\end{equation}
it is easy to see that  
\begin{equation}
\Lambda=\mu\exp\{-f(\alpha_X(\mu))\} 
\end{equation}
is RG invariant.  
Comparing this expression with the definition of the $\alpha$ coupling
(\ref{eq:alphacoupling}) we can see that provided
\begin{equation}
\xi=\frac{\beta_1}{\beta_0^2}=\frac{\delta-a}{2},
\label{cons0}  
\end{equation}
$\beta_{0}\alpha_{X}(h)=\alpha+O(\alpha^{3})$.
The relation (\ref{cons0}) is satisfied in all cases studied in this paper and
provides a bridge between the field theory perturbative approach and our
bootstrap based Wiener-Hopf treatment.

Using the perturbative
coefficients of $\hat{\mathcal{F}}$ from the Wiener-Hopf result,
one can match them to the coefficients of the $\alpha_X(h)$ expansion obtained
perturbatively in the field theory.
For the bosonic case this comparison gives 
\begin{equation}
\mathcal{F}=-h^{2}\left(\alpha_{X}^{-1}f_{0}+f_{1}+O(\alpha_{X})\right)=-h^{2}k_{1}^{2}\left(\alpha^{-1}+\frac{y_{1}+a-2}{2}+O(\alpha)\right)
\end{equation}
which means 
\begin{align}
k_{1}^{2} & =f_{0}\beta_{0}=\frac{H^{2}(0)}{4}, & y_{1} & =2\left(1+\frac{f_{1}}{\beta_{0}f_{0}}\right)-a.
\end{align}
The first equality, similarly to (\ref{cons0}), is a consistency relation
while the second expression gives the massgap relation through formula
(\ref{eq:y1}) for $y_{1}$:
\begin{equation}
\frac{m}{\Lambda}=\frac{\sqrt{2\pi}}{e}\frac{H(0)}{H(1)}\exp\left(-\frac{z_{1}+a\left(\gamma_{E}+3\ln2-1\right)}{2}-\frac{f_{1}}{\beta_{0}f_{0}}\right).
\end{equation}

For the fermionic models the free energy density is 
\begin{align}
\mathcal{F} & =-h^{2}\left(f_{0}+f_{1}\alpha_{X}+f_{2}\alpha_{X}^{2}+O(\alpha_{X}^{3})\right)\nonumber \\
 & =-h^{2}k_{0}^{2}\left(1-\frac{a\alpha}{2}+\frac{1}{8}a\alpha^{2}(3a+2y_{1}-4)+O\left(\alpha^{3}\right)\right)
\end{align}
and matching the expansions give the following relations
\begin{align}
k_{0}^{2} & =f_{0}=\frac{H^{2}(0)}{2\pi}, & \frac{f_{1}}{\beta_{0}f_{0}} & =-\frac{a}{2}, & y_{1} & =2\left(1-\frac{f_{2}}{\beta_{0}f_{1}}\right)-\frac{3a}{2}.
\end{align}
Finally, the latter gives the mass-gap as 
\begin{equation}
\frac{m}{\Lambda}=\frac{2}{e}\frac{H(0)}{H(1)}\exp\left(-\frac{z_{1}+a(\gamma_{E}+\ln2-3/2)}{2}+\frac{f_{2}}{\beta_{0}f_{1}}\right).
\end{equation}
The mass gap was calculated with this method in \cite{Hasenfratz:1990ab,Hasenfratz:1990zz}
for the O(N) models and for other relativistic models in \cite{Forgacs:1991nk,Forgacs:1991rs,Balog:1992cm,Evans:1994sv,Evans:1994sy,Evans:1995dn}. 

\section{Perturbative coefficients for bosonic and fermionic models\label{sec:Coefficients}}

In this Appendix we present the first few perturbative coefficients
for bosonic and fermionic models. 

\subsection{Bosonic models}

Volin originally calculated $2A_{1,1}$ for the $O(N)$ model with
the conventions $a=1-2\Delta,\,b=2\Delta(1-\ln\Delta)-(1+\ln2)$ and
$L=-b-4\Delta\ln2$. In his algorithm it is easy to trace back the
$\zeta_{k}$ expressions coming from the ${\cal A}(x)$ kernel. One
can then use the $O(N)$ relation $\zeta_{2k+1}=z_{2k+1}(2k+1)/(2\Delta^{2k+1}-2+2^{-2k})$
to replace these $\zeta_{2k+1}$ with $z_{2k+1}$ and $\Delta$ with
$(1-a)/2$. In this way we can obtain a result, which is valid for
the generic kernel (\ref{eq:calA}). Performing this calculation we
could easily obtain more than 20 terms analytically. In specific models
one can even go to few thousand terms numerically \cite{Abbott:2020qnl,Abbott:2020mba,Bajnok:2021zjm}.
For demonstration we present here the first few terms. For the observable
$A_{1,1}$ we found 
\begin{align}
2A_{1,1} & =1+\frac{v}{2}+\left(-\frac{5a}{4}+\frac{9}{8}\right)v^{2}+\left(\frac{10a^{2}}{3}-\frac{53a}{8}+\frac{57}{16}\right)v^{3}\\
 & +\frac{v^{4}}{384}\left(36a^{3}(21\zeta_{3}-94)+10924a^{2}-13344a+9(144z_{3}+625)\right)\nonumber \\
 & +\frac{v^{5}}{3840}\left(816156a^{2}-2400a(76z_{3}+327)+405(272z_{3}+705)\right)\nonumber \\
 & +\frac{v^{5}}{3840}\left(-160a^{4}(665\zeta_{3}-562)+140a^{3}(459\zeta_{3}-2882)\right)+O\left(v^{6}\right)\nonumber 
\end{align}
Now, we can use the differential equation (\ref{eq:de1-1}) for $\alpha=\beta=1$
to obtain
\begin{align}
a_{1} & =1+\frac{v}{4}+\left(-\frac{5a}{8}+\frac{9}{32}\right)v^{2}+\left(\frac{5a^{2}}{3}-\frac{53a}{32}+\frac{75}{128}\right)v^{3}\\
 & +\frac{v^{4}}{6144}\left(9(1152z_{3}+1225)+288a^{3}(21\zeta_{3}-94)+43696a^{2}-35160a\right)\nonumber \\
 & +\frac{v^{5}}{122880}\left(4304496a^{2}-120a(24320z_{3}+25683)+405(2176z_{3}+2205)\right)\nonumber \\
 & +\frac{v^{5}}{122880}\left(-2560a^{4}(665\zeta_{3}-562)+1120a^{3}(459\zeta_{3}-2882)\right)+O\left(v^{6}\right)\nonumber 
\end{align}
Then using (\ref{eq:de3-1}) for $\alpha=1$ we can determine the
perturbative part of the master function $f$
\begin{align}
	f=&-v^{2}+6av^{3}-26a^{2}v^{4}+v^{5}\left(-\frac{a^{3}}{4}(63\zeta_{3}-386)-27z_{3}\right)\\
	&+v^{6}\left(\frac{a^{4}}{6}(1757\zeta_{3}-1984)+502az_{3}\right)+O\left(v^{7}\right).\nonumber
\end{align}
By solving (\ref{eq:de3-1}) for other $\alpha$-s we get 
\begin{align}
a_{\alpha} & =1+\frac{v}{4\alpha}+\frac{v^{2}(-20a\alpha+9)}{32\alpha^{2}}+\frac{v^{3}\left(640a^{2}\alpha^{2}-636a\alpha+225\right)}{384\alpha^{3}}\\
 & \quad+\frac{v^{4}\left(288\alpha^{3}\left(-a^{3}(94-21\zeta_{3})+36z_{3}\right)+43696a^{2}\alpha^{2}-35160a\alpha+11025\right)}{6144\alpha^{4}}\nonumber \\
 & +\frac{v^{5}\left(4304496a^{2}\alpha^{2}-3081960a\alpha+893025\right)}{122880\alpha^{5}}\nonumber \\
 & +\frac{v^{5}\left(-2560\alpha^{4}\left(a^{4}(665\zeta_{3}-562)+1140az_{3}\right)\right)}{122880\alpha^{5}}\nonumber \\
 & +\frac{v^{5}\left(-160\alpha^{3}\left(-7a^{3}(459\zeta_{3}-2882)-5508z_{3}\right)\right)}{122880\alpha^{5}}+O\left(v^{6}\right).\nonumber 
\end{align}
One can check that $a_{\alpha}$ can be obtained directly from $a_{1}$
by the $v\to\frac{v}{\alpha}$, $a\to a\alpha$, $z_{2k+1}\to\alpha^{2k+1}z_{2k+1}$
replacements. This is merely an interesting observation, and we are
not aware of any derivation of it yet. For $a_{0}$ the overall constant
can be determined from the leading term in the explicit iterative
solution of the Wiener-Hopf integral equation \cite{Bajnok:2022rtu}.
The exceptional $a_{0}$ behaves as 
\begin{align}
a_{0} & =\frac{H(0)\sqrt{\pi}}{2\sqrt{v}}\biggl(1+\frac{av}{2}-\frac{5a^{2}v^{2}}{8}+\frac{1}{16}v^{3}\left(-a^{3}(7\zeta_{3}-15)-12z_{3}\right)\\
 & \qquad\qquad\qquad+\frac{1}{384}v^{4}\left(1484a^{4}\zeta_{3}-655a^{4}+2544az_{3}\right)+O(v^{5})\biggr)\nonumber 
\end{align}
The inclusion of $H(0)$ is crucial to get the $\lim_{\alpha\to0}w_{\alpha}=w_{0}$
correctly. 

Finally, by solving (\ref{eq:de1}) we obtain our basic building blocks
%\begin{align}
%A_{\alpha,\beta} & =\frac{1}{\beta+\alpha}+\frac{v}{4\beta\alpha}+\frac{v^{2}(-20a\beta\alpha+9\beta+9\alpha)}{32\beta^{2}\alpha^{2}}\\
% & \quad+\frac{v^{3}\left(\beta^{2}\left(640a^{2}\alpha^{2}-636a\alpha+225\right)+6\beta\alpha(-106a\alpha+39)+225\alpha^{2}\right)}{384\beta^{3}\alpha^{3}}\nonumber \\
% & \quad+\frac{v^{4}\left(\beta^{2}\alpha\left(43696a^{2}\alpha^{2}-36432a\alpha+11475\right)+15\beta\alpha^{2}(-2344a\alpha+765)+11025\alpha^{3}\right)}{6144\beta^{4}\alpha^{4}}\nonumber \\
% & \quad+\frac{v^{4}\left(\beta^{3}\left(288\alpha^{3}\left(-a^{3}(94-21\zeta_{3})+36z_{3}\right)+43696a^{2}\alpha^{2}-35160a\alpha+11025\right)\right)}{6144\beta^{4}\alpha^{4}}\nonumber \\
% & \quad+\frac{v^{5}\left(\beta^{4}\left(-2560\alpha^{4}\left(a^{4}(665\zeta_{3}-562)+1140az_{3}\right)-160\alpha^{3}\left(-7a^{3}(459\zeta_{3}-2882)-5508z_{3}\right)\right)\right)}{122880\beta^{5}\alpha^{5}}\nonumber \\
% & \quad+\frac{v^{5}\left(\beta^{4}\left(4304496a^{2}\alpha^{2}-3081960a\alpha+893025\right)+4\beta^{3}\alpha\left(-40\alpha^{3}\left(-7a^{3}(459\zeta_{3}-2882)-5508z_{3}\right)\right)\right)}{122880\beta^{5}\alpha^{5}}\nonumber \\
% & \quad+\frac{v^{5}\left(4\beta^{3}\alpha\left(-40\alpha^{3}\left(-7a^{3}(459\zeta_{3}-2882)-5508z_{3}\right)+1112376a^{2}\alpha^{2}-799110a\alpha+231525\right)\right)}{122880\beta^{5}\alpha^{5}}\nonumber \\
% & \quad+\frac{v^{5}\left(6\beta^{2}\alpha^{2}\left(717416a^{2}\alpha^{2}-532740a\alpha+155025\right)+1260\beta\alpha^{3}(-2446a\alpha+735)+893025\alpha^{4}\right)}{122880\beta^{5}\alpha^{5}}+O(v^{6})\nonumber 
%\end{align}
\begin{align}
A_{\alpha,\beta} & =\frac{1}{\beta+\alpha}+\frac{v}{4\beta\alpha}+\frac{v^{2}(-20a\beta\alpha+9\beta+9\alpha)}{32\beta^{2}\alpha^{2}}\\
 & \quad+\frac{v^{3}\left(\beta^{2}\left(640a^{2}\alpha^{2}-636a\alpha+225\right)+6\beta\alpha(-106a\alpha+39)+225\alpha^{2}\right)}{384\beta^{3}\alpha^{3}}\nonumber \\
 & \quad+\frac{v^{4}\left(\beta^{2}\alpha\left(43696a^{2}\alpha^{2}-36432a\alpha+11475\right)+15\beta\alpha^{2}(-2344a\alpha+765)+11025\alpha^{3}\right)}{6144\beta^{4}\alpha^{4}}\nonumber \\
 & \quad+\frac{v^{4}\left(\beta^{3}\left(288\alpha^{3}\left(-a^{3}(94-21\zeta_{3})+36z_{3}\right)+43696a^{2}\alpha^{2}-35160a\alpha+11025\right)\right)}{6144\beta^{4}\alpha^{4}}\nonumber \\
 & \quad-\frac{v^{5}\left(a^{4}(665\zeta_{3}-562)+1140az_{3}\right)}{48\beta\alpha}
  +\frac{v^{5}\left(7a^{3}(459\zeta_{3}-2882)+5508z_{3}\right)}{768\beta\alpha^{2}}\nonumber \\ 
 & \quad+\frac{v^{5}\left(4304496a^{2}\alpha^{2}-3081960a\alpha+893025\right)}{122880\beta\alpha^{5}}
 -\frac{v^{5}\left(-7a^{3}(459\zeta_{3}-2882)-5508z_{3}\right)}{384\beta^{2}\alpha}\nonumber \\
 & \quad+\frac{v^{5}\left(1112376a^{2}\alpha^{2}-799110a\alpha+231525 \right)}{122880\beta^{5}\alpha^{5}}
 +\frac{v^{5}\left(717416a^{2}\alpha^{2}-532740 a\alpha+155025\right)}{20480\beta^{3}\alpha^{3}} \nonumber \\
 & \quad+\frac{v^{5}\left(1260\beta\alpha^{3}(-2446a\alpha+735)+893025\alpha^{4}\right)}{122880\beta^{5}\alpha^{5}}+O(v^{6})\nonumber. 
\end{align}

The exceptional terms can be derived from the differential equation
(\ref{eq:de1-1}) 
\begin{align}
A_{\alpha,0} & =\frac{H(0)\sqrt{\pi}}{2\sqrt{v}}\Bigg\lbrace\frac{1}{\alpha}+\frac{v(2a\alpha-3)}{4\alpha^{2}}+\frac{v^{2}(4a\alpha(8-5a\alpha)-15)}{32\alpha^{3}}\\
 & -\frac{v^{3}\left(24a^{3}\alpha^{3}\left(7\zeta_{3}-15\right)+836a^{2}\alpha^{2}-858a\alpha+9\left(32\alpha^{3}z_{3}+35\right)\right)}{384\alpha^{4}}\nonumber \\
 & +\frac{v^{4}\left(2\alpha^{3}a^{4}\left(1484\zeta_{3}-655\right)+8\alpha^{2}a^{3}\left(491-126\zeta_{3}\right)-6549\alpha a^{2}\right)}{768\alpha^{4}}\nonumber \\
 & +\frac{v^{4}\left(8\alpha\left(6a\left(848\alpha^{3}z_{3}+915\right)-1728\alpha^{2}z_{3}\right)-14175\right)}{6144\alpha^{5}}+O\left(v^{5}\right)\Bigg\rbrace\nonumber 
\end{align}
\begin{align}
A_{0,0} & =\frac{H^{2}(0)\pi}{4v}\Bigg\lbrace\frac{1}{4v}+a-\frac{a^{2}v}{4}+\frac{1}{16}v^{2}\left(a^{3}\left(7\zeta_{3}-2\right)+12z_{3}\right)\nonumber \\
 & +v^{3}\left(\frac{1}{48}a^{4}\left(10-77\zeta_{3}\right)-\frac{11az_{3}}{4}\right)+O\left(v^{4}\right)\Bigg\rbrace
\end{align}
where the integration constant ($-a^{2}v/4$ term) and prefactors
were fixed from the $\mathcal{O}_{0,0}=\lim_{\alpha\to0}\mathcal{O}_{\alpha,0}$
limit. 

The capacity $C^{(+)}$ starts as\footnote{Note that the published version of
ref. \cite{Bajnok:2022rtu} contained errors in eqs. (4.29) and (4.30).}

\begin{align}
C_{0}^{(+)}(s) & =1+2s(\mathsf{L}-1)+s^{2}\left(\mathsf{L}^{2}-2\right)+\frac{1}{2}s^{3}\left(2\mathsf{L}^{2}-3\zeta_{3}-1\right)+O(s^{4})\\
C_{2}^{(+)}(s) & =2i+is\left(2\mathsf{L}+\frac{3}{2}\right)+is^{2}\left(2\mathsf{L}+\frac{17}{16}\right)+is^{3}\left(\frac{3}{2}\zeta_{3}+\left(\frac{15}{16}-\mathsf{L}\right)\mathsf{L}+\frac{161}{192}\right)+O(s^{4})\\
C_{4}^{(+)}(s) & =(1+4i)+s\left((1+4i)\mathsf{L}+\left(\frac{1}{2}+\frac{3i}{2}\right)\right)+s^{2}\left((1+4i)\mathsf{L}+\left(\frac{1}{4}+\frac{37i}{32}\right)\right)\\
 & +s^{3}\left(\frac{3}{4}\left(1+4i\right)\zeta_{3}-\mathsf{L}\left(\frac{1}{2}(1+4i)\mathsf{L}-\left(\frac{3}{4}+\frac{91}{32}i\right)\right)+\left(\frac{43}{96}+\frac{1301i}{768}\right)\right)+O\left(\delta^{4}\right)
\end{align}
where we defined $\mathsf{L}\equiv-\ln(s/8)$.

Observables with bar can be obtained from (\ref{eq:barAab}) as
\begin{align}
\bar{A}_{\alpha,\beta} & =\frac{1}{\alpha+\beta}-\frac{3v}{4\alpha\beta}-\frac{v^{2}(-44a\alpha\beta+15\alpha+15\beta)}{32\alpha^{2}\beta^{2}}\\
 & +\frac{v^{3}\left(\alpha^{2}(4a\beta(237-320a\beta)-315)+6\alpha\beta(158a\beta-45)-315\beta^{2}\right)}{384\alpha^{3}\beta^{3}}\nonumber \\
 & -\frac{3v^{4}\left(\alpha^{3}\left(8\beta\left(4a^{3}\beta^{2}\left(35\zeta_{3}-178\right)+858a^{2}\beta-645a+240\beta^{2}z_{3}\right)+1575\right)\right)}{2048\alpha^{4}\beta^{4}}\nonumber \\
 & -\frac{3v^{4}\left(\alpha^{2}\beta(16a\beta(429a\beta-283)+1365)+15\alpha\beta^{2}(91-344a\beta)+1575\beta^{3}\right)}{2048\alpha^{4}\beta^{4}}\nonumber \\
 & +\frac{v^{5}\left(960a^{4}\beta^{3}\left(343\zeta_{3}-342\right)+20a^{3}\beta^{2}\left(27674-4095\zeta_{3}\right)-687498a^{2}\beta\right)}{15360\alpha\beta^{4}}\nonumber \\
 & +\frac{v^{5}\left(8\beta\left(315a\left(1792\beta^{3}z_{3}+1517\right)-140400\beta^{2}z_{3}\right)-1091475\right)}{122880\alpha\beta^{5}}\nonumber \\
 & -\frac{v^{5}\left(40a^{3}\beta^{3}\left(4095\zeta_{3}-27674\right)+1224744a^{2}\beta^{2}-844290a\beta+675\left(416\beta^{3}z_{3}+357\right)\right)}{30720\alpha^{2}\beta^{4}}\nonumber \\
 & +\frac{v^{5}\left(-6\alpha^{2}\beta^{2}(4a\beta(229166a\beta-140715)+159075)\right)}{122880\alpha^{5}\beta^{5}}\nonumber\\
 & +\frac{v^{5}\left(1260\alpha\beta^{3}(3034a\beta-765)-1091475\beta^{4}\right)}{122880\alpha^{5}\beta^{5}}+O\left(v^{6}\right)\nonumber
\end{align}
and
\begin{align}
\bar{a}_{\alpha} & =1-\frac{3v}{4\alpha}+\frac{v^{2}(44a\alpha-15)}{32\alpha^{2}}+\frac{v^{3}(4a\alpha(237-320a\alpha)-315)}{384\alpha^{3}}\\
 & -\frac{3v^{4}\left(8\alpha\left(4\alpha^{2}a^{3}\left(35\zeta_{3}-178\right)+858\alpha a^{2}-645a+240\alpha^{2}z_{3}\right)+1575\right)}{2048\alpha^{4}}\nonumber \\
 & +\frac{v^{5}\left(960\alpha^{3}a^{4}\left(343\zeta_{3}-342\right)+20\alpha^{2}a^{3}\left(27674-4095\zeta_{3}\right)-687498\alpha a^{2}\right)}{15360\alpha^{4}}+O\left(v^{6}\right)\nonumber \\
 & +\frac{v^{5}\left(8\alpha\left(315a\left(1792\alpha^{3}z_{3}+1517\right)-140400\alpha^{2}z_{3}\right)-1091475\right)}{122880\alpha^{5}}+O\left(v^{6}\right)\nonumber. 
\end{align}	

For the free energy density we obtained 
%\begin{align}
%F_{0}(\alpha) & =1+\frac{1}{2}\alpha(a+y_{1}-2)-\frac{\alpha^{2}}{4}(a-1)(a+y_{1}-2)+O\left(\alpha^{3}\right)\\
%F_{1}(\alpha) & =-\frac{2(iS_{\kappa_{1}}+\hat{S}_{\kappa_{1}})}{\kappa_{1}(\kappa_{1}-1)}\Bigg\lbrace\frac{1}{\kappa_{1}-1}+\frac{\alpha\left(-2a\kappa_{1}+\kappa_{1}-2\kappa_{1}y_{1}-1\right)}{4\kappa_{1}}\nonumber \\
% & +\frac{\alpha^{2}\left(\kappa_{1}\left(\kappa_{1}^{2}(2a+2y_{1}-1)^{2}+\kappa_{1}(2a(4a+4y_{1}-5)-4y_{1}+5)+2a-1\right)+3\right)}{32\kappa_{1}^{2}}+O\left(\alpha^{3}\right)\Bigg\rbrace\\
%F_{2}(\alpha) & =\frac{2(iS_{\kappa_{1}}+\hat{S}_{\kappa_{1}})^{2}}{(\kappa_{1}-1)^{2}\kappa_{1}^{2}}\Bigg\lbrace1+\frac{\alpha\left(4\kappa_{1}\left(-2\kappa_{1}(a+y_{1})+a+\kappa_{1}+y_{1}-1\right)+2\right)}{8\kappa_{1}}\nonumber \\
% & +\frac{\alpha^{2}\left(\kappa_{1}\left(a(2a-3)\kappa_{1}+\kappa_{1}^{2}(2a+2y_{1}-1)^{2}-2a(a+y_{1}-1)+2\kappa_{1}-2\kappa_{1}y_{1}^{2}-1\right)+3\right)}{8\kappa_{1}}+O\left(\alpha^{3}\right)\Bigg\rbrace\nonumber \\
% & -\frac{iS_{\kappa_{2}}+\hat{S}_{\kappa_{2}}}{(2\kappa_{1}-1)\kappa_{1}}\Bigg\lbrace\frac{1}{2\kappa_{1}-1}-\frac{\alpha\left(\kappa_{1}(4a+4\text{y1}-2)+1\right)}{8\kappa_{1}}\nonumber \\
% & +\frac{\alpha^{2}\left(2\kappa_{1}\left(4\kappa_{1}^{2}(2a+2\text{y1}-1)^{2}+2\kappa_{1}(2a(4a+4\text{y1}-5)-4\text{y1}+5)+2a-1\right)+3\right)}{128\kappa_{1}^{2}}+O\left(\alpha^{3}\right)\Bigg\rbrace
%\end{align}
\begin{align}
F_{0}(\alpha) & =1+\frac{1}{2}\alpha(a+y_{1}-2)-\frac{\alpha^{2}}{4}(a-1)(a+y_{1}-2)+O\left(\alpha^{3}\right)\\
F_{1}(\alpha) & =-\frac{2(iS_{\kappa_{1}}+\hat{S}_{\kappa_{1}})}{\kappa_{1}(\kappa_{1}-1)}\Bigg\lbrace\frac{1}{\kappa_{1}-1}+\frac{\alpha\left(-2a\kappa_{1}+\kappa_{1}-2\kappa_{1}y_{1}-1\right)}{4\kappa_{1}}\nonumber \\
 & +\frac{\alpha^{2}\left(2a(4a+4y_{1}-5)-4y_{1}+5 \right)}{32}  \nonumber \\
 & +\frac{\alpha^{2}\left(\kappa_{1}\left(\kappa_{1}^{2}(2a+2y_{1}-1)^{2}+2a-1\right)+3\right)}{32\kappa_{1}^{2}}+O\left(\alpha^{3}\right)\Bigg\rbrace\\
F_{2}(\alpha) & =\frac{2(iS_{\kappa_{1}}+\hat{S}_{\kappa_{1}})^{2}}{(\kappa_{1}-1)^{2}\kappa_{1}^{2}}\Bigg\lbrace1+\frac{\alpha\left(4\kappa_{1}\left(-2\kappa_{1}(a+y_{1})+a+\kappa_{1}+y_{1}-1\right)+2\right)}{8\kappa_{1}}\nonumber \\
 & +\frac{\alpha^{2}\left(-2a(a+y_{1}-1)+2\kappa_{1}-2\kappa_{1}y_{1}^{2}-1\right)}{8}\nonumber \\
 & +\frac{\alpha^{2}\left(\kappa_{1}\left(a(2a-3)\kappa_{1}+\kappa_{1}^{2}(2a+2y_{1}-1)^{2}\right)+3\right)}{8\kappa_{1}}+O\left(\alpha^{3}\right)\Bigg\rbrace\nonumber \\
 & -\frac{iS_{\kappa_{2}}+\hat{S}_{\kappa_{2}}}{(2\kappa_{1}-1)\kappa_{1}}\Bigg\lbrace\frac{1}{2\kappa_{1}-1}-\frac{\alpha\left(\kappa_{1}(4a+4 y_1-2)+1\right)}{8\kappa_{1}}
   +\frac{\alpha^{2}\left( \kappa_{1} (2a+2 y_1-1)^{2} \right)}{16} \nonumber\\
 & +\frac{\alpha^{2}\left(2\kappa_{1}\left(2\kappa_{1}(2a(4a+4 y_1 -5)-4 y_1+5)+2a-1\right)+3\right)}{128\kappa_{1}^{2}}+O\left(\alpha^{3}\right)\Bigg\rbrace
\end{align}

\subsection{Fermionic models}

Similarly to the bosonic case, we can calculate the perturbative part
of the energy density via the appropriately modified Volin's method
. Here we used a code where we directly parametrized the fermionic
kernels in terms of the generic $z_{k}$ variables, and we determined
the energy density: 
\begin{align}
2A_{1,1} & =1+0\cdot v+av^{2}+\left(4a-\frac{19a^{2}}{6}\right)v^{3}+\frac{1}{8}a(a(73a-180)+144)v^{4}\\
 & +v^{5}\left(a^{4}\left(2\zeta_{3}-\frac{51}{2}\right)+\frac{557a^{3}}{6}-\frac{758a^{2}}{5}+24az_{3}+96a\right)\nonumber \\
 & -\frac{1}{6}av^{6}(139a-120)\left(a^{3}\zeta_{3}+12z_{3}\right)\nonumber \\
 & +\frac{1}{72}av^{6}\left(a(a(a(5061a-24254)+60264)-79344)+43200\right)+O\left(v^{7}\right)\nonumber 
\end{align}
Next, using the same steps as for the bosonic models, we determined
the boundary value $a_{1}$ from using (\ref{eq:de1-1}) for $\alpha=\beta=1$
\begin{align}
a_{1} & =1+0\cdot v+\frac{av^{2}}{2}+\left(a-\frac{19a^{2}}{12}\right)v^{3}+\frac{1}{16}a(a(73a-90)+48)v^{4}\\
 & +v^{5}\left(a^{4}\left(\zeta_{3}-\frac{51}{4}\right)+\frac{557a^{3}}{24}-\frac{253a^{2}}{10}+12az_{3}+12a\right)\nonumber \\
 & -\frac{1}{12}av^{6}(139a-60)\left(a^{3}\zeta_{3}+12z_{3}\right)\nonumber \\
 & +\frac{1}{144}av^{6}\left(a(a(a(5061a-12127)+20133)-19872)+8640\right)+O\left(v^{7}\right)\nonumber 
\end{align}
which then determines the universal function via (\ref{eq:de3-1})
for $\alpha=1$ as 
\begin{align}
f & =-4av^{3}+23a^{2}v^{4}-96a^{3}v^{5}+v^{6}\left(-20a^{4}\zeta_{3}+351a^{4}-240az_{3}\right)\\
 & \quad+v^{7}\left(298a^{5}\zeta_{3}-\frac{2389a^{5}}{2}+3576a^{2}z_{3}\right)+O\left(v^{8}\right).\nonumber 
\end{align}
Note that - in contrast to the bosonic case - for $a=0$ this function
vanishes, and all the moments are trivial. This case corresponds to
the $N\to\infty$ limit in the Gross-Neveu and chiral Gross-Neveu
cases. 

The differential equation (\ref{eq:de3-1}) then restores the $\alpha$-dependence
as 
\begin{align}
a_{\alpha} & =1+0\cdot v+\frac{av^{2}}{2\alpha}+\frac{av^{3}(12-19a\alpha)}{12\alpha^{2}}+\frac{av^{4}(a\alpha(73a\alpha-90)+48)}{16\alpha^{3}}\\
 & +\frac{av^{5}\left(a\alpha\left(5a\alpha\left(6a\alpha\left(4\zeta_{3}-51\right)+557\right)-3036\right)+1440\alpha^{3}z_{3}+1440\right)}{120\alpha^{4}}\nonumber \\
 & -\frac{av^{6}\alpha^{3}(139a\alpha-60)\left(a^{3}\zeta_{3}+12z_{3}\right)}{12\alpha^{5}}\nonumber \\
 & +\frac{av^{6}\left(a\alpha(a\alpha(a\alpha(5061a\alpha-12127)+20133)-19872)+8640\right)}{144\alpha^{5}}+O\left(v^{7}\right)\nonumber 
\end{align}
and this is again a formula that can be obtained also directly from
$a_{1}$ using the same rescalings $v\to\frac{v}{\alpha}$, $a\to a\alpha$,
$z_{2k+1}\to\alpha^{2k+1}z_{2k+1}$ as in the bosonic case. The perturbative
expansion of the exceptional boundary value for $\alpha=0$ looks
as
\begin{align}
a_{0} & =H(0)\Biggl\{1-\frac{av}{2}+\frac{5a^{2}v^{2}}{8}-\frac{19a^{3}v^{3}}{16}+v^{4}\left(-\frac{1}{4}a^{4}\zeta_{3}+\frac{323a^{4}}{128}-3az_{3}\right)\nonumber \\
 & \qquad\qquad+v^{5}\left(\frac{53a^{5}\zeta_{3}}{24}-\frac{4349a^{5}}{768}+\frac{53a^{2}z_{3}}{2}\right)+O\left(v^{6}\right)\Biggr\},
\end{align}
which then through (\ref{eq:de1}) determines the also exceptional
moments with $\beta=0$:
\begin{align}
A_{\alpha,0} & =H(0)\Biggl\{\frac{1}{\alpha}-\frac{av}{2\alpha}+\frac{av^{2}(5a\alpha-4)}{8\alpha^{2}}+\frac{av^{3}(a\alpha(80-57a\alpha)-48)}{48\alpha^{3}}\\
 & -\frac{v^{4}\left(a\left(a\alpha\left(a\alpha\left(3a\alpha\left(32\zeta_{3}-323\right)+1904\right)-2256\right)+1152\alpha^{3}z_{3}+1152\right)\right)}{384\alpha^{4}}\nonumber \\
 & +\frac{av^{5}\alpha^{3}(53a\alpha-24)\left(a^{3}\zeta_{3}+12z_{3}\right)}{24\alpha^{5}}+O\left(v^{6}\right)\nonumber \\
 & +\frac{av^{5}\left(a\alpha(100608-5a\alpha(a\alpha(4349a\alpha-10904)+19216))-46080\right)}{3840\alpha^{5}}+O\left(v^{6}\right)\Biggr\}\nonumber 
\end{align}
and $\alpha,\beta=0$:

\begin{align}
A_{0,0}=H(0)^{2}\Biggl\{\frac{1}{2v}+0-\frac{a^{2}v}{4}+\frac{3a^{3}v^{2}}{8}+v^{3}\left(\frac{1}{48}a^{4}\left(4\zeta_{3}-29\right)+az_{3}\right)\nonumber \\
+v^{4}\left(\frac{1}{192}a^{5}\left(209-100\zeta_{3}\right)-\frac{25a^{2}z_{3}}{4}\right)+O\left(v^{5}\right).\Biggr\}
\end{align}
Finally the same differential equation for generic $\alpha,\beta$
determines 
\begin{align}
A_{\alpha,\beta} & =\frac{1}{\alpha+\beta}+0\cdot v+\frac{av^{2}}{2\alpha\beta}+\frac{av^{3}(\alpha(12-19a\beta)+12\beta)}{12\alpha^{2}\beta^{2}}\\
 & +\frac{av^{4}\left(\alpha^{2}(a\beta(73a\beta-90)+48)+6\alpha\beta(8-15a\beta)+48\beta^{2}\right)}{16\alpha^{3}\beta^{3}}\nonumber \\
 & +\frac{av^{5}\left(1440\alpha^{2}+\beta^{2}(a\alpha(2785a\alpha-3036)+1440)+144\alpha\beta(10-21a\alpha)\right)}{120\alpha^{4}\beta^{3}}\nonumber \\
 & +\frac{av^{5}\left(a\beta\left(5a\beta\left(6a\beta\left(4\zeta_{3}-51\right)+557\right)-3036\right)+1440\beta^{3}z_{3}+1440\right)}{120\alpha\beta^{4}}\nonumber \\
 & -\frac{av^{6}(\alpha(139a\beta-60)-60\beta)\left(a^{3}\zeta_{3}+12z_{3}\right)}{12\alpha^{2}\beta^{2}}\nonumber \\
 & +\frac{av^{6}\left(\alpha^{2}(a\beta(2237a\beta-2200)+960)+96\alpha\beta(10-23a\beta)+960\beta^{2}\right)}{16\alpha^{5}\beta^{3}}\nonumber \\
 & +\frac{av^{6}(a\beta(a\beta(19998-12127a\beta)-19800)+8640)}{144\alpha^{2}\beta^{4}}\nonumber \\
 & +\frac{av^{6}(a\beta(a\beta(a\beta(5061a\beta-12127)+20133)-19872)+8640)}{144\alpha\beta^{5}}+O\left(v^{7}\right)\nonumber 
\end{align}
using solely the expression of $a_{\alpha}$. The capacity $C^{(-)}$ starts as 
\begin{align}
C_{0}^{(-)} & =1+s(\mathsf{L}+1)+\frac{1}{2}s^{2}(2\mathsf{L}+1)+\frac{1}{4}s^{3}(1-2(\mathsf{L}-1)\mathsf{L})\nonumber \\
 & +\frac{1}{24}s^{4}\left(-24\zeta_{3}+8(\mathsf{L}-3)\mathsf{L}^{2}+3\right)+O\left(s^{5}\right)\\
C_{1}^{(-)} & =2i+2is-is^{2}(2\mathsf{L}+1)+\frac{1}{3}is^{3}\left(6\mathcal{\mathsf{L}}^{2}+1\right)\nonumber \\
 & -\frac{1}{6}is^{4}(-36\zeta_{3}+6\mathsf{L}(2(\mathsf{L}-1)\mathsf{L}+1)+13)+O\left(s^{5}\right)\\
C_{2}^{(-)} & =1+s+\frac{1}{2}s^{2}(1-2\mathsf{L})+\frac{1}{3}s^{3}(3(\mathsf{L}-2)\mathsf{L}-5)\nonumber \\
 & +\frac{1}{24}s^{4}(72\zeta_{3}-24\mathsf{L}((\mathsf{L}-4)\mathsf{L}-3)+89)+O\left(s^{5}\right)\\
C_{3}^{(-)} & =i+is-\frac{1}{2}is^{2}(2\mathsf{L}+1)+\frac{1}{6}is^{3}\left(6\mathsf{L}^{2}+7\right)\nonumber \\
 & -\frac{1}{6}is^{4}(-18\zeta_{3}+3\mathsf{L}(2(\mathsf{L}-1)\mathsf{L}+7)+32)+O\left(s^{5}\right)\\
C_{4}^{(-)} & =2+2s+\frac{1}{2}s^{2}(1-4\mathsf{L})+\frac{1}{3}s^{3}\left(6\mathsf{L}^{2}-9\mathsf{L}+1\right)\nonumber \\
 & +\frac{1}{12}s^{4}(72\zeta_{3}-6\mathsf{L}(\mathsf{L}(4\mathsf{L}-13)+8)-59)+O\left(s^{5}\right)
\end{align}
where $\mathsf{L}\equiv-\ln(2s)$ for the $^{(-)}$ case.

Observables with bar are as follows:
\begin{align}
\bar{A}_{\alpha,\beta} & =\frac{1}{\alpha+\beta}-\frac{av^{2}}{2(\alpha\beta)}+\frac{av^{3}(\alpha(17a\beta-12)-12\beta)}{12\alpha^{2}\beta^{2}}\\
 & +\frac{av^{4}\left(\alpha^{2}(a\beta(86-63a\beta)-48)+2\alpha\beta(43a\beta-24)-48\beta^{2}\right)}{16\alpha^{3}\beta^{3}}\nonumber \\
 & -\frac{v^{5}a\left(\alpha^{3}\left(30a^{3}\beta^{3}\left(4\zeta_{3}-43\right)+2605a^{2}\beta^{2}-2964a\beta+1440\left(\beta^{3}z_{3}+1\right)\right)\right)}{120\alpha^{4}\beta^{4}}\nonumber \\
 & -\frac{v^{5}a\left(\alpha^{2}\beta(a\beta(2605a\beta-2976)+1440)+12\alpha\beta^{2}(120-247a\beta)+1440\beta^{3}\right)}{120\alpha^{4}\beta^{4}}\nonumber \\
 & +\frac{av^{6}\left(-3\alpha^{2}\beta^{2}(a\beta(6463a\beta-6552)+2880)\right)}{144\alpha^{5}\beta^{5}}\nonumber \\
 & +\frac{av^{6}\left(180a^{3}\alpha^{3}\beta^{3}\zeta_{3}(\alpha(9a\beta-4)-4\beta)+576\alpha\beta^{3}(34a\beta-15)\right)}{144\alpha^{5}\beta^{5}}\nonumber \\
 & +\frac{av^{6}\left(a\beta(a\beta(-1399a\beta(3a\beta-8)-19389)+19584)+2160\beta^{3}z_{3}(9a\beta-4)-8640\right)}{144\alpha\beta^{5}}\nonumber \\
 & +\frac{av^{6}\left(2\alpha^{3}\beta\left(a\beta(a\beta(5596a\beta-9759)+9828)-4320\beta^{3}z_{3}-4320\right)-8640\beta^{4}\right)}{144\alpha^{5}\beta^{5}}+O\left(v^{7}\right)\nonumber 
\end{align}
and
\begin{align}		
\bar{a}_{\alpha} & =1-\frac{av^{2}}{2\alpha}+\frac{av^{3}(17a\alpha-12)}{12\alpha^{2}}+\frac{av^{4}(a\alpha(86-63a\alpha)-48)}{16\alpha^{3}}\\
 & -\frac{v^{5}\left(a\left(30a^{3}\alpha^{3}\left(4\zeta_{3}-43\right)+2605a^{2}\alpha^{2}-2964a\alpha+1440\left(\alpha^{3}z_{3}+1\right)\right)\right)}{120\alpha^{4}}\nonumber \\
 & +\frac{av^{6}\left(3a^{4}\alpha^{4}\left(540\zeta_{3}-1399\right)+8a^{3}\alpha^{3}\left(1399-90\zeta_{3}\right)-19389a^{2}\alpha^{2}\right)}{144\alpha^{5}}\nonumber \\
 & +\frac{av^{6}\left(144a\alpha\left(135\alpha^{3}z_{3}+136\right)-8640\left(\alpha^{3}z_{3}+1\right)\right)}{144\alpha^{5}}+O\left(v^{7}\right)\nonumber,
\end{align}
while the free energy density in the running coupling looks like:
\begin{align}
F_{0}(\alpha) & =1-\frac{a\alpha}{2}+\frac{1}{8}a\alpha^{2}(3a+2y_{1}-4)+O\left(\alpha^{3}\right)\\
F_{1}(\alpha) & =-\frac{2(iS_{\kappa_{1}}+\hat{S}_{\kappa_{1}})}{\kappa_{1}(\kappa_{1}-1)}\left\{ \frac{1}{\kappa_{1}-1}+\frac{a}{2}\alpha+\frac{1}{8}\alpha^{2}a\left(a\left(\kappa_{1}-3\right)-2y_{1}+2\right)+O\left(\alpha^{3}\right)\right\} \\
F_{2}(\alpha) & =\frac{2(iS_{\kappa_{1}}+\hat{S}_{\kappa_{1}})^{2}(2\kappa_{1}-1)}{\kappa_{1}^{2}(\kappa_{1}-1)^{2}}\Bigg\lbrace\frac{1}{2\kappa_{1}-1}+\frac{a\alpha}{2}\nonumber \\
 & +\frac{1}{8}a\alpha^{2}\left(-\frac{1}{\kappa_{1}(2\kappa_{1}-1)}+a\left(2\kappa_{1}-3\right)-2y_{1}+2\right)+O\left(\alpha^{3}\right)\Bigg\rbrace+\nonumber \\
 & -\frac{iS_{\kappa_{2}}+\hat{S}_{\kappa_{2}}}{(2\kappa_{1}-1)\kappa_{1}}\left\{ \frac{1}{2\kappa_{1}-1}+\frac{a\alpha}{2}+\frac{1}{8}a\alpha^{2}\left(a\left(2\kappa_{1}-3\right)-2y_{1}+2\right)+O\left(\alpha^{3}\right)\right\} .
\end{align}

\bibliographystyle{utphys}
\bibliography{paper}

\end{document}